%% file: cas-manuscript.tex
\documentclass[a4paper,fleqn]{cas-dc}

\usepackage[numbers]{natbib}
\usepackage{amsmath}

\def\tsc#1{\csdef{#1}{\textsc{\lowercase{#1}}\xspace}}
\tsc{WGM}
\tsc{QE}
\tsc{EP}
\tsc{PMS}
\tsc{BEC}
\tsc{DE}
\begin{document}
\let\WriteBookmarks\relax
\def\floatpagepagefraction{1}
\def\textpagefraction{.001}
\shorttitle{Co-exploration of racetrack memory based CNN inference}
\shortauthors{Choong et. al.}

\title [mode = title]{Hardware-software co-exploration with racetrack memory based in-memory computing for CNN inference in embedded systems}

\address[1]{Department of Electrical and Computer Engineering, National University of Singapore, 4 Engineering Drive 3, Singapore 117583.}
\address[2]{Institute of High Performance Computing, Agency for Science, Technology and Research. 1 Fusionopolis Way, \#16-16 Connexis, Singapore 138632.}
\address[3]{Institute of Computing Technology, Chinese Academy of Sciences. 6 Kexueyuan South Road, Zhongguancun, Haidian District, Beijing 100190, China.}
\address[4]{School of Computing, National University of Singapore. COM1, 13, Computing Dr, Singapore 117417.}
\address[5]{Hong Kong University of Science and Technology, Clear Water Bay, Kowloon, Hong Kong.}

\author[1]{Benjamin Chen Ming Choong}
\author[2]{Tao Luo} \cormark[1] \ead{leto.luo@gmail.com}
\author[3]{Cheng Liu}
\author[4]{Bingsheng He}
\author[5]{Wei Zhang}
\author[2]{Joey Tianyi Zhou}

\cortext[cor1]{Corresponding author}

\begin{abstract}
Deep neural networks generate and process large volumes of data, posing challenges for low-resource embedded systems. In-memory computing has been demonstrated as an efficient computing infrastructure and shows promise for embedded AI applications. Among newly-researched memory technologies, racetrack memory is a non-volatile technology that allows high data density fabrication, making it a good fit for in-memory computing. However, integrating in-memory arithmetic circuits with memory cells affects both the memory density and power efficiency. It remains challenging to build efficient in-memory arithmetic circuits on racetrack memory within area and energy constraints. To this end, we present an efficient in-memory convolutional neural network (CNN) accelerator optimized for use with racetrack memory. We design a series of fundamental arithmetic circuits as in-memory computing cells suited for multiply-and-accumulate operations. Moreover, we explore the design space of racetrack memory based systems and CNN model architectures, employing co-design to improve the efficiency and performance of performing CNN inference in racetrack memory while maintaining model accuracy. Our designed circuits and model-system co-optimization strategies achieve a small memory bank area with significant improvements in energy and performance for racetrack memory based embedded systems. 
\end{abstract}


\maketitle

%% INTRODUCTION
\input{tex/1_Introduction}

%% BACKGROUND
\input{tex/2_Background}

%% PROPOSED ARITHMETIC CIRCUITS
\input{tex/3_Proposed_Arithmetic_Circuits}

%% IN-MEMORY ACCELERATOR ARCHITECTURE
\input{tex/4_In-Memory_Accelerator_Architecture}

%% EVALUATION
\input{tex/5_Experimental_Results}

%% RELATED WORKS
\input{tex/6_Related_Works}

%% CONCLUSION
\input{tex/7_Conclusion}

\bibliographystyle{cas-model2-names}

\bibliography{cas-refs}

\end{document}

%% file: tex/1_Introduction.tex
%%%%%%%%%%%%%%%%%%%%%%%%%%%%%%%%%%%%%%%%%%%%%%
\section{Introduction}

%Deep neural networks (DNNs) have proven highly effective at modeling complex data, demonstrating state-of-the-art capabilities in multiple domains \cite{Hannun2019, ResNet20, LeCun2015DeepLearning, Silver2016Go}. 
%However, achieving high performance and accuracy requires large amounts of data to be created, processed, and transferred, typically in the order of tens to hundreds of megabytes (MB) or more \cite{Chen2017}. 
%As a result, DNN training and inference are computationally expensive and memory intensive processes. 
Deep neural network (DNN) training and inference are computationally expensive and memory intensive processes. 
The large volume of data generated during an inference pass incur significant latency and energy costs, posing a obstacle to implementation on memory- and power-constrained devices. 
Conventionally, DNN accelerators are constrained by a tight on-chip memory budget, and thus require an off-chip memory such as cache or DRAM for intermediate storage. However, off-chip data transfers largely dominate latency and power costs \cite{Chen2017}, motivating new optimizations in both model and hardware domains to reduce the memory accesses required for inference.

Model optimizations such as parameter pruning \cite{Kang2020Prune, EnergyAwarePruning2017}, parameter repetition \cite{UCNN2018}, and model quantization \cite{Jacob2018, LogNet2017} adjust the network structure and precision to produce smaller, less complex models. Smaller size and precision can reduce the amount of data transfers and improve computational efficiency, but at the cost of accuracy loss. Furthermore, the savings from model optimizations can only be realized when executed on appropriate hardware platforms that exploit model sparsity and reduced precision as well.

On the other hand, hardware optimizations aim for efficient, high-performance systems to perform model inferences. At the circuit level, computing engines are designed to efficiently process reduced or varying data bit-widths \cite{kim2020exploiting, moons2016energy, sharma2018bit}. System-level design strategies include improving data reuse through dataflow mapping \cite{Chen2017, ESSA2020, kwon2021heterogeneous}, and compression sparse data for off-chip transfers \cite{aimar2018nullhop, Chen2017,  zhu2020efficient}. These DNN accelerator designs improve the memory bottleneck and computing overhead significantly, but inevitably require the use of off-chip memory. For example, with less than 1 MB of on-chip memory, the accelerator chips Eyeriss \cite{Chen2017} and ESSA \cite{ESSA2020} still require several MB of costly off-chip transfers even after aggressive compression and data reuse. Moreover, on-chip memory technologies are volatile and require power to correctly retain their data.   

In order to mitigate the above overhead costs, \textit{in-memory computing} has emerged as a promising computing infrastructure for big data processing~\cite{chen2018gflink,chen2018flinkcl, chen2017gpu}. With in-memory computing, computation is executed in memory, avoiding data transfers to and from the processor for simple arithmetic operations. Such capabilities can dramatically reduce the latency and power consumption of many memory-intensive applications including DNN inference. 

As the medium for both storage and logic, the memory technology used has critical influence over the performance of an in-memory computing system. Racetrack memory (RM) is an emerging technology with great promise due to its high data density, high speed, low power, and non-volatility \cite{Parkin2008}. In addition to its potential for adoption across all levels of the memory hierarchy, racetrack memory also demonstrates CMOS compatibility \cite{Lin2009} and can be applied to in-memory logic design. However, there is a lack of efficient general arithmetic circuits using racetrack memory. Furthermore, the direct application of conventional memory structures and data layouts to racetrack memory leads to sub-optimal results: racetrack memory achieves high storage density using a shifting access mechanism, but the access patterns of convolutional neural networks (CNN) result in a great degree of redundant shifting, degrading performance \cite{wang2020automatic}. Hence, the targeted design of RM-based logic and systems would present opportunities for exploiting the intrinsic properties of RM technology for improved performance and efficiency.  

In this work, we present an in-memory CNN accelerator architecture based on racetrack memory. We propose RM-based arithmetic units for in-memory computation, and study the design space of model and system architectures for accurate and efficient CNN inference. Our findings suggest that the co-optimization of the model and system for racetrack memory can enable the memory technology to be exploited for significant energy and latency savings.    

First, we build in-memory multipliers for Booth multiplication and shift-based multiplication using RM-based basic arithmetic units.
Then, we present a novel energy optimization technique for RM-based logic. On the observation that the write operation consumes an order of magnitude greater energy than the shift operation, we introduce a novel energy optimization by transforming operand write operations to shift operations in our arithmetic circuits. This write-shift transformation reduces energy consumption of Booth multiplication by up to 94.4\%. 

Lastly, we demonstrate how these units are used to accelerate multiply-and-accumulate (MAC) operations, which are dominant in convolution and fully-connected layers of CNN models. We exploit the structural layout of the RM cell unit to maximize data reuse and parallelism. Furthermore, we utilize the shifting nature of racetrack memory to accelerate logarithmically-quantized (shift-based) models. We find that combining the optimizations yields up to 83.5$\times$ improved energy efficiency and 1.68$\times$ better performance for 8-bit CNN models. 

The paper is organized as follows. In Section 2, we provide the background of racetrack memory and convolutional neural networks. Section 3 presents the proposed circuit designs for RM based adders and multipliers. Section 4 details the accelerator architecture, including the mapping of CNN operations and model data. Section 5 presents and analyzes experimental results, and discusses the hardware-software design space of the RM system. Section 6 reviews related works, whereas Section 7 concludes the paper.

%% file: tex/2_Background.tex
\section{Background}
\subsection{Racetrack Memory} \label{RacetrackMemory}
Racetrack memory is a non-volatile memory technology developed by a team led by Stuart Parkin in early 2008 \cite{Parkin2008}. In racetrack memory, data is stored in strips of nanowire, which are like racetracks of data. Each nanowire comprises short, magnetized units termed domains, each of which store a bit of information. 

\begin{figure}
	\centering
		\includegraphics[scale=.24]{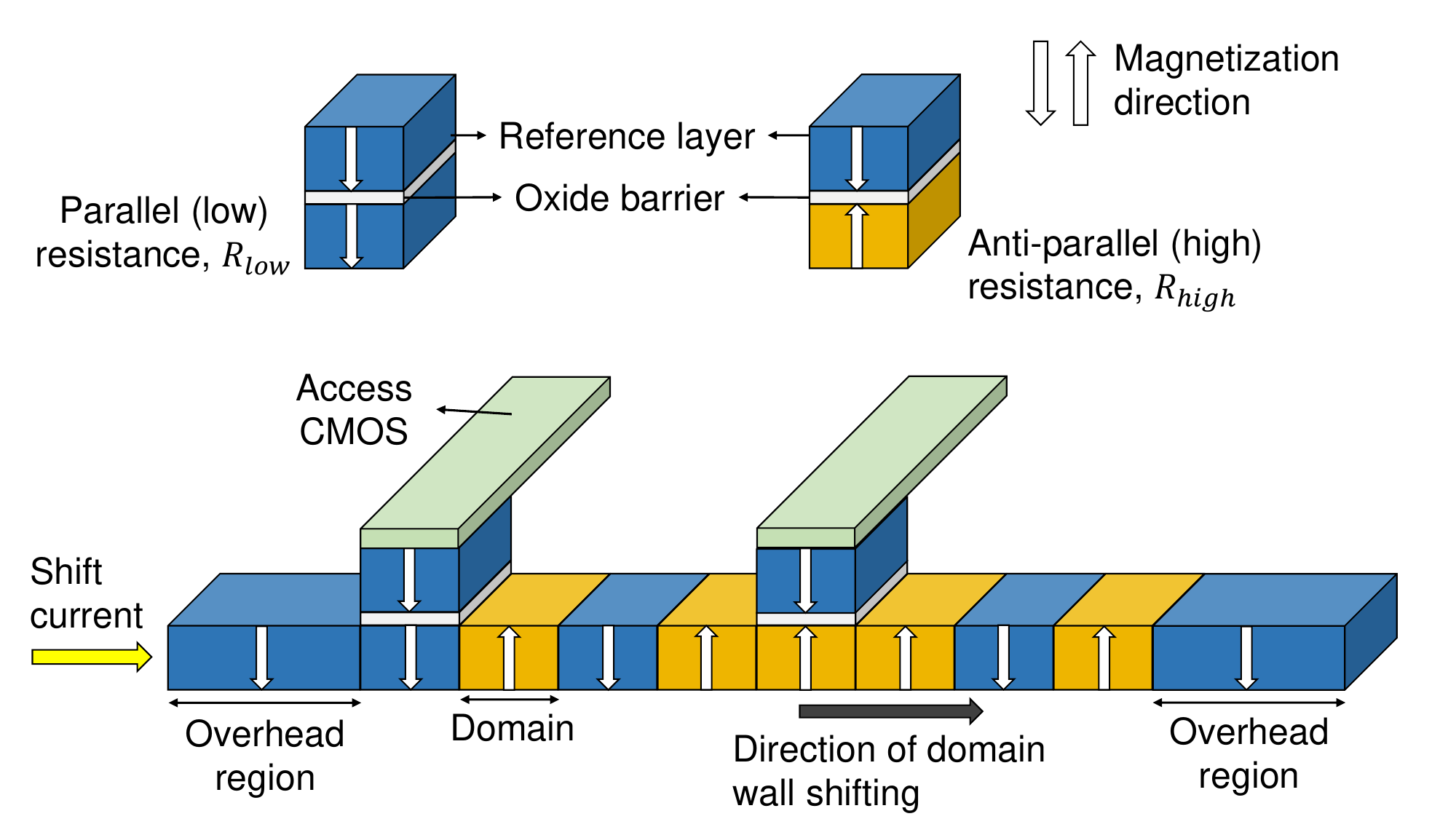}
	\caption{Basic structure of vertical magnetic tunnel junction (top) and magnetic nanowire of racetrack memory (bottom).}
	\label{fig:1}
\end{figure}

Fig. \ref{fig:1} shows the structure of a magnetic strip of racetrack memory. As shown in the figure, each domain is magnetized in a certain direction. The magnetization direction can be sensed or changed using magnetic tunnel junctions (MTJ), which comprise two ferromagnetic layers separated by an oxide barrier. One of the layers is the reference layer, which has a fixed magnetization direction, while the other layer has a free direction. When the magnetization directions of the two layers are the same, the MTJ has low electrical resistance (denoted as $R_{low}$ in this paper); otherwise, the resistance is high ($R_{high}$). In this manner, the magnetization direction of the free layer can be used to denote two distinct states. In racetrack memory, the domains form the free ferromagnetic layer. Thus, the magnetization directions of the domains are used to represent the "0" and "1" state in a computer system for data storage. For example, if $R_{low}$ represents "0" and $R_{high}$ represents "1", the strip in Fig. \ref{fig:1} stores the bits "01011101". 

The domains in a nanowire are divided by non-magnetic regions called domain walls (DW). By applying pulses of current at the end of the magnetic stripe, the domain walls can be moved in the direction of the current, shifting the domains along the strip. In effect, a racetrack memory strip acts as a shift register \cite{Parkin2008}. With very short domain widths (45-100nm), racetrack memory achieves high storage density, showing promise as an efficient non-volatile storage medium.

Racetrack memory typically performs three basic operations: read, write, and shift. A combination of these fundamental operations can be used to access data or even facilitate arithmetic operations involving shifting. Fig. \ref{fig:1} depicts a racetrack memory strip with two access ports, with each port comprising one MTJ and an access CMOS. The write operation is performed by applying a high current through the the domain under the access CMOS to set its magnetization direction. Conversely, the read operation reads a bit value by measuring the resistance of the domain below the CMOS. The shift operation involves applying a shift current to move the domain walls along the direction of the current. While racetrack memory achieves high storage density, the domains must be shifted below an access port before it can be read or written, incurring overhead in access latency and energy. Hence, careful design and data mapping strategies should therefore be employed to reduce the costs of this sequential access mechanism.

Furthermore, additional overhead regions must be added to both ends of the racetrack strip to facilitate shifting. These overhead domains (depicted in Fig. \ref{fig:1}) do not store data; instead, data-storing domains can be shifted into these overhead regions during shifting, avoiding data loss. The length of the overhead regions required is determined by the number of domains sharing an access port. When more domains share an access port, the number of access ports is reduced but the overhead region length needed increases, presenting an additional design trade-off.

\begin{figure}
	\centering
		\includegraphics[scale=.24]{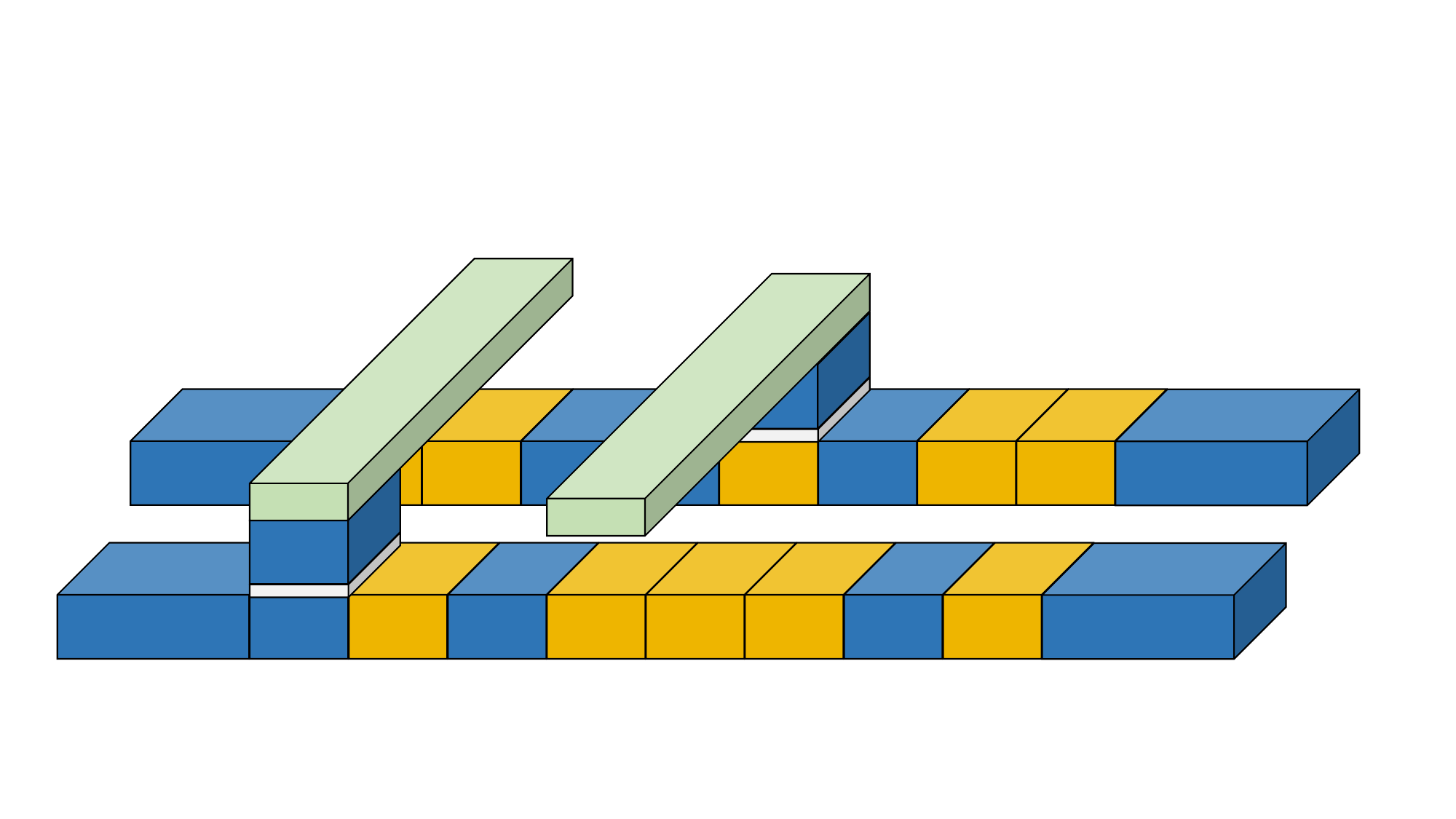}
	\caption{The Macro Unit (MU) structure which interleaves racetrack memory strips and access CMOS for better area utilization. Here, $N_{dom} = 8$, $N_{ports} = 2$ and $N_{tracks} = 2$.}
	\label{fig:2}
\end{figure}

In addition to shifting overhead, the access ports and shift transistors can also result in poor area utilization, as these transistors are typically wider than the nanowire. To improve area efficiency, \citet{Chao2015} proposed a racetrack memory layout called the Macro Unit (MU) which overlaps and interleaves racetrack memory strips and CMOS transistors for more compact placement.  An MU can be characterised by three parameters: the number of racetrack strips ($N_{tracks}$), number of ports ($N_{ports}$), and the number of domains ($N_{dom}$) in each racetrack. Fig. \ref{fig:2} shows the layout of an MU comprising two racetracks and two ports, with each racetrack having eight domains. 

Several prior works \cite{DWMAcc2019, Hu2016} have used the MU or similar layouts as the building block for memory systems. \citet{Hu2016} investigated the effects of MU parameters on a proposed racetrack memory-based main memory, and found that maximising the number of racetracks and number of domains leads to better MU layout efficiency. The number of ports in an MU is selected to balance between access transistor size and overhead region size. In this study, we adopt the MU configuration proposed in \cite{Hu2016} with $N_{dom} = 64$, $N_{ports} = 16$, and $N_{tracks} = 4$ that reduces access overheads with minimal power and area costs incurred.   

\subsection{Convolutional Neural Networks (CNN)} \label{ConvolutionalNeuralNetworks}

\subsubsection{Computation in CNN}

\begin{figure}
	\centering
		\includegraphics[scale=.24]{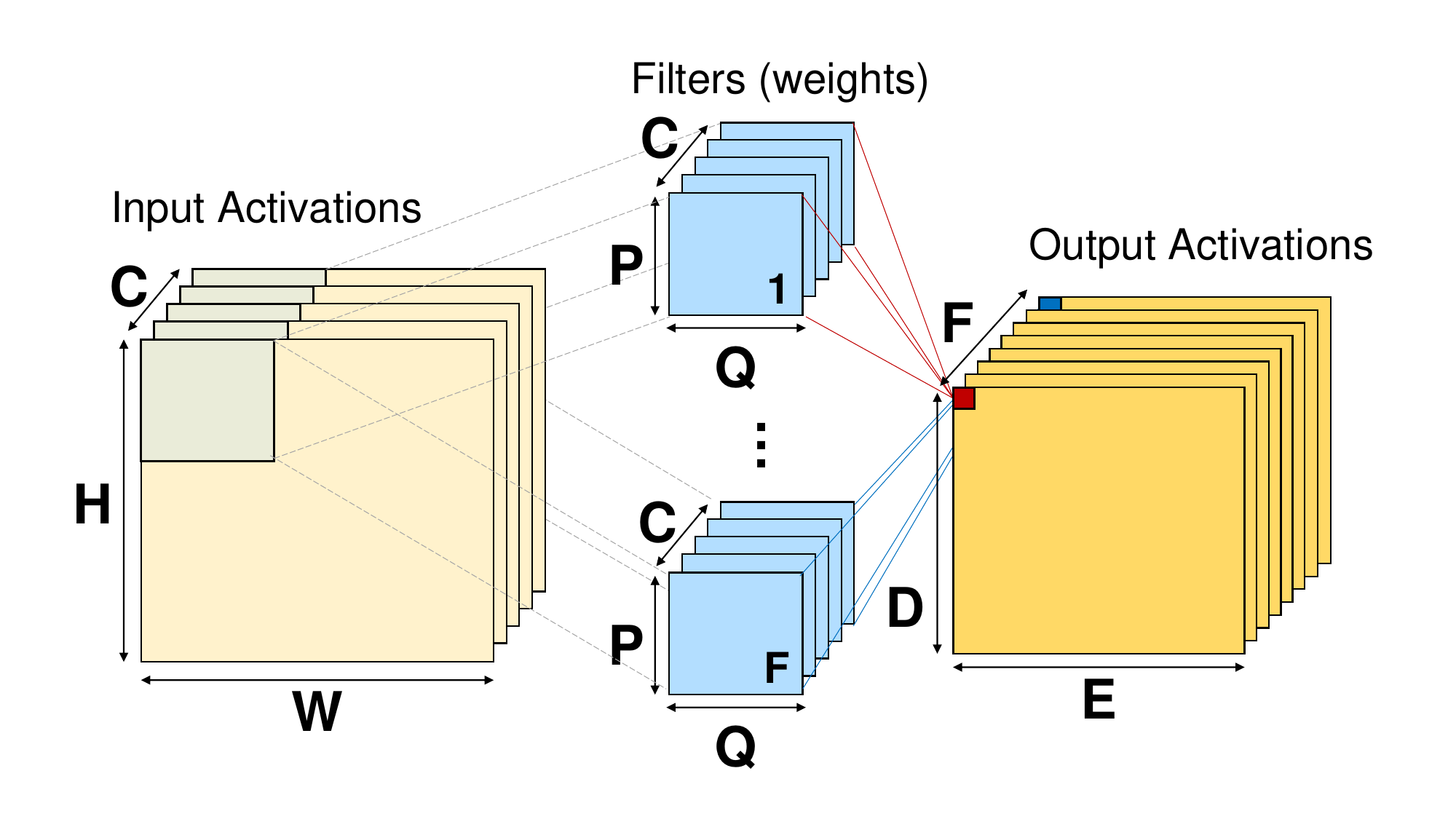}
	\caption{Visualization of a convolution layer. The result of MAC operations of the Filter 1 at the highlighted input position is the red output activation in output channel 1, whereas that of Filter F is the blue output activation in output channel F.}
	\label{fig:3}
\end{figure}

The largest and most computation-intensive layers of a CNN are typically the convolution layers, in which high-dimensional convolutions are performed. Fig. \ref{fig:3} visualizes a convolution layer. In the layer depicted in Fig. \ref{fig:3}, the 3D input has width of $W$, height of $H$ and $C$ channels. Thus, the number of activations in the input is \(C*W*H\). The input is convolved with different filters, each filter of which has width of $Q$, height of $P$ and equal number ($C$) of channels. Each filter thus has \(C*Q*P\) weights. A filter is overlaid with the input at a position, and each weight in the filter is multiplied with the input value it overlaps. All the products are added together along with a bias value, and passed through a non-linear function (such as ReLU) to yield a single output activation. This filter then slides along the input activations to produce multiple outputs in a single output channel (shaded in Fig. \ref{fig:3}). An output channel depicted in Fig. \ref{fig:3} has a width of $E$ activations and a height of $D$. The input activation is convolved with $F$ filters to produce $F$ output channels. The computation of this convolutional layer is described in Eq. \ref{eq:1}, in which $O$, $B$, $I$ and $W$ are matrices of output activations, filter bias values, input activations and filters respectively.

\begin{equation} \label{eq:1}
\small
\begin{split}
&\mathbf{O}[\text{f}][d][e] = \\
&\text{ReLU}\bigg(\mathbf{B}[\text{f}] + \sum_{c=0}^{C-1} \sum_{p=0}^{P-1} \sum_{q=0}^{Q-1}
    \mathbf{I}[c][Ud+p][Ue+q]\; \times\\
    &\mathbf{W}[\text{f}][c][p][q] \bigg),\\
   &0\leq\text{f}<F,\;\; 0\leq d<D,\;\; 0\leq e<E,\\
   &D=\frac{(H-P+U)}{U},\;\; E=\frac{(W-Q+U)}{U}
\end{split}
\end{equation}

\subsubsection{Linear and Logarithmic Quantization}
As observed in Eq. \ref{eq:1}, the majority of operations in processing a CNN layer are multiplications, which are more costly in terms of performance and computational resources. In view of this, various works have proposed optimizations to reduce computational complexity. Model quantization is a widely used strategy in which the precision of weights and/or activations are reduced. Reducing precision can simplify arithmetic computation as fewer bits must be processed in during multiplication or addition. Furthermore, data with smaller bit-width also require less storage and memory transfers, reducing access energy costs and allowing more data to be stored in memory. 

CNN models are typically developed in floating-point representation, where data is encoded as a fixed number of significant digits and an exponent. However, floating-point computation is more resource-costly than fixed-point representation, in which all bits of the data is encoded in the same number base (base 2 for binary) without exponent bits. Linear quantization converts model parameters and activations to fixed-point format, and can be done post-training, or during model training itself \cite{han2016deep, Jacob2018}. Today, 16-bit fixed point and 8-bit fixed point data representations have demonstrated sufficient accuracy for modern CNNs and are widely adopted for acceleration.

In addition, a class of studies on quantization exploit the fact that multiplication by power-of-two values can be reduced to shift operations, which are computationally less expensive. In two's complement representation, binary numbers can be multiplied by power-of-two values by shifting the bits by $Exp_{2}$ positions to the right or left, where $Exp_{2}$ is the power. For example, multiplying -3 (binary form: "11111101") by 4 ($Exp_{2} = 2$) can be achieved by shifting the binary word two positions to the left and shifting in zeroes, resulting in "11110100" which is -12. In another example, multiplying -4 (binary form: "11111100" by 0.5 ($Exp_{2} = -1$) is equal to a right shift by 1 position. Importantly, for signed numbers, the most significant bit should be shifted in from the left instead of zero to preserve the correct sign. Hence, performing a right shift by one position on -4 will yield "11111110", which is -2.   

In \cite{LogNet2017}, weights are logarithmically quantized, constraining all weight values to powers of two. An example is shown in Eq. \ref{eq:2}, in which the filter matrix is \(W = (2^{-2}, 2^0, 2^3) \). Multiplication with these three weight values can be converted to right shift by two, no shift applied, and left shift by three respectively. In \cite{LightNN2018, INQ2017}, weight values are constrained to be a combination of one or two power-of-two values, hence multiplication is similarly reduced to a limited number of shift-and-add operations. 
Across these works, quantizing weights with widths of 4 bits (corresponding to $\pm7$ shift range) or greater could achieve high inference accuracy, with top-5 accuracy loss of less than 5\% from that of full-precision floating point models. 
Such shift-based neural networks therefore present opportunities for improving performance and energy efficiency while maintaining accuracy. 

\begin{align}
\small
I \otimes W = \sum_{i=0}^{2}(I_i \times W_i) & = (I_0 \times 2^{-2}) + I_1 + (I_2 \times 2^{3}) \nonumber \\
 & = (I_0 >> 2) + I_1 + (I_2 << 3)
\label{eq:2} 
\end{align}

In order to gain computational savings, these shift-and-add operations must be executed on shifting logic rather than full-precision multipliers. As racetrack memory offers natural support for shift operations, this work explores an efficient implementation of shift-based models as well. The shifting circuitry of racetrack memory can align data under the read port such that the correctly-shifted product is obtained during access. In this manner, products can be accumulated immediately upon access. Our proposed shift-based MAC unit is presented in Section \ref{Shift-based Multiplier}.

%% file: tex/3_Proposed_Arithmetic_Circuits.tex
\section{Proposed Arithmetic Circuits} \label{ProposedArithmeticCircuits}
In this section, we detail the proposed designs of general arithmetic circuits on racetrack memory. The proposed circuits are building blocks for implementing MAC operations with high performance and efficiency on racetrack memory. First, we illustrate energy-efficient MTJ-based half adder and full adder circuits~\cite{luo2020energy, luo2016racetrack}. These circuits are used to implement both bit-serial and ripple carry adder units. Next, we utilize the adder designs to develop bit-serial Booth multiplication and a shift-based MAC units. 

\subsection{Racetrack Memory Based Adders}\label{Adder}
\begin{figure}
	\centering
		\includegraphics[scale=.24]{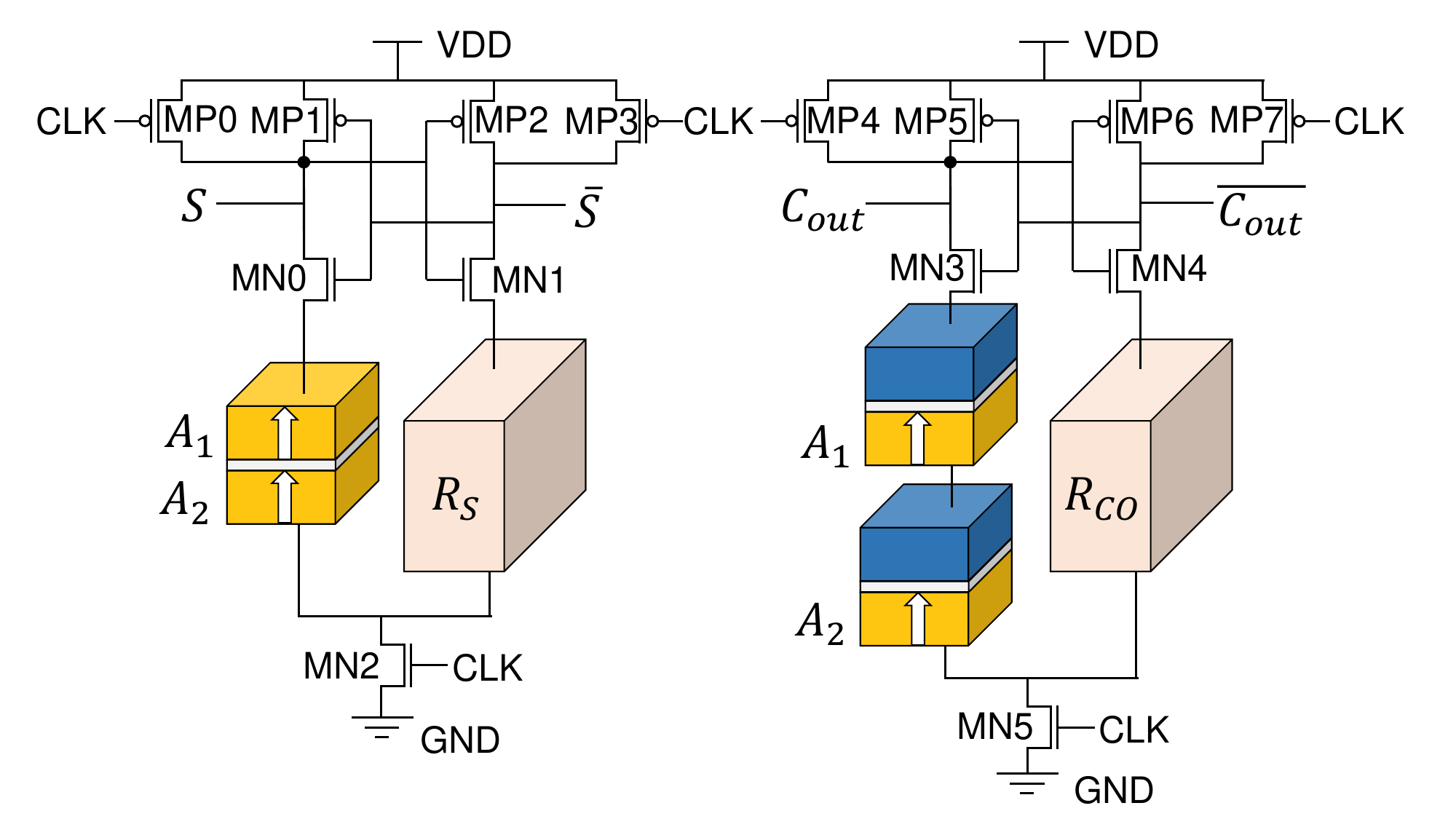}
	\caption{Schematic of the proposed half adder.}
	\label{fig:4}
\end{figure}

\begin{figure}
	\centering
		\includegraphics[scale=.22]{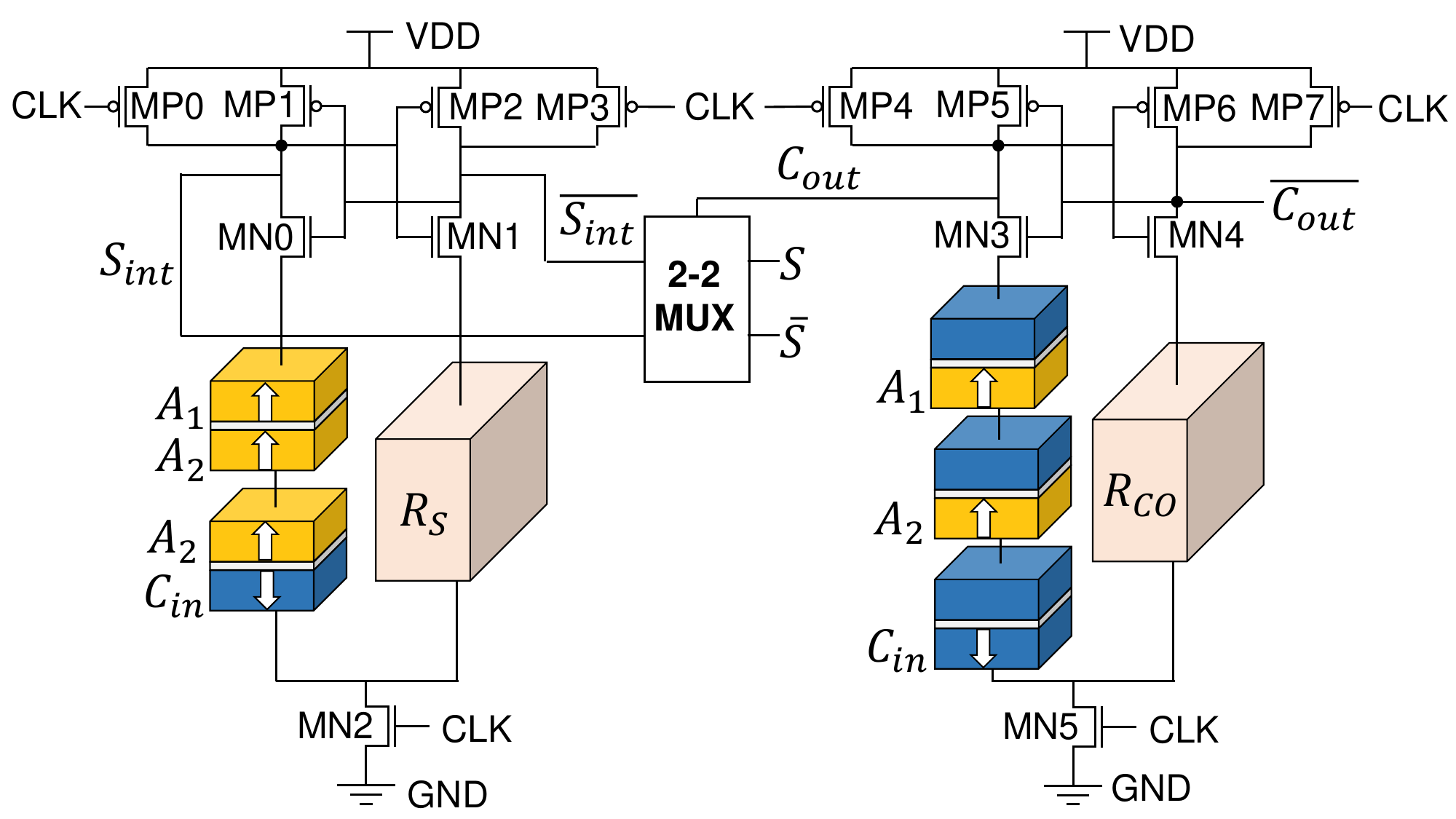}
	\caption{Schematic of the proposed full adder.}
	\label{fig:5}
\end{figure}

Adders are the fundamental building components of arithmetic operators. Thus, the area and energy efficiency of adder units are vital to in-memory computing efficiency. Circuits of the single-bit half adder and full adder are shown in Fig. \ref{fig:4} and Fig. \ref{fig:5}, respectively. The logic functions of a half adder are given by Eq. \ref{eq:3} and \ref{eq:4}, where $A_1$ and $A_2$ are the two addends, $S$ is the result bit, and $C_{out}$ is carry-out signal.

\begin{align} 
\small
S = A_1 \oplus A_2 \label{eq:3} \\
C_{out} = A_1 \cdot A_2 \label{eq:4}
\end{align}

These circuits are designed to reduce the number of MTJs in the adder, and thus the number of costly MTJ write operations needed to transfer inputs. Compared to a magnetic full adder (MFA) of a previous work~\cite{Trinh2013}, our full adder circuit reduces the MTJ writes in an addition from 16 to 7.

\begin{figure}
	\centering
		\includegraphics[scale=.24]{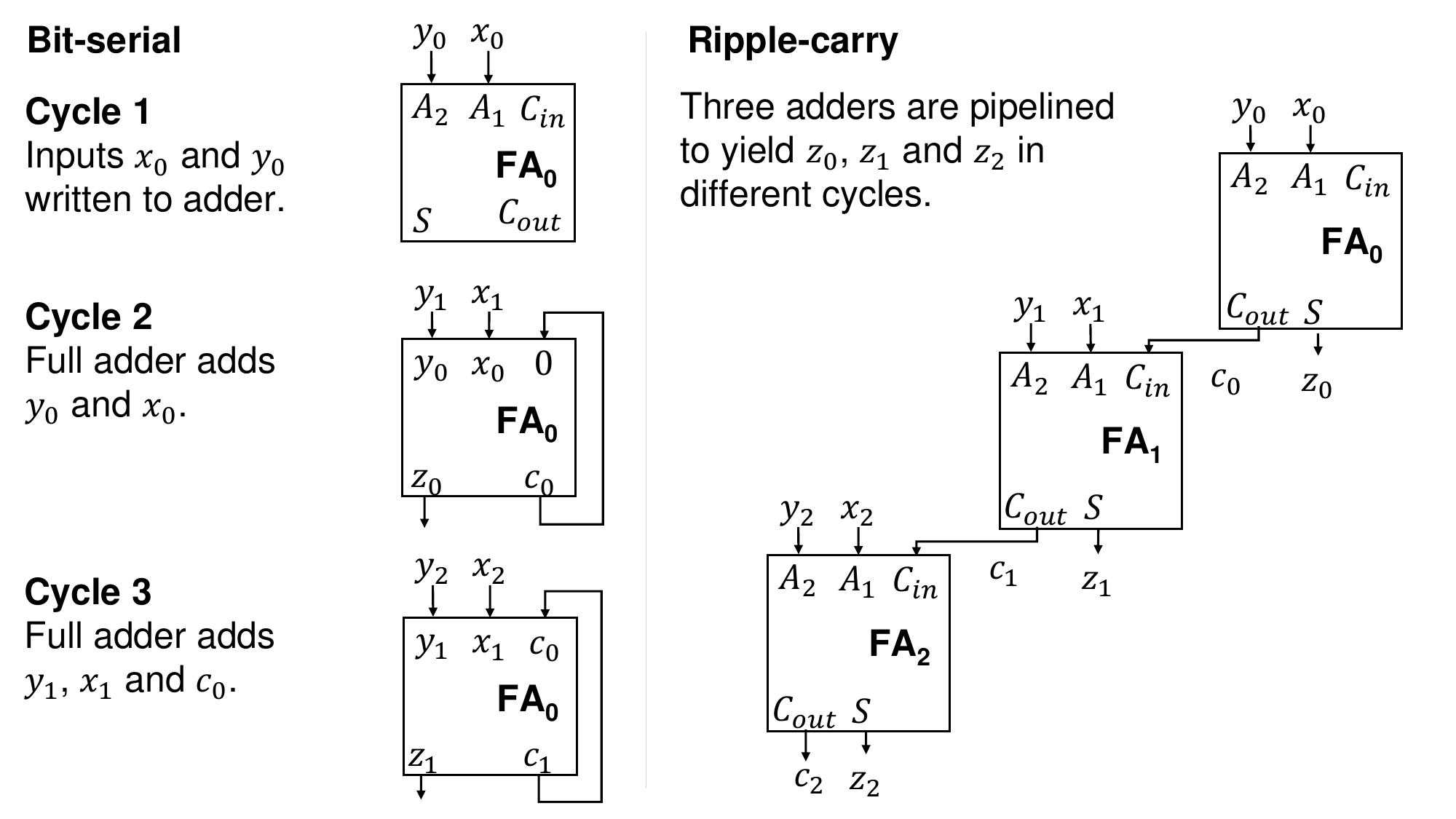}
	\caption{Behavioural depiction of implemented bit-serial adder (left) and ripple-carry adder (right). The adders are processing addends $x$ and $y$, to yield sum $z$ and carry bits $c$. Subscripts represent bit indices.}
	\label{fig:6}
\end{figure}

The single-bit half adder and full adder can be used to implement different types of adder units, such as bit-serial adders and ripple-carry adders. Fig. \ref{fig:6} shows behavioural diagrams of these different adder unit types, where FA represents our racetrack memory based full adder. In the figure, the adders are given two addends $x$ and $y$, where $x_i$ and $y_i$ represent the $i$\textsuperscript{th} bit of $x$ and $y$ respectively. The adders yield sum bits $z_i$ and carry bits $c_i$.  

In our CNN accelerator architecture, only the bit-serial adder unit is utilized as it has lower latency in accumulating many partial products and partial sums during convolution.
%%%%%%%%%%%%%%%%%%%%%%%%%%%%%%%%%%%%%%%%%%%%%%%%%%%%%%%%%%%%%%%
\subsection{Booth Multiplier}
In digital systems, multiplication of two numbers (termed multiplier and multiplicand) involves the generation of partial product terms from the two inputs, and accumulating partial products to obtain the final product. Among various algorithms, Booth multiplication \cite{Booth1951} is adopted for its high speed. The algorithm reduces the number of partial products generated by recoding the multiplier with a different radix. Hence, Booth multipliers are conventionally distinguished by radix used such as radix-2, radix-4, and radix-8. Among these, radix-4 Booth multipliers are especially efficient for hardware implementation, and are implemented in this work for multiplication in racetrack memory.

\begin{figure}
	\centering
		\includegraphics[scale=.24]{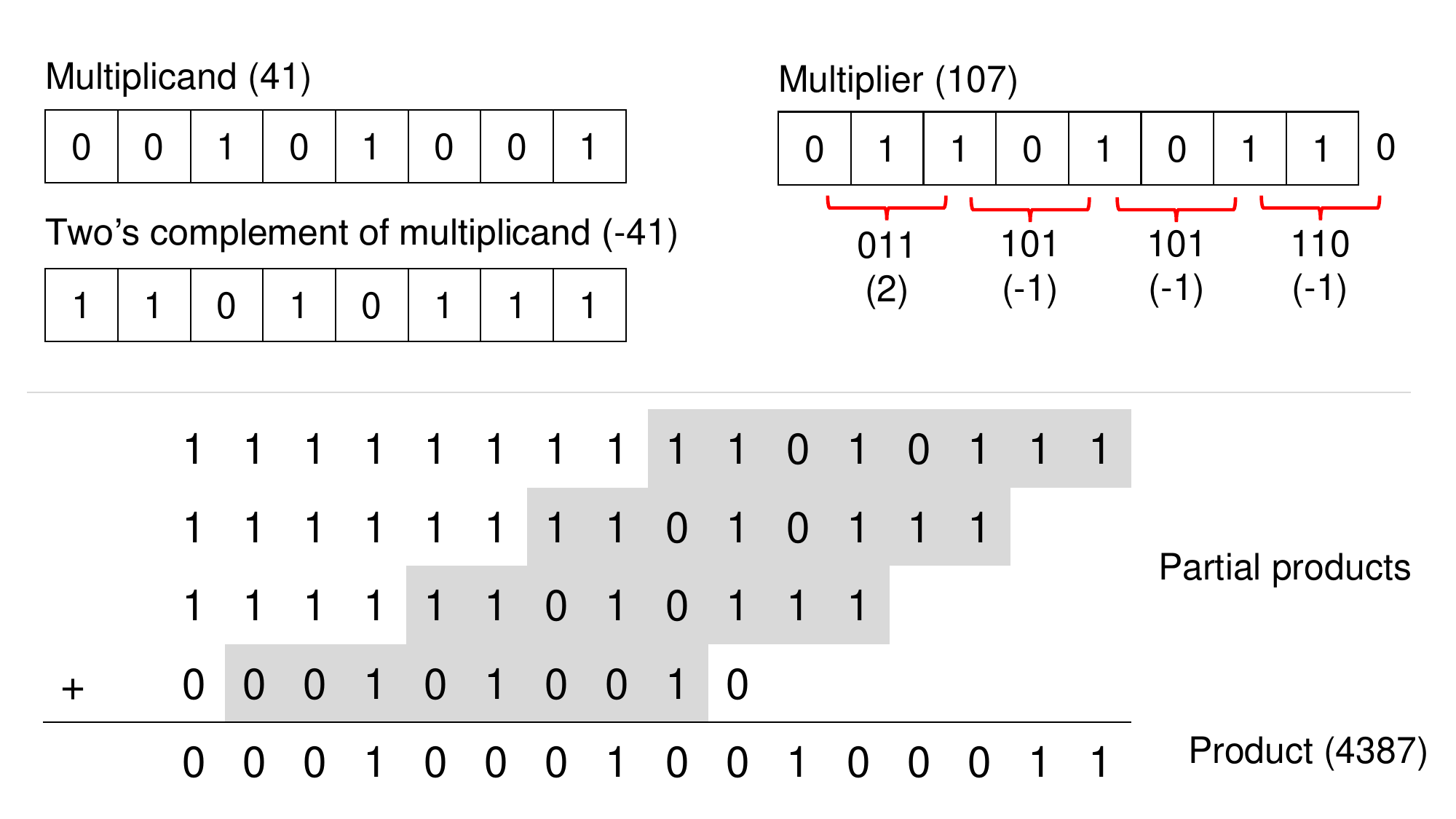}
	\caption{Example of Booth multiplication algorithm.}
	\label{fig:7}
\end{figure}

Booth multiplication can be divided into two steps: partial product generation, and partial product accumulation. We illustrate the workings of the radix-4 Booth algorithm in Fig. \ref{fig:7}, which considers the multiplication of two 8-bit numbers. First, a zero bit is appended to the least significant bit of the multiplier. The multiplier is then divided into three-bit blocks, with each block overlapping its neighbouring blocks by one bit. In the example of Fig. \ref{fig:7}, the multiplier “01101011” is recoded into four blocks: “110”, “101”, “101”, and “011”. These blocks are used to perform transformations on the multiplicand to generate the partial products. Table \ref{table:5} tabulates the encoded multiplier blocks and their corresponding transformation on the partial products. In the example given, the four blocks correspond to a multiplication of the multiplicand with $-1$, $-1$, $-1$ and $2$ respectively. The four partial products generated are then correctly aligned and accumulated to obtain the final product as the result. The alignment and accumulation of partial products in our example is depicted in Fig. \ref{fig:7} as well. 

\begin{table}
\small
\begin{center}
\caption{Encoded multiplier blocks and corresponding transformation.}
\label{table:5}
\begin{tabular}{|c|c|}
\hline
Block & Partial Product \\
\hline
000 & $0 \times Multiplicand$ \\
\hline
001 & $1 \times Multiplicand$ \\
\hline
010 & $1 \times Multiplicand$ \\
\hline
011 & $2 \times Multiplicand$ \\
\hline
100 & $-2 \times Multiplicand$ \\
\hline
101 & $-1 \times Multiplicand$ \\
\hline
110 & $-1 \times Multiplicand$ \\
\hline
111 & $0 \times Multiplicand$ \\
\hline
\end{tabular}
\end{center}
\end{table}

Next, we describe our implementation of both partial product generation and accumulation with racetrack memory.

\subsubsection{Partial Product Generation Logic}
As shown in Table \ref{table:5}, the transformations performed on the multiplicand include multiplications with five values: $-2$, $-1$, $0$, $1$ and $2$. These transformations can be performed as a combination of one or more operations: "remain", "set-to-zero", "complement", "increment", and "left-shifting". First, the three-bit blocks are decoded to select the correct transformation to be applied. Let the three bits of the block be $B_2$, $B_1$ and $B_0$ respectively, where $B_2 = 0$, $B_1 = 1$ and $B_0 = 1$ forms block "011" for multiplication by $2$. The transformations are selected based on the logic equations Eq. \ref{eq:8} to Eq. \ref{eq:11}, in which $ZERO$, $COMP$, $INCR$ and $LS$ represent "set-to-zero", "complement", "increment" and "left-shifting" respectively. These logic functions are implemented using CMOS logic. To simplify the logic for $COMP$, the function $ZERO$ is implemented to have higher priority than $COMP$: if both $ZERO$ and $COMP$ are $1$, the "set-to-zero" operation will be selected.

\begin{flalign}
\small
&ZERO = \big(B_2 \cdot B_1 \cdot B_0\big) +  \big(\overline{B_2} \cdot \overline{B_1} \cdot \overline{B_0}\big) \label{eq:8} \\
&COMP = B_2 \label{eq:9} \\
&INCR = COMP \cdot \overline{ZERO}  \label{eq:10}\\
&LS = \big(B_2 \cdot \overline{B_1} \cdot \overline{B_0}\big) + \big(\overline{B_2} \cdot B_1 \cdot B_0\big) \label{eq:11} \\ \nonumber
\end{flalign}

Multiplication with $1$ involves “remain”, in which the multiplicand bits are simply copied identically as the partial product. Hence, "remain" is executed when logic functions of all four logic functions (Eq. \ref{eq:8} to \ref{eq:11}) return $0$. Multiplication with $0$ involves the “set-to-zero” operation, as all output bits are zero. Any multiplications with magnitude of two ($2$ or $-2$) involve “left-shifting”. As explained in Section \ref{ConvolutionalNeuralNetworks}, a binary number can be multiplied by magnitude of two by shifting the number left by one bit position. Lastly, any multiplication with a negative value ($-1$ or $-2$) involves performing a two’s complement on the multiplicand value. A two's complement is performed by combining "complement" - in which every bit is inverted - and "increment" which adds the value of 1 to the complemented output. The "complement" of the input bit is implemented by multiplexing between the input bit and its inverted bit, whereas a bit-serial adder unit proposed in Section \ref{Adder} is used to add 1 for "increment". 

\begin{figure}
	\centering
		\includegraphics[scale=.24]{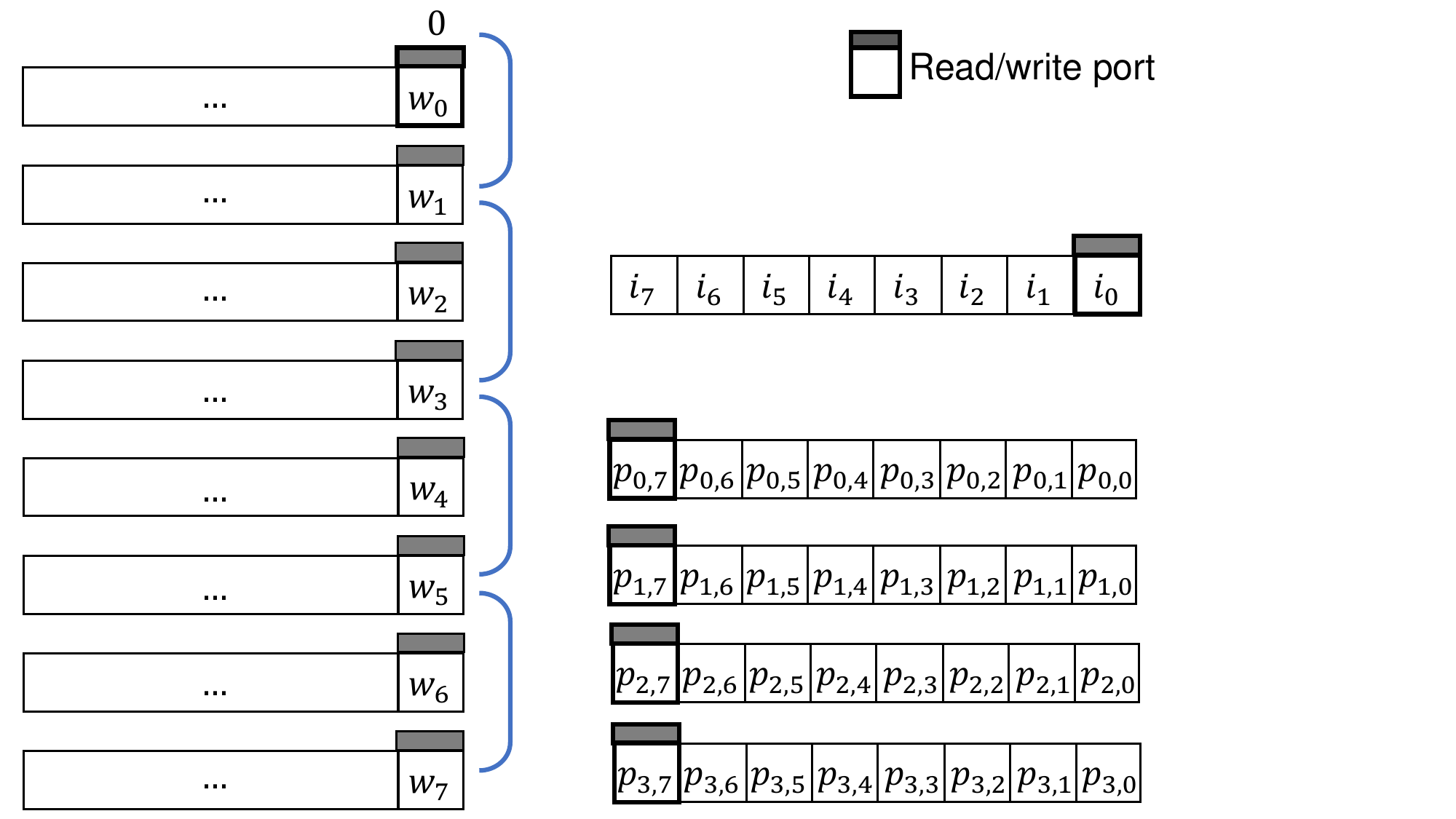}
	\caption{Data organization of multiplicand, multiplier and partial product terms.}
	\label{fig:8}
\end{figure}

To maintain high performance, we generate partial products in parallel, which requires that all multiplier bits be simultaneously accessible. Fig. \ref{fig:8} shows the data organization of multiplier and multiplicand in racetrack memory strips, where $w_b$ is the $b$\textsuperscript{th} bit of the multiplier, $i_b$ is the $b$\textsuperscript{th} bit of the multiplicand, and $p_{T,b}$ is the $b$\textsuperscript{th} bit of the $T$\textsuperscript{th} partial product term. For parallel generation, we adopt an asymmetrical storage of multiplier and multiplicand: the multiplier ($w$) is stored in bit-parallel format across several strips, allowing all bits of the multiplier to be available for decoding in the same cycle. Conversely, the multiplicand ($i$) is stored in a bit-serial format, in which all bits are stored in the same strip and are read one bit at a time each cycle. For DNN applications, we map model weights as multiplier bits, as the weights are known before inference and can be transposed to bit-parallel format beforehand. 

As the four partial products are generated in bit-serial fashion, the number of cycles to generate all partial products is equal to the number of bits in the multiplicand. Both the resources used (CMOS logic and racetracks) and number of cycles for partial product generation scale linearly with operand bit-width. By optimizing the area-performance trade-off, our design is scalable and can be applied to higher-bit multiplication as well (such as 64-bit).

Upon generation, the partial products are each stored in separate racetracks to be accessed during partial product accumulation. 

\subsubsection{Partial Product Accumulation}
Partial product accumulation comprises two stages: the alignment stage, and addition stage. As shown in Fig. \ref{fig:7}, the partial products must be properly aligned for correct accumulation to yield the product. Specifically, the partial product for each multiplier block should be left-shifted two positions from the partial product of the previous multiplier block. Hence, during the alignment stage, the racetrack memory tracks storing the partial products are shifted to the correct positions.

\begin{figure}
	\centering
		\includegraphics[scale=.24]{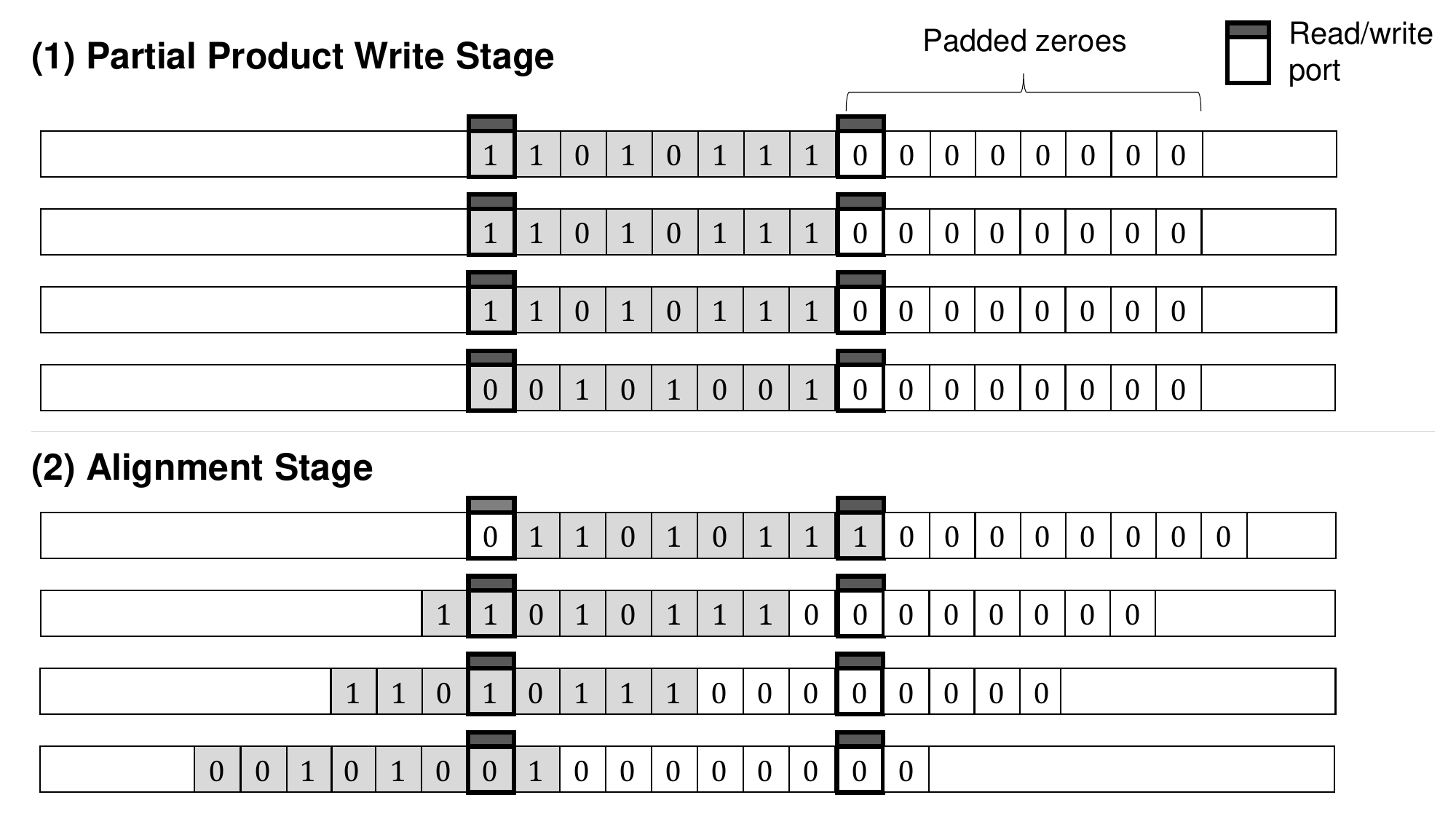}
	\caption{Write and alignment stages of partial product accumulation. The partial products are shaded in grey.}
	\label{fig:9}
\end{figure}

Fig. \ref{fig:9} shows the write stage and alignment stage of four partial products generated for 8-bit multiplication from our previous example of Fig. \ref{fig:7}. During Booth multiplication, these four tracks are dedicated for storing partial sums. As such, the four tracks can be zero-padded on the right side to support left-shifting of partial sums.

\begin{figure}
	\centering
		\includegraphics[scale=.24]{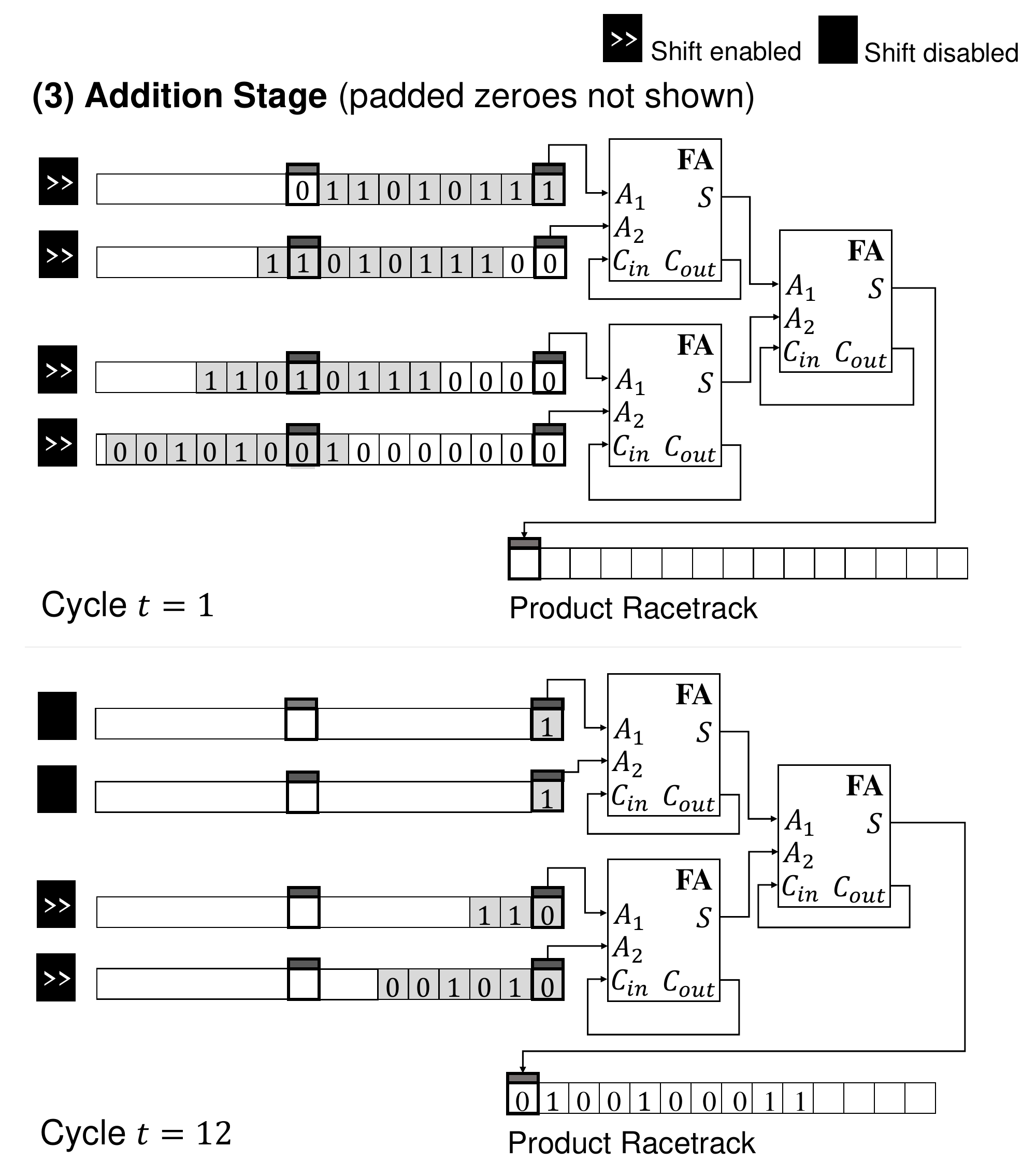}
	\caption{Write and alignment stages of partial product accumulation. The partial products are shaded in grey.}
	\label{fig:10}
\end{figure}

Fig. \ref{fig:10} depicts the addition stage. During this stage, all four tracks are shifted right while they are accessed at the read port one bit at a time. The accessed bits are transferred as inputs to an adder tree, made of the proposed full adders in Section \ref{Adder}. The full adders accumulate the partial products in bit-serial fashion: the $S$ bit is transferred to the next stage adder and the carry-out signal $C_{out}$ is transferred to its own carry-in $C_{in}$ for the next bit.

Sign extension is supported by disabling the shifting circuitry once the MSB of a partial product is reached. For example, in Fig. \ref{fig:10} (bottom), the MSB of the first partial product is under the read port. Hence, the shifting circuitry is disabled for this racetrack memory for the remaining cycles, allowing the sign bit ($1$) to be repeatedly read. For a fixed data bit-width (such as 8-bit multiplication), the enabled/disabled state of shift circuitry is fixed and is asserted by control circuitry. The control of shifting circuitry is further elaborated in Section \ref{Shift-based Multiplier}.

%%%%%%%%%%%%%%%%%%%%%%%%%%%%%%%%%%%%%%%%%%%%%%%%%%%%%%%%%%%%%%%
\subsection{Shift-based Multiplier} \label{Shift-based Multiplier}
As elaborated in Section \ref{ConvolutionalNeuralNetworks}, shift-based neural network models have great potential to improve performance while reducing computational costs of CNN inference. However, execution of shift-based models on architectures optimized for regular multiplication typically results in sub-optimal performance. Consider the proposed 8-bit Booth multiplier accelerating a shift-based model. As weights are constrained to powers-of-two values, each weight would be a one-hot encoded number in which at most one of the eight bits is $1$. If mapped as the multiplier, most multiplier blocks would be $000$, resulting in all but one of the partial products generated being zero. These zero values are still written to racetracks and accumulated together, wasting resources and energy. Furthermore, unnecessary transformations would be applied to the multiplicand only to yield the same binary sequence shifted several positions. 

\begin{figure}
	\centering
		\includegraphics[scale=.24]{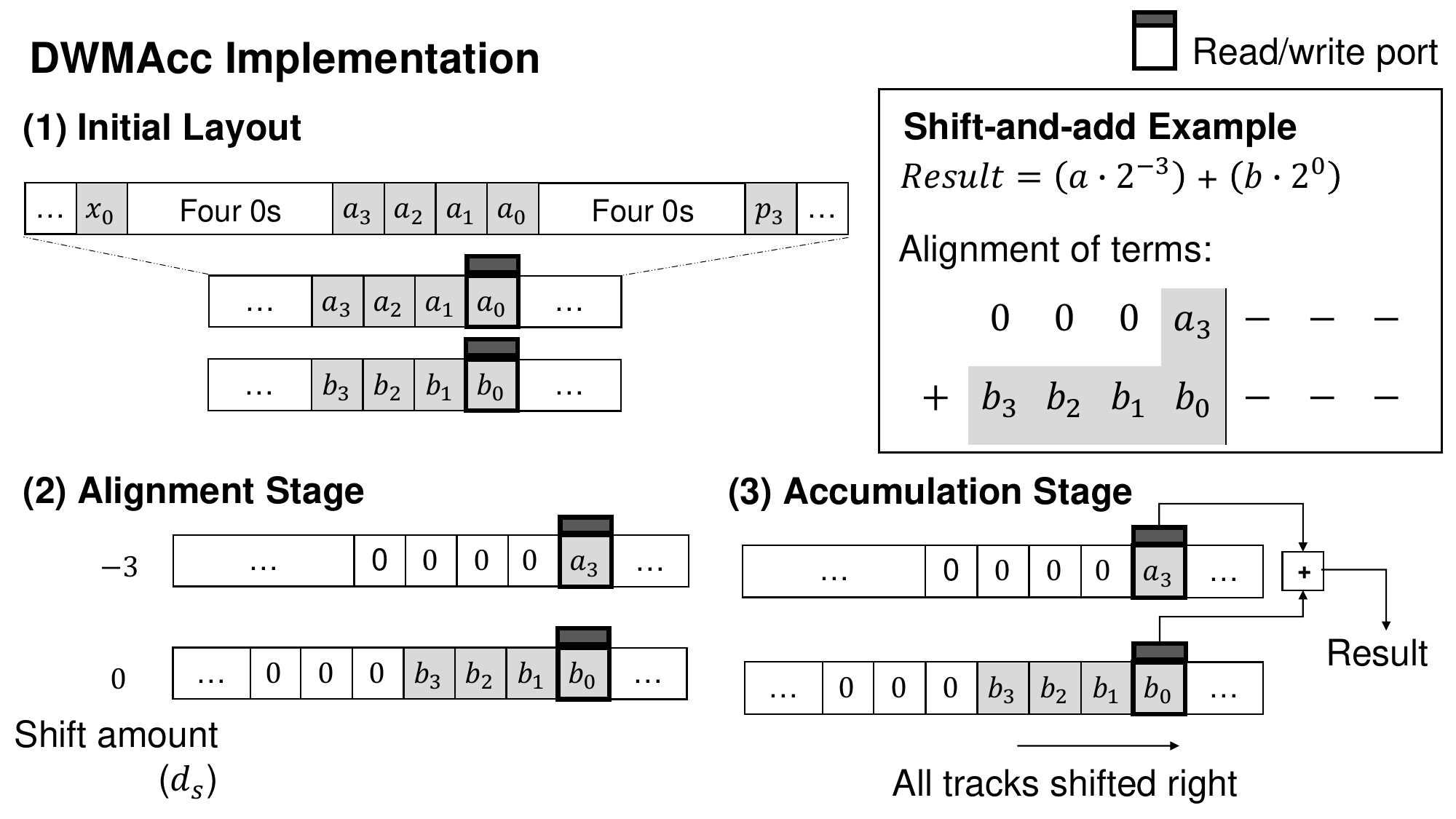}
	\caption{Implementation of shift-based multiplier of DWMAcc}
	\label{fig:11}
\end{figure}

With characteristics of a shift register, racetrack memory is especially suitable for accelerating shift operations. \citet{DWMAcc2019} proposed a shift-based DNN architecture named DWMAcc which performs multiple shift operations in parallel, allowing a shift-and-add operation to be completed in $O(n)$ time complexity. Fig. \ref{fig:11} depicts the shift-and-add implementation of DWMAcc in a group of two tracks. For ease of illustration, we consider an example of 4-bit activations ($a$, and $b$) with shift distance $-3 \leq d_{s} \leq 0$. In their approach, activations are stored in bit-serial manner, where all bits of an activation are stored in the same track. To support shifting operations, activations in the same track are separated by bits of $0$; the number of zeroes used for separation is equal to $\max(|d_{s}|) + 1$. In Fig. \ref{fig:11}, 4-bit activations in the same track are separated by four $0$ bits. 

DWMAcc divides the shift operation into an alignment stage and an access stage. During the alignment stage, each track is shifted to different positions according to its weight value. This shifting is facilitated by comparing two registers in the track's Location Unit (LU) containing the current position and target position respectively. Upon completion of alignment, all tracks are accessed together and shifted in the downstream direction by one domain each cycle. The accessed bits are written to an adder tree which completes the shift-and-add computation.

While DWMAcc effectively exploits the shifting nature of racetrack memory for acceleration, the alignment-access approach has several limitations:

\begin{itemize}
    \item DWMAcc requires padding of $0$ bits which reduces storage density. The number of padded bits scales linearly with shift distance supported. While they propose a zero-sharing scheme, the padded zeroes still occupy significant proportion of the racetrack memory, occupying one-third of memory in the case of 8-bit activations with 8-bit shift distance.
    \item The alignment stage results in additional latency (cycle) costs, as DWMAcc must wait a full seven cycles to ensure all alignment is complete before access and accumulation can begin.
    \item Due to the padded zeroes, the approach of DWMAcc does not support sign extension and is appropriate only for unsigned numbers.
    \item When the number of padded zeroes is equal to the data bit-width, DWMAcc only support shifting in one direction (such as right shifting) at a time. If accumulated terms are shifted in opposite directions, other activations beyond the zero-padding may incorrectly be accumulated together. This constrains the range of power-of-two values supported. However, as CNN weight values can be constrained to values smaller than 1, it is reasonable to support right-shifting only for CNN acceleration. 
\end{itemize}

\begin{figure}
	\centering
		\includegraphics[scale=.24]{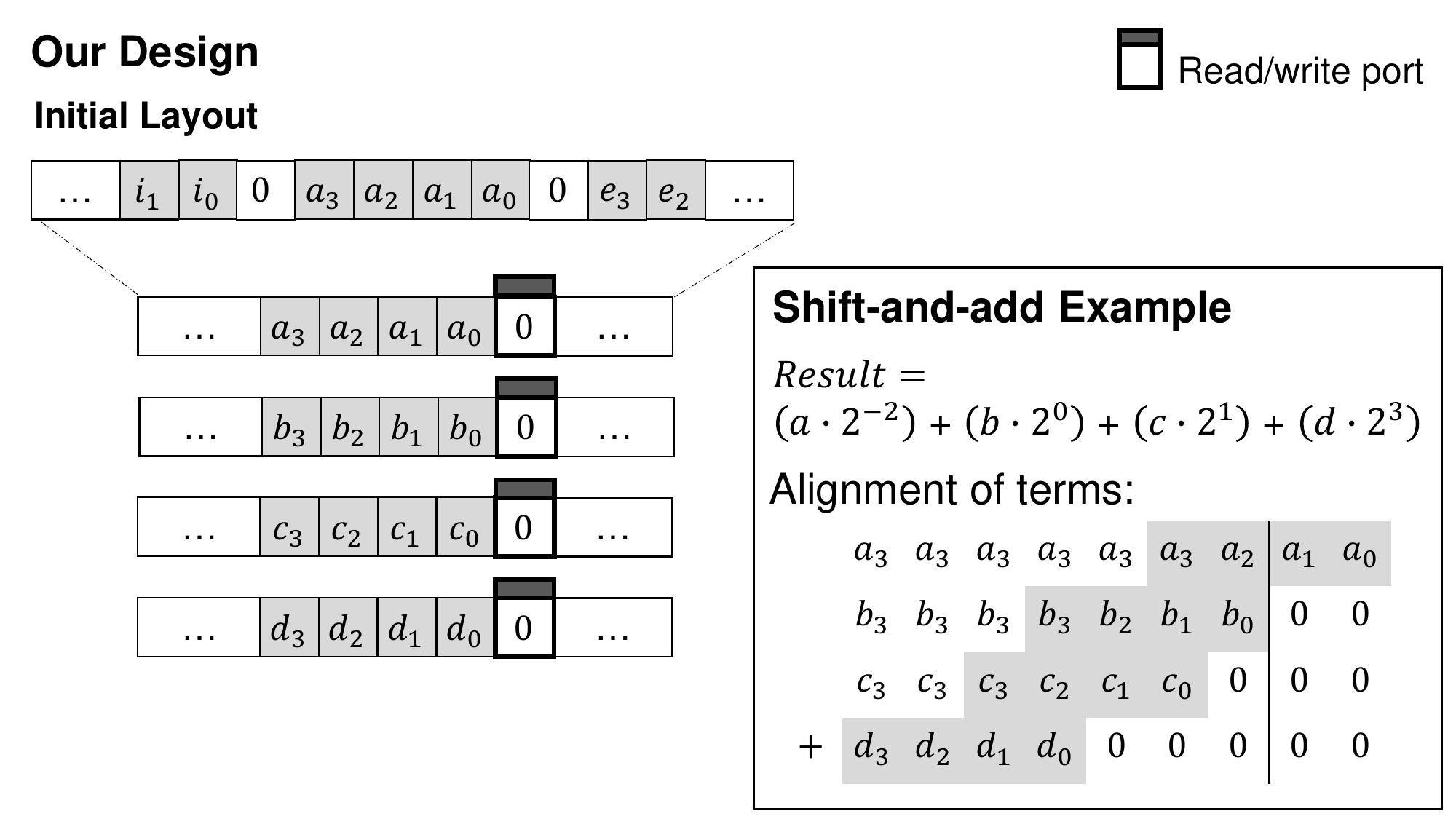}
	\caption{Initial data organization for shift-based adder, and shift-and-add example.}
	\label{fig:12}
\end{figure}

We propose a shift-based multiplier design that similarly uses the shifting capabilities of racetrack memory, but targets higher storage density, low latency, and support for sign extension and varying shift directions. Our design achieves these targets by generating shifted products by \emph{selectively enabling/disabling the shifting circuitry} of each track instead. Fig. \ref{fig:12} (left) shows the data layout of terms before shifting, with each letter ($a$ to $i$) representing a term for accumulation. Similar to DWMAcc, activations are stored in a bit-serial manner, but with only a single $0$ bit separating activations in the same track. We illustrate the mechanism of our shift-based adder using a similar example of 4-bit activations ($a$, $b$, $c$ and $d$) with shift distance $-3 \leq d_{s} \leq 3$, with positive $d_{s}$ for left-shifting as well. The considered shift-and-add example is shown in Fig. \ref{fig:12} (right). 

\begin{figure}
	\centering
		\includegraphics[scale=.24]{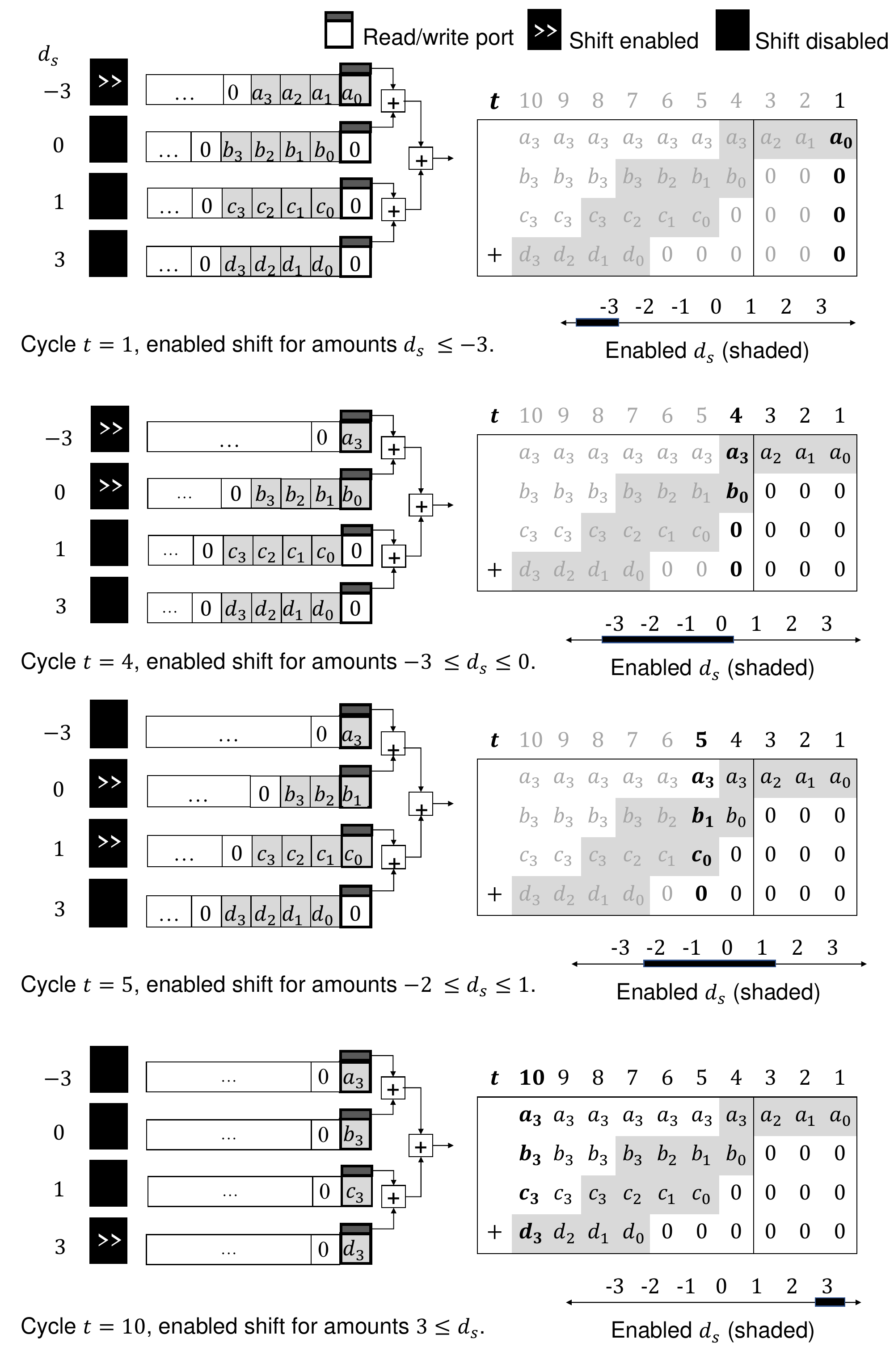}
	\caption{Execution of shift-and-add by the proposed shift-based multiplier.}
	\label{fig:13}
\end{figure}

Fig. \ref{fig:13} depicts the access mechanism of our shift-based multiplier design for the example considered. The figure also shows the correct alignment of terms for accumulation and the multiplier's progress in different cycles. In order to execute shift-and-add, the four activations in racetracks are accessed and transferred down an adder tree comprising bit-serial full adders as shown in Section \ref{Adder}. 

Our design enables or disables the shift circuitry based on the shift amount ($d_{s}$) in different cycles. In cycle $t=1$, only the racetrack strip with the smallest $d_{s}$ of $-3$ is first shift-enabled, while other tracks remain with the padded $0$ bit under the access port. In following cycles, tracks with incrementally larger $d_{s}$ values are shift-enabled to facilitate correct alignment. In cycle $t=2$, the enabled shift range is $-3 \leq d_{s} \leq -2$, and in the next cycle $t=3$, the enabled shift range is $-3 \leq d_{s} \leq -1$, and so on. However, in cycle $t=4$, the MSB of the first track ($d_{s} =-3$) is aligned under the access port. Hence, we disable shifting for this track from cycle $t=5$, keeping the MSB aligned with the access port for sign extension. In effect, the currently-enabled $d_s$ values can be represented as four integer values moving up a number line, which is depicted as the black bar on the number line in Fig. \ref{fig:13}. By the last cycle $t=10$, the MSB of the tracks with largest supported $d_{s}$ has been accessed. At this stage, all racetrack strips are aligned at their MSB, permitting a simple, standardized reset of position for the next access without irregular data alignments between tracks, not requiring positional tracking with Location Units as in DWMAcc.  

An efficient control mechanism is needed to selectively enabled the shift circuitry of tracks. For $n_b$-bit activations, the control mechanism should:
\begin{itemize}
    \item enable shifting for most negative $d_{s}$ first, and
    \item disable shifting after a track has been shifted $n_b$ places for sign extension.
\end{itemize}

\begin{figure}
	\centering
		\includegraphics[scale=.24]{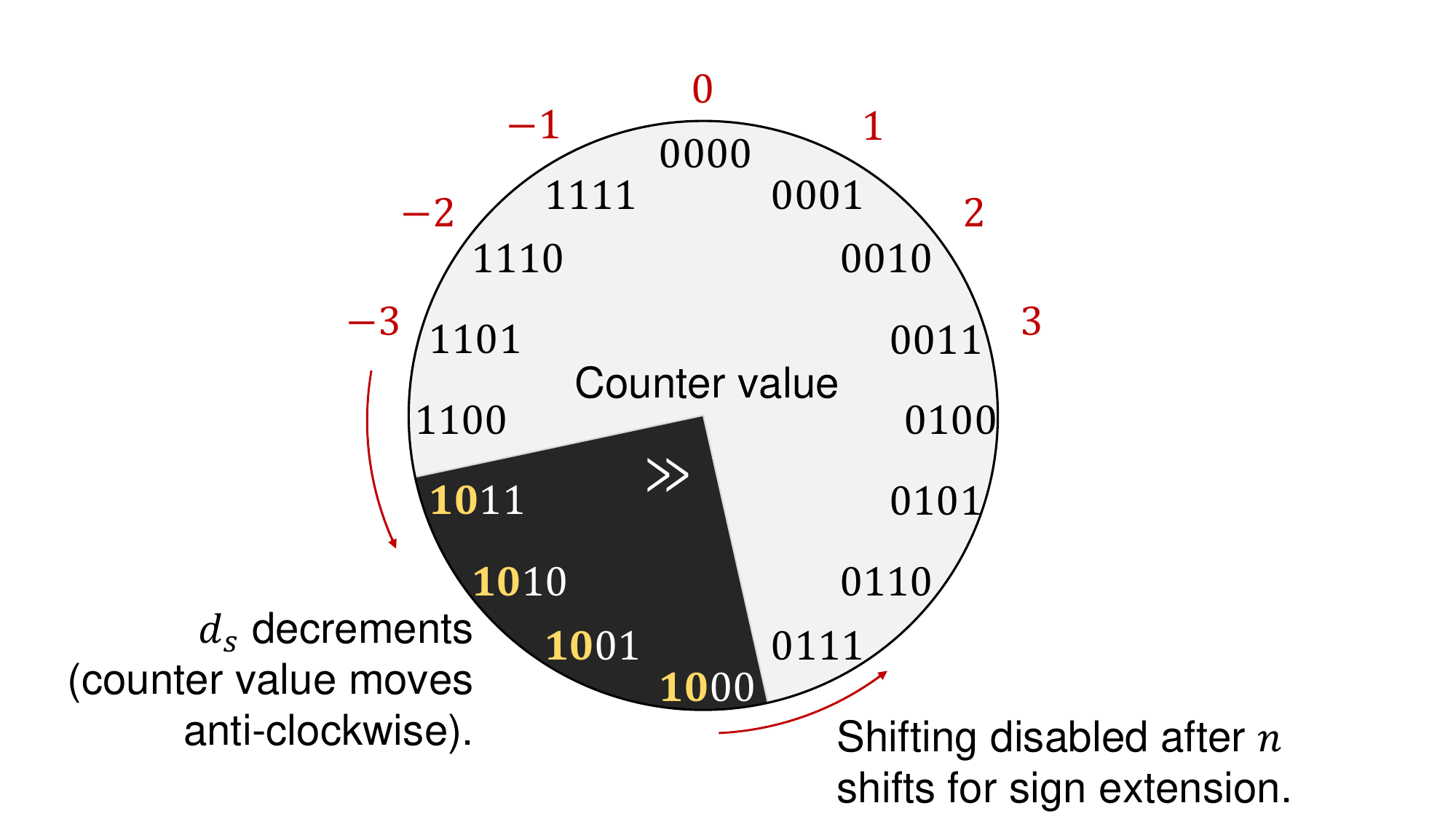}
	\caption{State of shift enable/disable across the range of counter values.}
	\label{fig:14}
\end{figure}

We implement the control mechanism using a CMOS decrementing counter. The shift amount (weight values in CNNs) are first written to the counter, which decrements the value by 1 every cycle. Fig. \ref{fig:14} depicts the range of values of the counter, where initial $d_{s}$ values are shown in red outside the circle. Each cycle, the counter value decrements, hence the values move counterclockwise about the circle. The shifting circuitry is enabled when the counter values are in the shaded region. As the values move counterclockwise, the most negative weights will be shift-enabled first, fulfilling the first control criteria listed above. Next, the enabled $d_{s}$ range is exactly $n_b$ bits, disabling shifting once the counter value exits the shaded range to perform sign extension. Lastly, the shaded shift-enabled region can be chosen for enable logic to be simple: in the example of Fig. \ref{fig:14}, the shaded range corresponds to when the two most significant bits are "$10$", which can be implemented with a single CMOS logical AND gate. Therefore, the decrementing counter allows shift-and-add to be controlled in a simple but effective manner.

For correctness, the decrementing counter range should be at least $n_b + 2d_{max}$ bits in width, where $d_{ max} = \text{max}(|d_{s}|) + 1$. For the regular case of $n = d_{max}$, this corresponds to one bit wider than signed weight values. In the example of Fig. \ref{fig:14}, $n_b = 4$ and $d_{max} = 4$, hence we use a 4-bit decrementing counter and sign-extend the 3-bit weights by one bit.

%%%%%%%%%%%%%%%%%%%%%%%%%%%%%%%%%%%%%%%%%%%%%%%%%%%%%%%%%%%%%%%
\subsection{Energy Optimization: Write-shift Transformation}
Among the operations of racetrack memory technology (read, write, shift), the write operation is most costly in both energy consumption and latency. A racetrack memory write operation consumes approximately $20\times$ more energy than a shift operation, and has a latency $10\times$ that of the shift operation as well (exact parameter values are listed in Table \ref{table:6} below). The high energy is incurred from the high current used to set the magnetization direction of the free layer of the MTJ. While our proposed RM-based arithmetic circuits minimizes the write operations needed compared to baseline designs \cite{Trinh2013}, we observe that the energy consumed by writing inputs still dominates. For example, the writing of seven input MTJs of the full adder circuit in Fig. \ref{fig:5} is more than $99\%$ of the total energy consumption, as write energy is an order of magnitude higher. Hence, we propose a novel energy optimization that converts write operations in RM-based arithmetic circuits to shift operations for greater efficiency. 

\begin{figure}
	\centering
		\includegraphics[scale=.24]{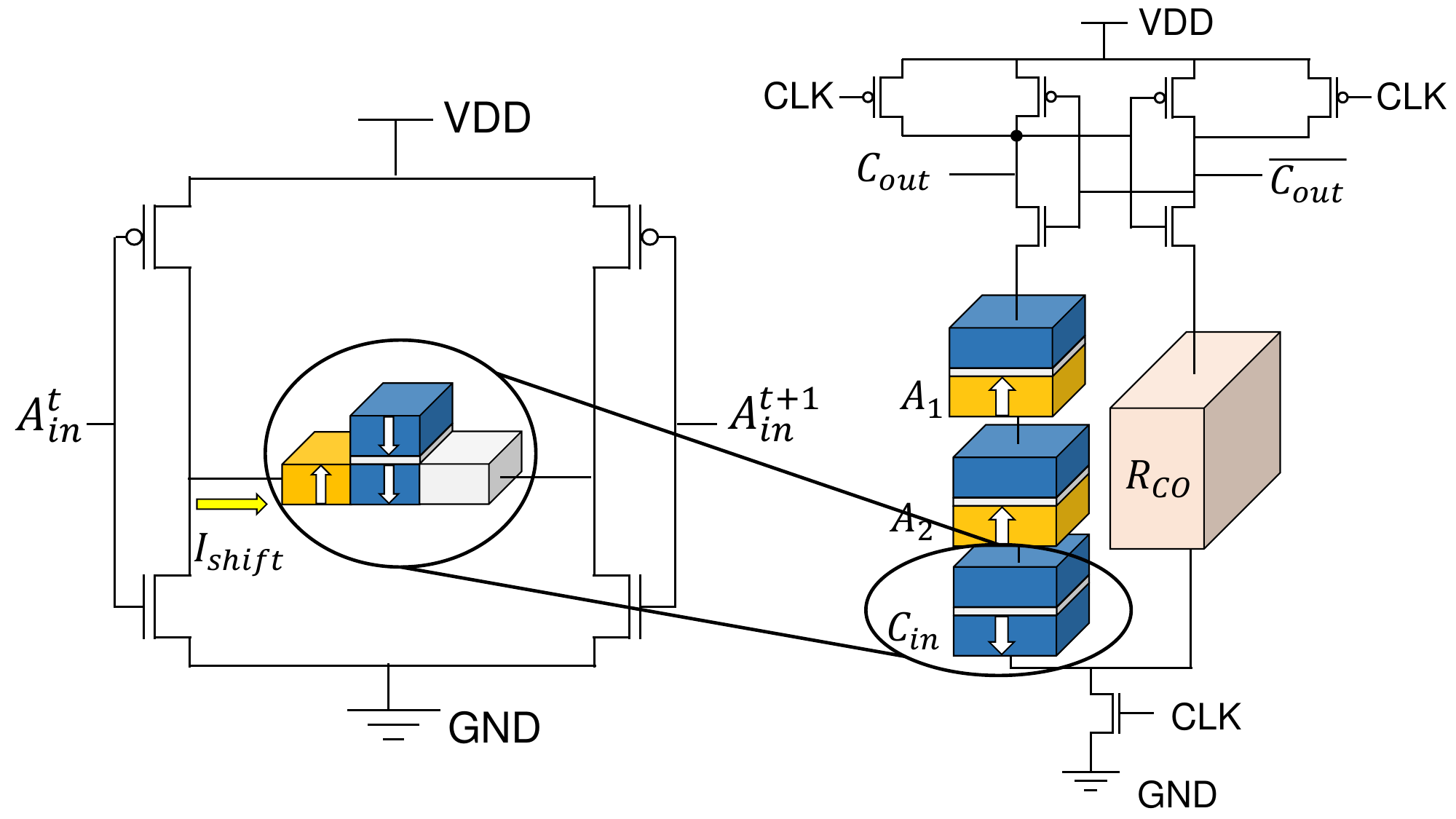}
	\caption{Control circuitry implementing write-shift transformation.}
	\label{fig:15}
\end{figure}

As the MTJs in the arithmetic logic circuits are used for computation rather than storage, we can shift the required bit value ("1" or "0") to the input MTJs, rather than performing a write to change their magnetization directions. Fig. \ref{fig:15} demonstrates our proposed method to achieve the write-shift transformation in MTJs of the full adder carry-out circuitry. Each input MTJ is equipped with a three-bit track, on which magnetization directions corresponding to bits "1" and "0" are already stored. When the input bit required is "1", the domains in the track are shifted such that the "1" bit is aligned with the MTJ, consuming shift energy rather than that of write operations. The same process is used to shift "0" to the MTJ when needed. Furthermore, if the the newly-required bit is the same as the previous bit aligned with the MTJ, we can skip the shift operation entirely and consume no additional energy. 

While the write-shift transformation can greatly reduce energy costs, additional resources are needed to control the shift current of each input MTJ. Adopting CMOS technology for the shift currents would cost large area and energy. Hence, we propose efficient control circuitry shown in Fig. \ref{fig:15} to execute the shift operation according to the current state of the MTJ. In the figure, $A^{t}_{in}$ is the bit aligned with the MTJ in the current cycle, whereas $A^{t+1}_{in}$ is the bit required in the next cycle. If the two bits $A^{t}_{in}$ and $A^{t+1}_{in}$ are the same, there is no shift current from the controlling transistors, and so no shift operation is performed. If the two bits are different, the shift current is applied in the correct direction to move the domains. In effect, the write operations of the seven MTJs are transformed to shift operations. 

In our full adder circuit, this optimization reduces energy consumption from 7.019 pJ to 0.392 pJ (94.4\%). However, this optimization comes at the cost of area for shift control circuitry, increasing the area from 1.14 $\mu m^2$ to 7.53 $\mu m^2$, which remains comparable to that of a CMOS full adder implementation. While the write-shift transformation also reduces latency from write delay (5 ns) to shift delay (0.5 ns), the full adder is synchronous and depends on pulses of the system clock, which remains bound by write operations in MU storage. Therefore, the write-shift transformation is mainly an energy optimization.  

%% file: tex/4_In-Memory_Accelerator_Architecture.tex
\section{In-memory Accelerator Architecture}
The bit-serial adder, along with both the Booth and shift-based multiplier units proposed in Section \ref{ProposedArithmeticCircuits}, are integrated with racetrack memory to implement the in-memory CNN accelerator. In this section, we describe the system-level architecture of the accelerator and the dataflow mapping of CNN layers on the accelerator.

%%%%%%%%%%%%%%%%%%%%%%%%%%%%%%%%%%%%%%%%%%%%%%%%%%%%%%%
\subsection{Chip Organization} 
NVSim \cite{NVSim2012} is a circuit-level model for non-volatile memory (NVM) chip design exploration. Given specific NVM cell characteristics, the model estimates chip area, access latency and energy consumption for different chip organizations. In addition, it searches the design space to provide optimal configurations for a given target metric.

\begin{figure}
	\centering
		\includegraphics[scale=.24]{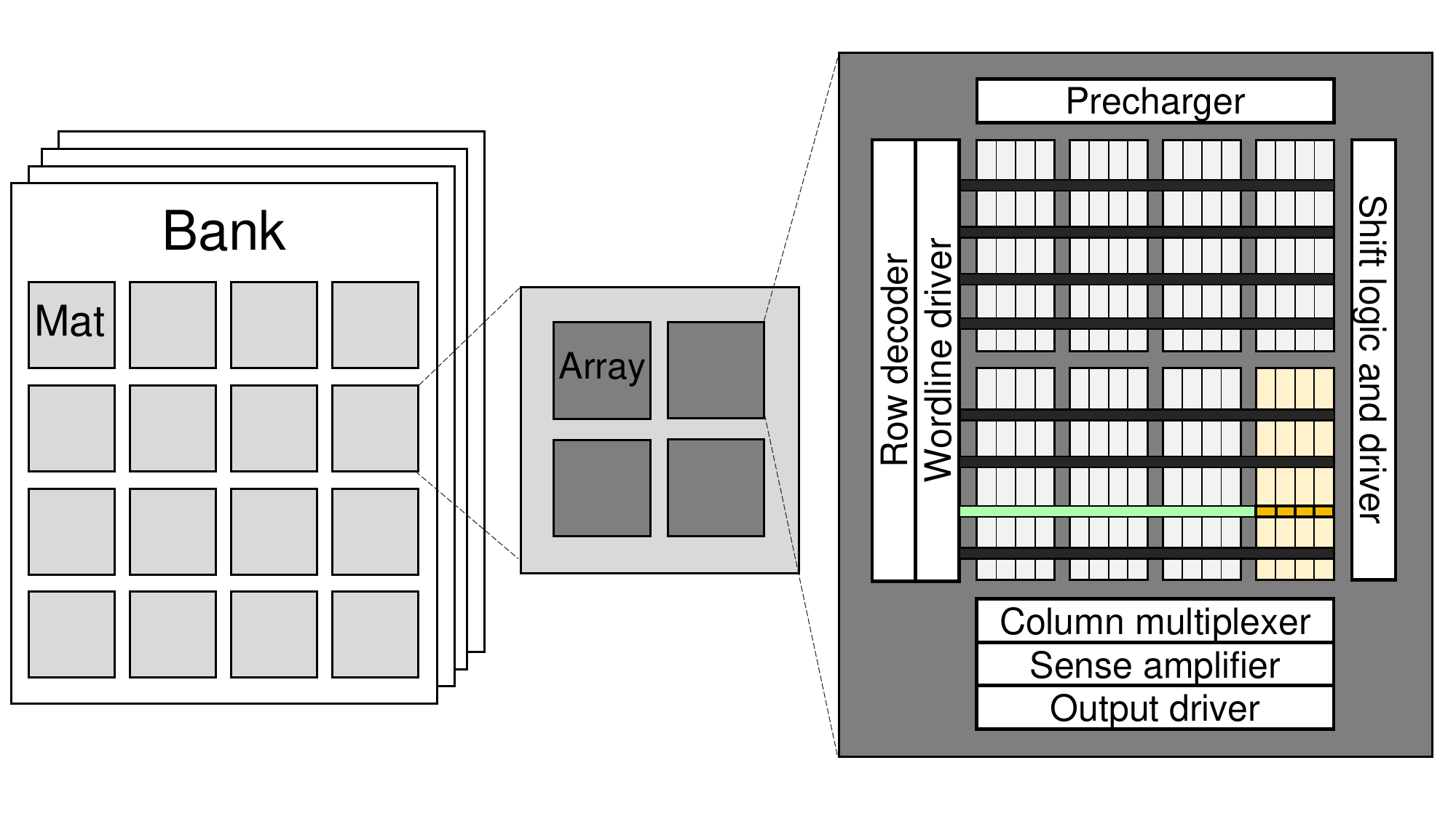}
	\caption{Memory organization simulated in NVSim.}
	\label{fig:16}
\end{figure}

NVSim organizes a memory chip with three levels of hierarchy: banks, mats, and subarrays. Fig. \ref{fig:16} illustrates the levels in memory array organization used. The bank is the top-level unit modelled by NVSim. Each bank comprises multiple mats, while each mat is constructed using subarrays as the building block. Each subarray is an array of memory cells along with additional access circuitry including row decoders, column multiplexers, sense amplifiers and drivers.  

\subsubsection{Subarray of Macro Units} \label{subarray of MU}
We follow a similar approach to that of \cite{Chao2015} and \cite{Hu2016} in simulating a memory architecture using the racetrack memory macro unit (MU) as the elementary memory cell unit. As stated in Section \ref{RacetrackMemory}, we adopt the MU configuration of $N_{dom} = 64$, $N_{ports} = 16$ and $N_{tracks} = 4$. The parameters and dimensions of the racetrack memory and MU are tabulated in Table \ref{table:6} and Table \ref{table:7} respectively. Each racetrack strip in an MU can be accessed via four dedicated access ports, with each port accessing four bits at a time (one bit per track).

At the subarray level, the array of MUs is divided into rows and columns. During a data access, the row decoder and wordline driver of the subarray selects a row of access ports (colored green in Fig. \ref{fig:16}), whereas the column multiplexers select a single column of MUs. Data is read or written from a single port across all four tracks during an access. This access is highlighted in orange in the accessed MU of Fig. \ref{fig:16}. As proposed in \cite{Chao2015}, additional shift decoders and shift drivers are added to facilitate domain wall shifting. In order to simplify decoding logic, we shift all four tracks in a selected MU together. Hence, the bit positions across all four tracks of an MU are always aligned.

\begin{figure}
	\centering
		\includegraphics[scale=.24]{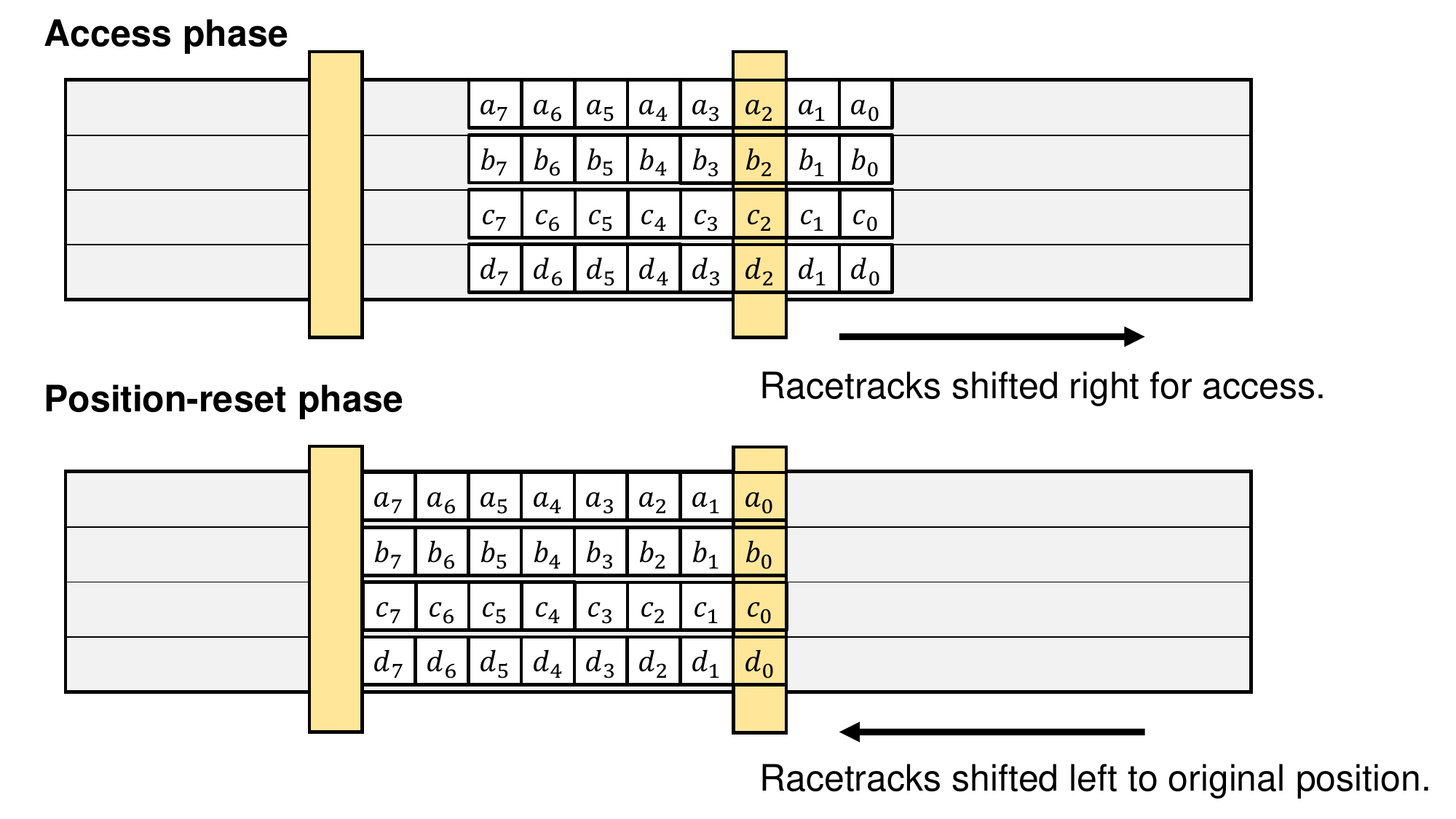}
	\caption{Access phase and position-reset phase of four racetracks in an MU. The gold block represents an access port which reads/writes the four bits aligned with it.}
	\label{fig:17}
\end{figure}

Fig. \ref{fig:17} depicts the parallel access cycles of 8-bit words from an MU. For ease of illustration, we depict eight bits sharing an access port in the MU of Fig. \ref{fig:17}. The access process can be divided into the \textit{access phase} and \textit {position-reset phase}. During the first 8 cycles, the tracks are shifted one position each cycle to access the words in a bit-serial manner. After the access is complete, another 8 cycles are used to shift all tracks to their original position, such that the least significant bit of each word is aligned to the access ports. The position-reset phase is required before words from other access ports of the same MU can be read.

\begin{table}
\small
\begin{center}
\caption{Parameters and dimensions of a single racetrack (RT) memory track.}
\label{table:6}
\begin{tabular}{|c|c|}
\hline
Parameter & Value \\
\hline
RT width & 1F \\
\hline
Spacing between RTs & 1F \\
\hline
RT length & 128F \\
\hline
Domain length & 2F \\
\hline
No. of domains per RT & 64 \\
\hline
RT thickness & 6nm \\
\hline
RT write energy & 1pJ \\
\hline
RT write latency & 5ns \\
\hline
RT shift energy & 0.051pJ \\
\hline
RT shift latency & 500ps \\
\hline
\end{tabular}
\end{center}
\end{table}

\begin{table}
\small
\begin{center}
\caption{Parameters and dimensions of the Macro Unit.}
\label{table:7}
\begin{tabular}{|c|c|}
\hline
No. of domains per RT & 64 \\
\hline
No. of RTs & 4 \\
\hline
No. of MU ports & 16 \\
\hline
No. of ports per RT & 4 \\
\hline
MU access CMOS width & 10F \\
\hline
MU access CMOS length & 4F \\
\hline
Spacing between access CMOS & 1F \\
\hline
MU Width & 10F \\ 
\hline
MU Length & 128F \\
\hline
Total MU storage & 32B \\
\hline
Total RT storage & 8B \\
\hline
\end{tabular}
\end{center}
\end{table}

\subsubsection{Architecture Overview}
In this section, we provide an overview of the accelerator architecture, including memory organization used and arithmetic logic insertion. First, we present the configuration of memory arrays at the bank, mat, and subarray level. Following this, we integrate the RM-based arithmetic units proposed in Section \ref{ProposedArithmeticCircuits} and demonstrate how addition and multiplication operations are performed in a group of mats.

We use NVSim to simulate a 2 MB accelerator bank using the 45 nm technology node. The accelerator bank is modular and can be replicated for a multi-bank memory organization according to application needs. For example, 16 of our simulated banks can be combined to achieve the 32 MB storage capacity comparable to that of previous works \cite{DWMAcc2019,  Eckert2018}. With a considerably large bank capacity of 2 MB, only several banks are needed to store all parameters of smaller CNN models meant for embedded system applications, such as MobileNet \cite{MobileNet2017} and ResNet \cite{ResNet20} models. As write energy is the dominant energy consumed in RM accesses, we configure NVSim to optimize for the write energy-delay product. Additionally, we follow \cite{Hu2016} in targeting low operating power devices which minimizes dynamic power, as the non-volatile nature of racetrack memory prevents data loss when peripheral circuitry is off. 

Our proposed RM-based arithmetic logic is integrated at the mat and bank levels of the design. Hence, while NVSim is capable of optimizing the number of subarrays in a mat, we add constraints for a minimum number of subarrays per mat to support the bandwidth requirements of the arithmetic logic. Similarly, we set a minimum number of mats in a bank to ensure sufficient system throughput.

Each accelerator bank comprises 256 mats. Every block of 16 mats is grouped into a \textit{mat group}, as mats in a mat group will share multiplication circuitry. Each bank therefore contains 16 mat groups, with each mat group containing 16 mats.

\begin{figure}
	\centering
		\includegraphics[scale=.24]{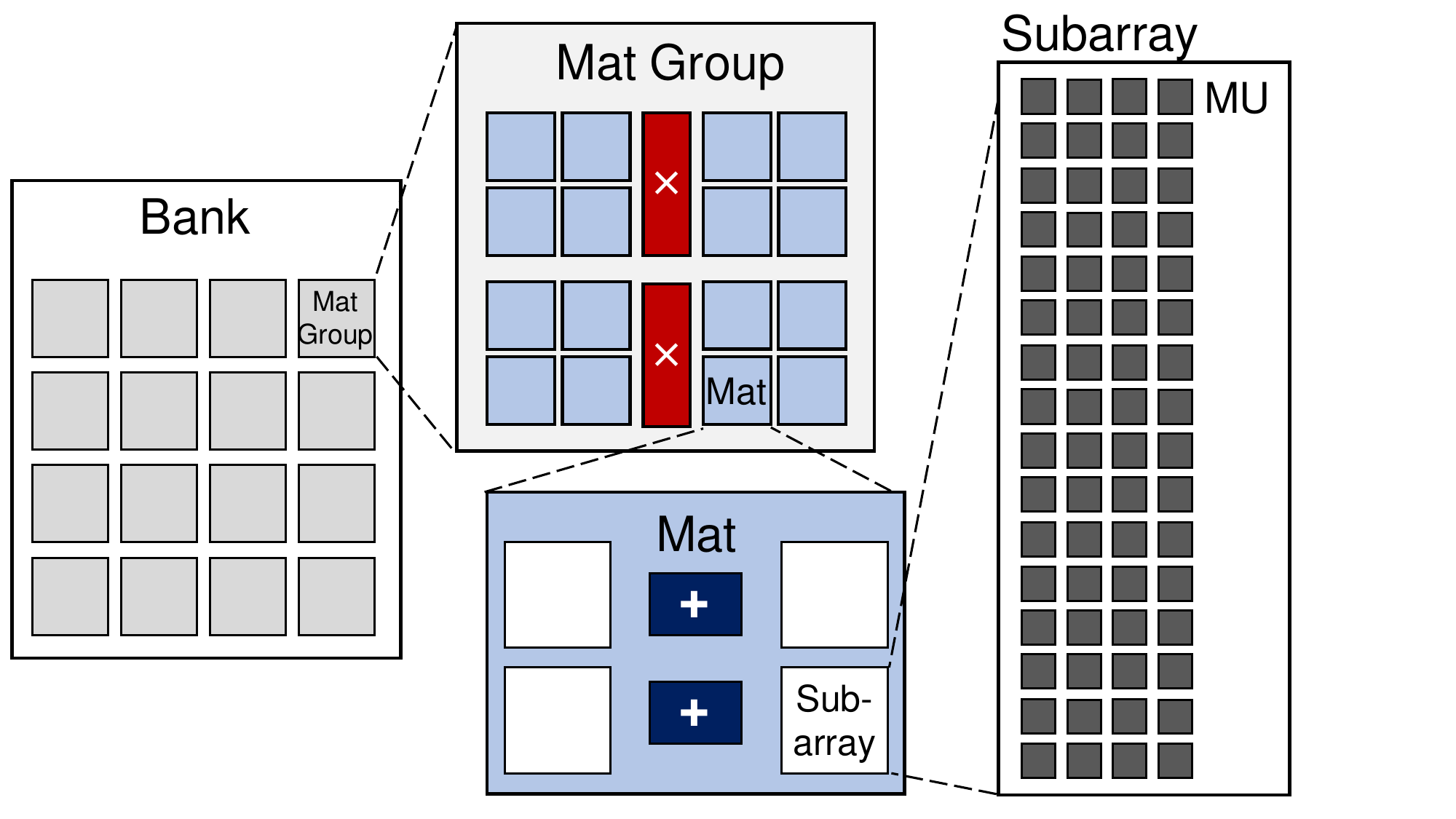}
	\caption{Memory organization of accelerator, as simulated in NVSim. The dark red blocks represent our proposed multiplier circuitry, whereas the dark blue blocks are adder unit blocks.}
	\label{fig:18}
\end{figure}

Each mat is built from four subarrays. Each subarray comprises an array of MUs, where only one MU of the subarray is accessed at a time. From the results of NVSim simulation, we organize a subarray with 16 rows and 4 columns of MUs. The memory organization detailed above is presented in Table \ref{table:8} and illustrated in Fig. \ref{fig:18}.

\begin{table}
\small
\begin{center}
\caption{Memory organization and parameters of in-memory accelerator.}
\label{table:8}
\begin{tabular}{|c|c|}
\hline
Bank capacity & 2MB \\
\hline
Mat groups/bank & 16 \\
\hline
Mat groups capacity & 256KB\\
\hline
Mats/mat group & 16 \\
\hline
Mat capacity & 16KB \\
\hline
Subarrays/mat & 4 \\
\hline
Subarray size (MUs) &  $16\times4$ \\
\hline
Subarray capacity &  2KB \\
\hline
\end{tabular}
\end{center}
\end{table}

As elaborated in Section \ref{subarray of MU}, all four tracks of an MU are accessed and shifted together. Hence, we implement arithmetic circuitry in replicated sets of four to exploit available parallelism. Let $A_0$, $A_1$, $A_2$ and $A_3$ be the four tracks of an MU in subarray $A$, and $B_0$, $B_1$, $B_2$ and $B_3$ be the four tracks of an MU in subarray $B$. For accumulation operations, we use four bit-serial adder units to perform four additions in parallel ($A_0$ with $B_0$, $A_1$ with $B_1$, $A_2$ with $B_2$, and $A_3$ with $B_3$). To reduce routing complexity, we do not allow accumulation to be performed across different tracks of the MUs (for example, $A_0$ cannot be accumulated with $B_1$). For multiplication with weights, all four tracks in an MU share the same weight and are shifted together, exploiting opportunities for weight reuse in CNNs.

Bit-serial adder units are integrated into selected mats. In convolution and fully-connected layers of CNN models, accumulation operations are performed on activations and bias terms, whereas weights are only used in multiplication. In our design, we use an asymmetrical storage of activations and weights: \textit{activation mats} are dedicated to storing activations in bit-serial format, whereas \textit{weight mats} are dedicated towards storing weights in bit-parallel format (to be used as the multiplier in Booth multiplication, or the shift distance in shift-based multiplication). As weight data is not involved in accumulation, we only insert bit-serial adders into activation mats.

\begin{figure}
	\centering
		\includegraphics[scale=.24]{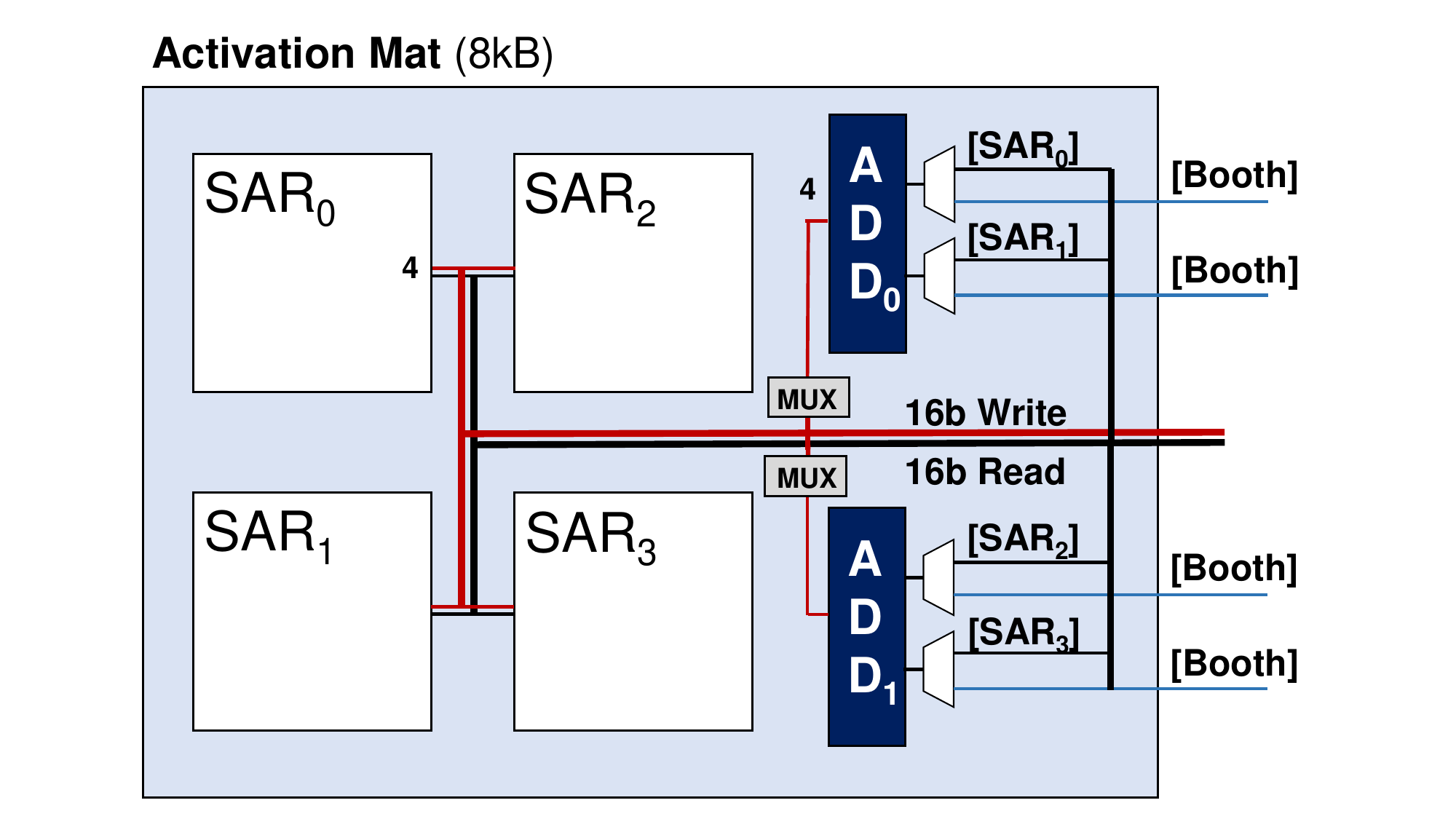}
	\caption{An activation mat consisting of four subarrays and two bit-serial adder units.}
	\label{fig:19}
\end{figure}

Fig. \ref{fig:19} depicts an activation mat. The activation mat comprises four subarrays (labelled $\text{SAR}_0$ to $\text{SAR}_3$) and two bit-serial adder units (labelled $\text{ADD}_0$ and $\text{ADD}_1$). Subarray accesses are performed using a set of 16 read wires and a set of 16 write wires (four for each subarray). Each adder unit can receive inputs from two subarrays: $\text{ADD}_0$ receives inputs from $\text{SAR}_0$ and $\text{SAR}_1$, whereas $\text{ADD}_1$ receives inputs from $\text{SAR}_2$ and $\text{SAR}_3$. The adder units then write the output sum to one of the four subarrays. Both adder units have write access to all four subarrays. Alternatively, the adder units can receive partial products from the Booth multiplier unit located externally, acting as the final stage of Booth partial product accumulation before products are written to subarrays.

\begin{figure}
	\centering
		\includegraphics[scale=.24]{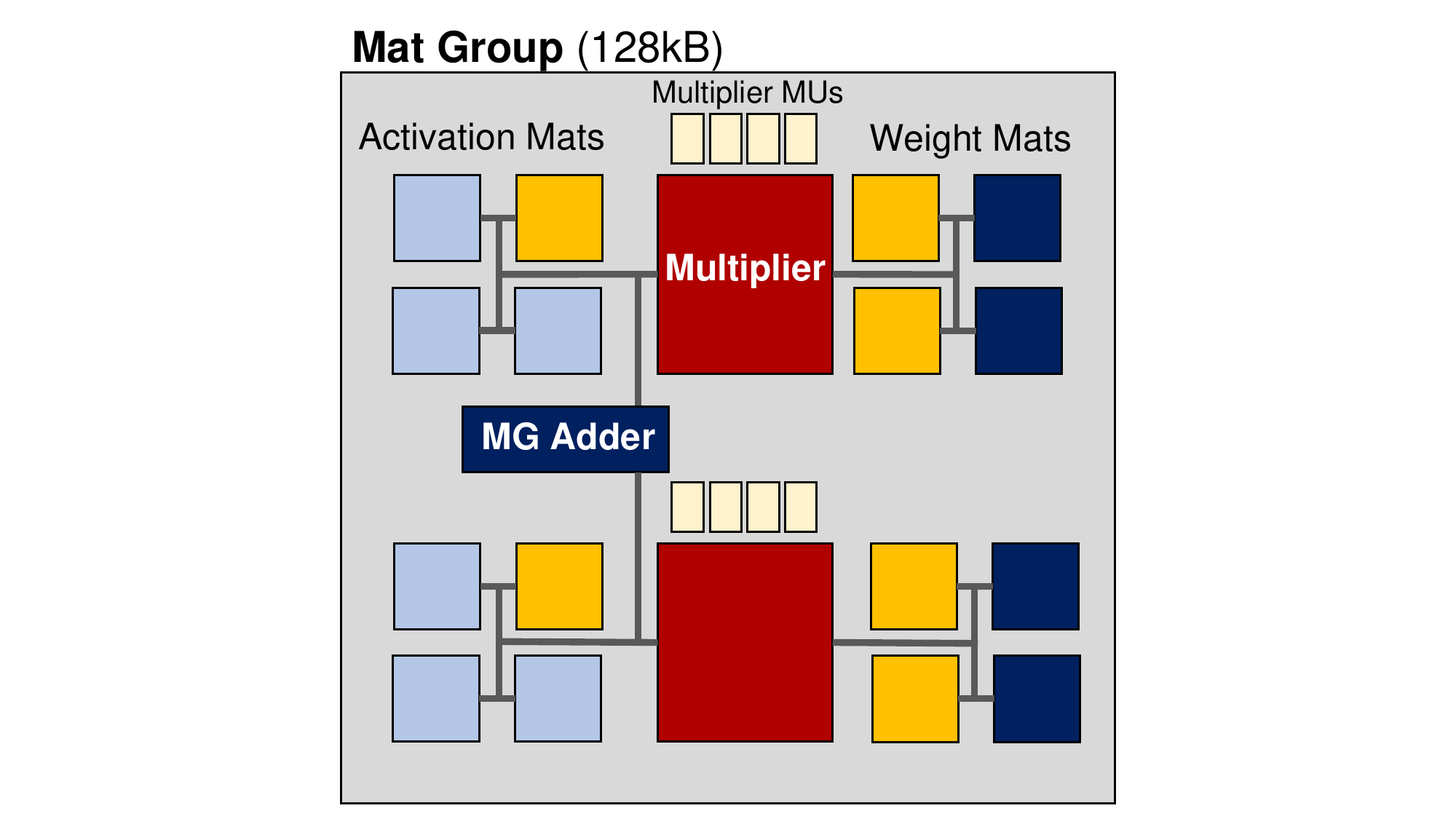}
	\caption{A mat group (MG) containing eight activation mats (light blue, left), eight weight mats (dark blue, right), two multiplier blocks and one adder unit. The mats accessed by the multiplier block during a multiplication are highlighted in orange. For simplicity, the connection of mats with the I/O H-tree is not shown.}
	\label{fig:20}
\end{figure}

Multiplication units are placed at the mat group level. Fig. \ref{fig:20} depicts a mat group. Each mat group has two multiplier blocks, 8 weight mats (right side of Fig. \ref{fig:20}) and 8 activation mats (left side of Fig. \ref{fig:20}). Each multiplier block comprises a Booth multiplier, a shift-based multiplier, and four MUs dedicated to storing generated Booth partial products. During multiplication, a multiplier block receives the model weights from a weight mat and performs relevant transformations on the activations received from an activation mat. Each multiplier block accesses activations from only one activation mat at a time, during which the MU of the mat accessed is in the access phase. When the multiplication is complete, the multiplier block then accesses activations from another mat, allowing the first mat to enter cycles of the position-reset phase. In this manner, we effectively hide the sequential access latency of racetrack memory by interleaving accesses across multipler mats. Furthermore, as the shape and order of operations of CNN inference are fixed beforehand, the correct mapping of data for such interleaving mechanism can be determined beforehand and controlled by software. More details on data mapping is explained in Section \ref{CNNMapping}.

Lastly, adder trees using the bit-serial adder units are implemented to accumulate activations across mat groups. Each bank has a 4-level adder tree implemented which receives an input from each the 16 mat groups. To support very large CNN models, additional adder trees can also be implemented to accumulate data across multiple banks, although at higher data access costs between banks.

%%%%%%%%%%%%%%%%%%%%%%%%%%%%%%%%%%%%%%%%%%%%%%%%%%%%%%%
\subsection{CNN Dataflow Mapping} \label{CNNMapping}
In this section, we demonstrate the mapping and computation of various layers of a CNN using the proposed in-memory accelerator. As convolution layers are typically the most intensive layers, we focus our discussion more on convolution layers.

The two key strategies employed in computing convolution layers include:

\begin{itemize}
    \item weight reuse across MU tracks, and
    \item immediate local accumulation of partial sums.
\end{itemize}   

Firstly, weight values are spatially reused: the four activations accessed from an MU share the same weight multiplier in a given cycle. As the convolution window moves across input activations during convolution, each weight is multiplied with several or all input activations (depending on the convolution stride). Hence, with a single weight access, the weight value can be reused for multiplication with several activations in parallel. The four tracks of an MU naturally support weight reuse as the tracks are accessed and shifted together. Weight reuse across tracks is especially important for shift-based multiplication as the four tracks share shift controls.
    
Secondly, partial sums are accumulated with newly generated products and reduced immediately to maintain small activation storage size. As shown in the mat group of Fig. \ref{fig:20}, our accelerator design dedicates half the memory capacity towards filter weight storage. As in-memory computing provides larger storage capacity located near computation logic, we choose to prioritize weight storage to allow more weights to fit in memory. If all model weights can fit in the dedicated storage capacity, the weights need not be transferred from off-chip storage for subsequent inferences. In order to dedicate more capacity for weights, we generate one partial sum for each output activation at a time and accumulate it immediately, reducing the capacity needed for partial sum storage.   

\subsubsection{Convolution Layers: Booth Multiplication}
\begin{figure}
	\centering
		\includegraphics[scale=.25]{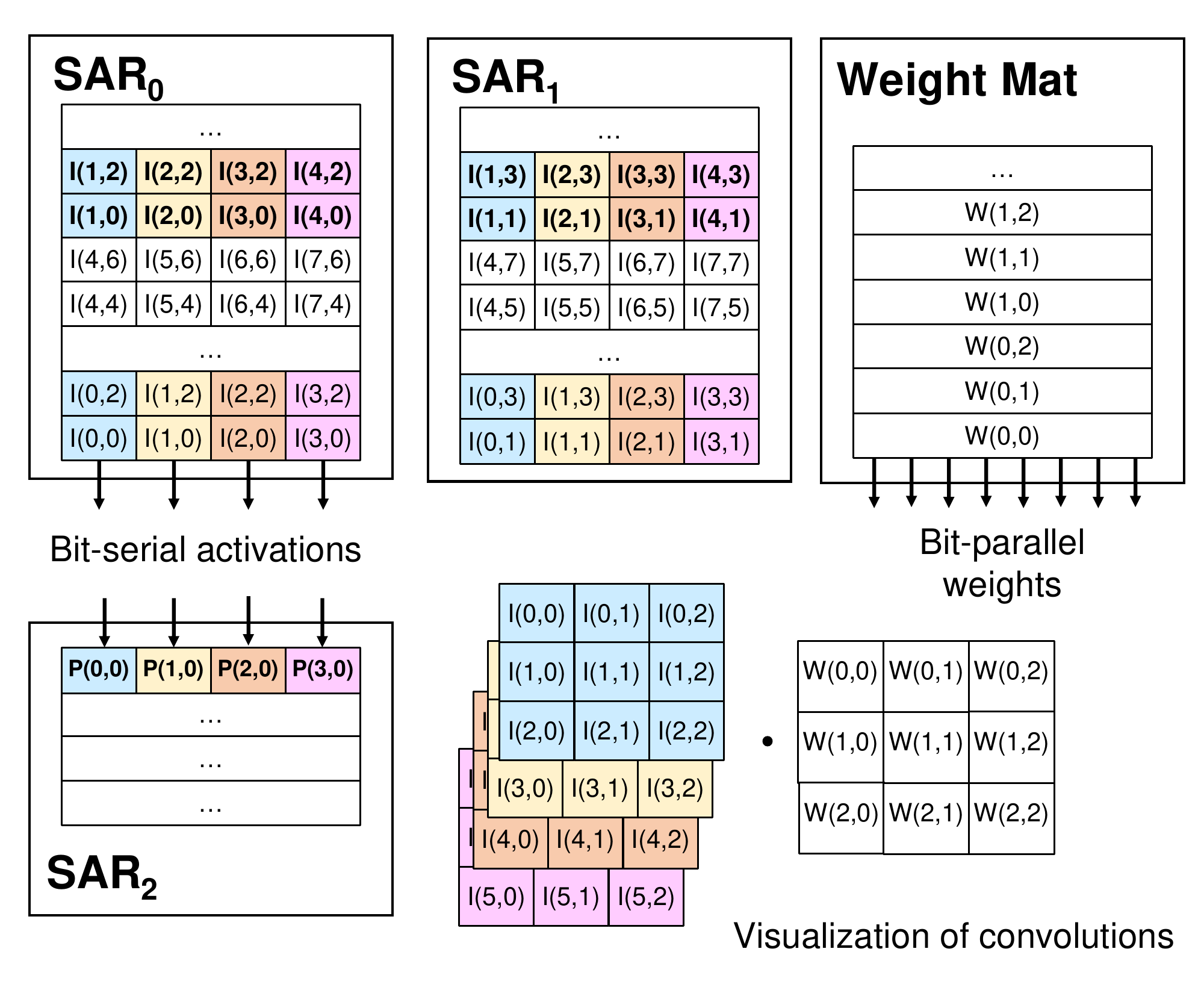}
	\caption{Data placement of activations and weights within subarrays.}
	\label{fig:21}
\end{figure}

Consider the example of an $8\times8$ input activation with 32 channels convolved with a $3\times3$ filter with a stride of 1. Fig. \ref{fig:21} illustrates the data placement of one input channel across two subarrays in a mat group, where $I(r,c)$ represents the activation at the $r^{th}$ row and $c^{th}$ column of the input channel. The figure also depicts the first filter channel stored in a weight mat of the same mat group, where $W(r,c)$ similarly represents the weight at the $r^{th}$ row and $c^{th}$ column of the filter channel. As weights are stored in bit-parallel format, each weight is distributed across several racetracks. 

As shown in Fig. \ref{fig:21}, input activations belonging to the different rows of the same column are placed together in the same positions across the four tracks of an MU. Moreover, activations of even-numbered columns are placed in the same subarray ($\text{SAR}_0$), whereas those of odd-numbered columns are placed in a separate subarray ($\text{SAR}_1$). In this convolution example, we use the term \textit{Booth multiplication} to refer to one pass of the Booth multiplier, where one weight is multiplied with four activations from an MU in parallel.

The first three Booth multiplications use the first row of weights ($W(0,0)$ to $W(0,2)$) for computation. During the first Booth multiplication, four activations belonging to the first four rows of column 0 are multiplied with $W(0,0)$. In effect, the Booth multiplier computes the first product of four different output products (denoted in different colors in Fig. \ref{fig:21}) in parallel. The four products ($P(0,0)$ to $P(4,0)$) are written to $\text{SAR}_2$. After the first Booth multiplication, $\text{SAR}_0$ enters the position-reset phase to correctly align domains for the subsequent access of activations in column 2. Meanwhile, activations of the first four rows of column 1 are accessed from $\text{SAR}_1$ and multiplied with $W(0,1)$ for the second Booth multiplication. This alternating access pattern prevents delay cycles due to the position-reset phase. The outputs products ($P(0,1)$ to $P(4,1)$) are stored in another activation mat $\text{SAR}_3$.

After completion of the second Booth multiplication, activations of column 2 are accessed for the third multiplication. While the Booth multiplier is generating partial product terms, the adders in the activation mat are not used by the Booth multiplier. Hence, the mat adder accumulates the output products from the first two Booth multiplications (currently stored in $\text{SAR}_2$ and $\text{SAR}_3$), writing the resulting partial sum back to $\text{SAR}_2$. In this manner, only a single partial sum per output is kept in memory at a time. At the end of the third Booth multiplication, output products of column 2 ($P(0,2)$ to $P(4,2)$) have been generated and written to $\text{SAR}_3$. These output products are similarly accumulated with the partial sums in $\text{SAR}_2$ during the fourth Booth multiplication cycle.

For the fourth to sixth Booth multiplications, the second row of weights ($W(1,0)$ to $W(1,2)$) are processed to generate products for the same outputs being accumulated in $\text{SAR}_2$. Eq. \ref{eq:12} shows the first four products generated from the first activation track for accumulation.

\begin{align} 
\small
    O(0,0) &= I(0,0) \cdot W(0,0) + I(0,1) \cdot W(0,1) \nonumber \\
    &+ I(0,2) \cdot W(0,2) + I(1,0) \cdot W(1,0) +... \label{eq:12}   
\end{align}

From Eq. \ref{eq:12}, the product generated from $I(1,0)$ is accumulated with that generated from $I(0,0)$. However, the two activations are placed in different racetracks of the MU, which cannot be accumulated due to their separate (parallel) arithmetic paths. Therefore, we store duplicate activations (bolded in Fig. \ref{fig:21}) to facilitate smooth convolution, allowing their products to be accumulated with the previous partial sums.

\subsubsection{Convolution Layers: Shift-based Multiplication}
The data placement of activations and weights for shift-based multiplication is the same as that of Booth multiplication. For simplicity of explanation, we describe only operations of one of the four tracks (blue track in Fig. \ref{fig:21}); however, the same operation is replicated across the four tracks in parallel. 

During a shift-based multiplication, two weight values ($W(0,0)$ and $W(0,1)$) are first written to decrementing counters controlling the shift circuitry of subarrays. Following this, two activations ($I(0,0)$ and $I(0,1)$) are accessed simultaneously, with each activation being accessed from a different subarray ($\text{SAR}_0$ and $\text{SAR}_1$ respectively). The decrementing counters control the shifting of each subarray during access to execute the shift-multiplication. Upon access, the shifted activation values are transferred to the adder unit of the mat, where they are accumulated and written to a third subarray ($\text{SAR}_2$), effectively completing a shift-and-add operation. While one bit-serial adder in the mat array accumulates shifted products, the other adder accumulates partial sums returned to subarrays by the first adder.

Considering the shift-and-add process above, the operation improves performance and efficiency over Booth multiplication due to several reasons. Firstly, data movement and computation is performed entirely within a single mat, reducing inter-mat data accesses which cost more energy than subarray accesses. Secondly, unlike Booth multiplication, shift-based multiplication generates no partial products, avoiding additional writes for storing and aligning these partial products. Lastly, throughput increases as shift-based multiplication consumes comparatively less resources and bandwidth, allowing more data to be processed in parallel. 

\subsubsection{Convolution Layers: Accumulation}
\begin{figure}
	\centering
		\includegraphics[scale=.27]{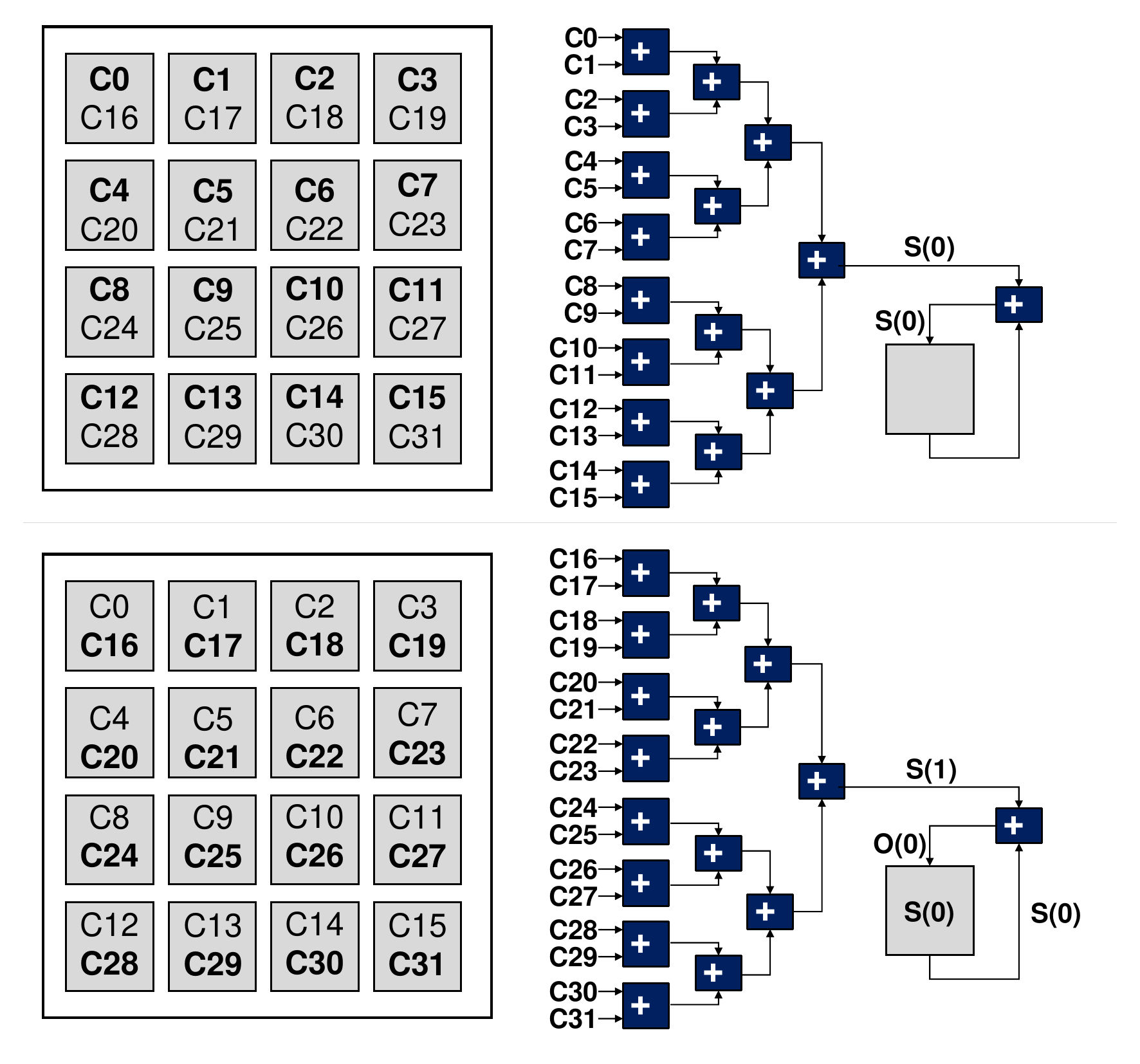}
	\caption{Accumulation between mat groups using the bank-level adder tree. The top half refers to the first accumulation pass, followed by that in the bottom half. Each mat group has two stored channels; the bolded one is accessed.}
	\label{fig:21a}
\end{figure}

In the example above, partial sums of a single channel are computed in a mat group. We distribute different input channels across the 16 mat groups for parallel processing. In an example involving 32 channels, each mat group processes 2 input channels. After local accumulation of partial sums is complete, the bank-level adder trees are used to accumulate partial sums across the 16 mat groups. Fig. \ref{fig:21a} depicts this accumulation, where $CN$ is channel $N$, $S(i)$ is partial sum $i$, and $O(0)$ is the output activation. As the adder tree processes 16 inputs at a time, accumulating across more than 16 channels will require more than one adder tree pass. For our 32-channel example, the first 16 channels ($C0$ to $C15$) are first accumulated using the adder tree and written to a vacant subarray in the destination mat of the output activation. Following this, the next 16 channels ($C16$ to $C31$) will first be accumulated by the adder tree, and the resulting partial sum ($S(1)$) accumulated with partial sum present in the destination mat ($S(0)$). The bit-serial adder within the mat adds the two numbers together to yield the output activation ($O(0)$).

For very large models unable to fit within a bank, the channels are distributed across multiple banks. Partial products across multiple banks are accumulated using the inter-bank adder tree in a similar manner. However, as data transfer energy costs are higher at the bank level, the accelerator performs accumulation at lower levels (mat or mat group level) as much as possible. 

\subsubsection{Other CNN Layers}
Fully-connected layers are commonly used as the final layers of CNN models to perform classification. In fully-connected layers, each output activation is a linear combination of the input activation, given by Eq. \ref{eq:13}.

\begin{equation} \label{eq:13}
\small
\begin{split}
&\mathbf{O}[n] =  
    ReLU\big(\sum_{m=0}^{M-1} \mathbf{I}[m] \times  \mathbf{W}[n][m] \big) \\
   &0\leq m<M, 0\leq n<N,\\
\end{split}
\end{equation}

The computation of fully-connected layers similarly involves MAC operations, and is performed in the same manner as convolution layers. However, unlike convolution layers, each weight value not reused across activations in fully-connected layers, providing no opportunity for parallel reuse of weights. As activation sizes of fully-connected layers are typically much smaller than that of convolution layers, we map activations to only one racetrack in an MU instead of 4. When computing fully-connected layers, the arithmetic circuitry of the other unused racetracks are switched off to save energy.

Batch normalization is another layer often used in state-of-the-art CNNs to accelerate model training and perform regularization. First proposed by \citet{BatchNorm2015}, batch normalization is applied to each channel in the activations as given in Eq. \ref{eq:14}, where $\mu$ is the channel mean, $Var$ is the channel variance, and $\gamma$ and $\beta$ are trainable parameters. These four parameters are determined for each channel during model training.

\begin{equation} \label{eq:14}
\small
\begin{split}
&\mathbf{O}[c][d][e] = \gamma[c] * \big(\frac{\mathbf{I}[c][d][e]-\mu[c]}{\sqrt{Var[c]}} \big) + \beta[c] \\
&0\leq c<C, 0\leq d<D, 0\leq e<E,\\
\end{split}
\end{equation}

During inference, the operation in Eq. \ref{eq:14} is applied to each activation. This computation is equivalent to one multiplication and two accumulations performed on each activation with the four parameters. The accelerator iterates over activations, first subtracting the channel mean $\mu$ from each activation using the mat adder unit. Next, each mean-subtracted activation is multiplied by $(\gamma / \sqrt{Var})$ of the corresponding channel using the Booth multiplier units, and accumulated with parameter $\beta$ within the mat.    

The last typical layer type in CNNs are pooling layers, such as average pooling and max pooling. In pooling layers, activations are divided into pooling windows, where each pooling window returns one output value. In average pooling layers, the output value returned is the mean of activations in the window. Using the example of a $2\times2$ pooling window, the four activations in the window are first accumulated using the mat adder unit, and then right shifted two positions to divide the number by 4. 

In max pooling layers, the output value returned is the maximum of the activations in the window. To compare the magnitude of values, we reuse the negation circuitry of the Booth multiplier and adder units to perform subtraction. For example, to compare the magnitudes of $I_0$ and $I_1$, we compute $I_0 - I_1$; if the result is positive, $I_0$ is greater than $I_1$, whereas $I_1$ is greater if the result is negative. The result value is used to select the appropriate activations for the max pooling output.  

%% file: tex/5_Experimental_Results.tex
\section{Experimental Results}
In this section, we present the experimental results of the proposed in-memory CNN accelerator. We first evaluate the energy consumption of proposed RM-based arithmetic units by varying designs parameters including operand bit-widths and architectures. Subsequently, we explore the interplay between CNN models and system architecture, and discuss the design space and trade-offs between inference accuracy, performance, and energy efficiency of the in-memory accelerator system.       

\subsection{Circuit-level Evaluation (Arithmetic Logic)}
We perform SPICE simulation of the RM-based arithmetic circuits proposed using a CMOS 45 nm design kit \cite{NanGate45}, and a model of perpendicular magnetic anisotropy racetrack memory based on the CoFeB/MgO structure from \cite{ZhangPMA2012}. The parameters of the model used are provided in Table \ref{table:6}. In addition, we utilize a racetrack memory model from \cite{Chao2015} to perform circuit-level evaluation of the performance and overheads of our RM-based multiplier units.

\subsubsection{Adders}
The performance and resource overheads of the proposed full adder (FA) and half adder (HA) circuits are tabulated in Table \ref{table:9}, along with those of a magnetic full adder of a previous work \cite{Trinh2013} for comparison. 
\begin{table}
\small
\begin{center}
\caption{Comparison of delay, energy and area of adder circuits.}
\label{table:9}
\begin{tabular}{|c|c|c|c|}
\hline
  & MFA \cite{Trinh2013} & Our FA & Our HA \\
\hline
Logic delay ($ps$)  & 180 & 240 & 153 \\
\hline
Energy for logic (fJ)  & 7.6 & 19 & 16.1 \\
\hline
MTJ Writes  & 16 & 7 & 4 \\
\hline
Area ($\mu m^2$) & 3.36 & 1.142 & 0.992\\
\hline
\end{tabular}
\end{center}
\end{table}
From the experiment results, our proposed adder units have a fewer number of MTJ writes and comparatively low area overhead, consuming 66.0\% less area than the MFA of \cite{Trinh2013}. The small area allows multiple full adders to be implemented for a higher degree of parallelism in computation. 

\subsubsection{Multipliers} \label{EvaluationMultipliers}
In this subsection, we first present the energy per bit of our proposed Booth multiplier unit at varying degrees of precision (number of bits). We compare the energy consumption between the Booth multiplier with write-shift transformation to that without. Next, we evaluate the energy per bit of the shift-based multiplier across different bit-widths as well as different shift distances. Lastly, we present and compare the latencies of both Booth and shift-based multiplier units.

\begin{figure}
	\centering
		\includegraphics[scale=.25]{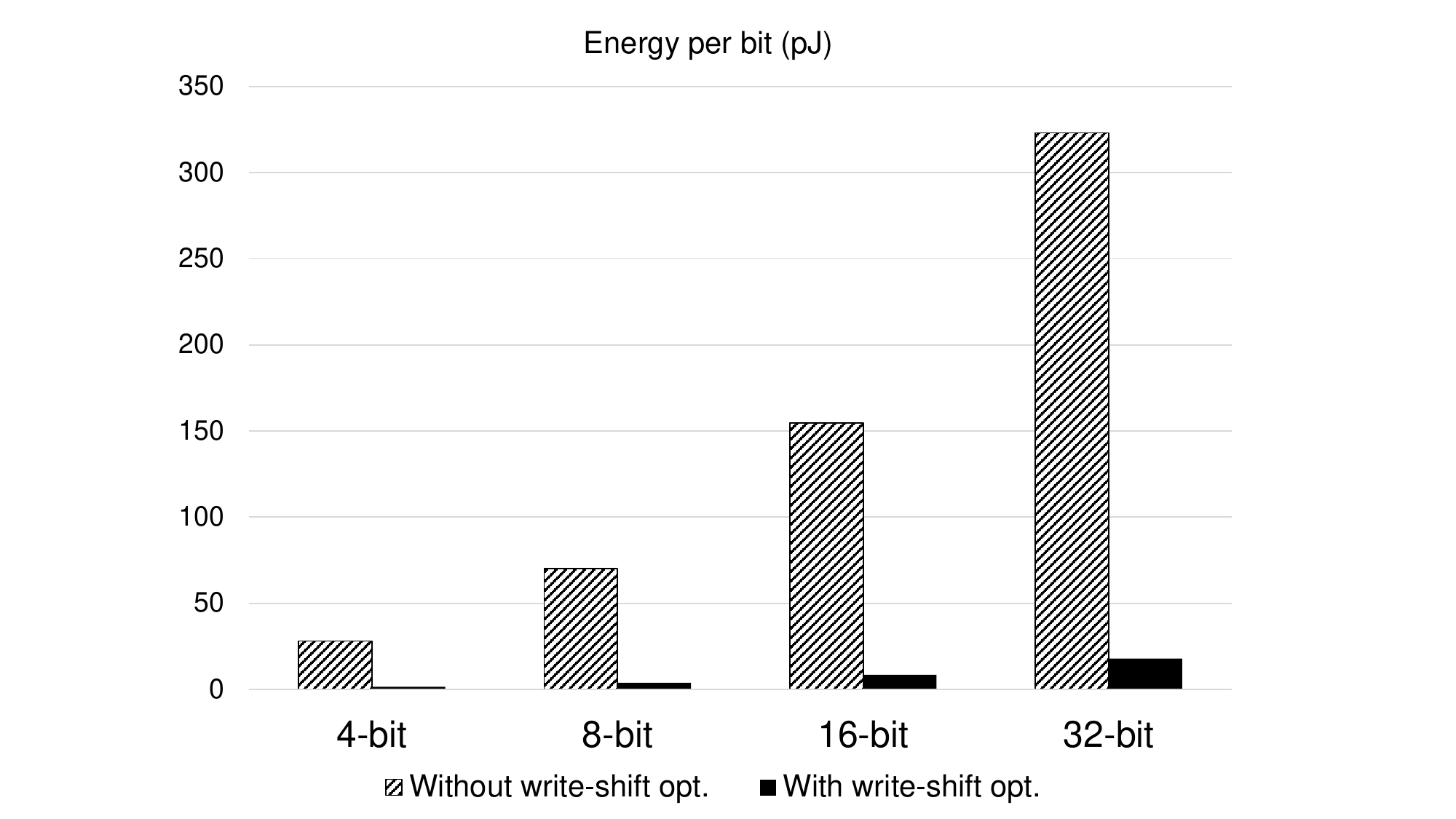}
	\caption{Energy per bit (pJ) of Booth multiplier at varying bit-widths, with and without the write-shift energy optimization.}
	\label{fig:BoothEnergy}
\end{figure}

Fig. \ref{fig:BoothEnergy} shows the energy consumption per bit of the proposed Booth multiplier unit at varying input bit-widths. The energy estimates in the plot account for only the computation logic of the Booth multiplier, including partial product generation and accumulation but not memory accesses. 

From Fig. \ref{fig:BoothEnergy}, we observe that the energy consumed per bit scales with the input bit-width: as the operand bit-width doubles, the energy per bit approximately doubles as well. This observation applies to both with and without the write-shift energy optimization applied. The reason for the doubling of energy is due to hardware replication for a greater number of partial products. The number of partial products generated is half the number of bits of the multiplier (e.g. for an 8-bit weight, four partial products are generated); when the bit-width doubles, so does the number of partial products. As the partial products are generated and accumulated in parallel, our design replicates a set of Booth CMOS control circuitry and full adder units for each partial product, resulting in greater energy consumed. 

Secondly, the write-shift transformation energy optimization reduces the energy consumed dramatically. We observe this dramatic reduction as the write operation costs the most energy, consuming 1 pJ per MTJ write. In contrast, the energy costs of other logic operations and the racetrack shift operation are at the fJ level. Furthermore, the energy consumed by the Booth multiplier is dominated by that of full adder units used in both partial product generation and accumulation. Hence, performing the energy optimization on the adder units can contribute significantly to the energy efficiency of the in-memory computation logic.

Following this, we evaluate the different overheads of the shift-based multiplier unit. We evaluate these overheads for both different multiplicand (activation) bit-widths, as well as different shift-distances $\pm d_{s}$ to investigate the effect of these two parameters on performance and efficiency. For this evaluation, we consider the operation of shifting two activations of $N_b$ bits and adding them together, as the shifting and accumulation process occur together.

\begin{figure}
	\centering
		\includegraphics[scale=.24]{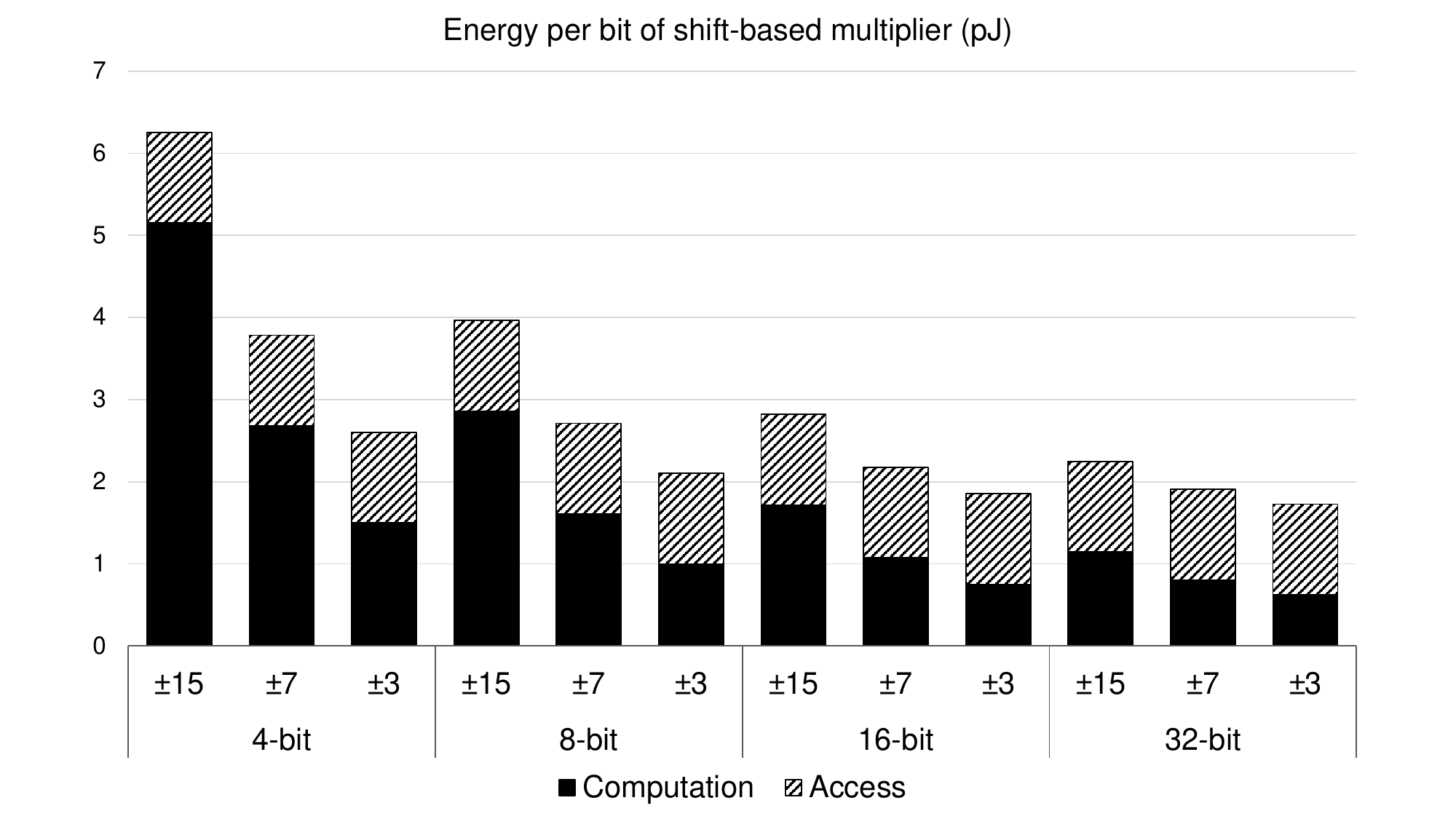}
	\caption{Energy per bit (pJ) of the proposed shift-based multiplier.}
	\label{fig:ShiftEnergy}
\end{figure}

Fig. \ref{fig:ShiftEnergy} presents the energy consumed per bit of the shift-based multiplier. We divide the energy into \textit{computation energy} which includes the shift control circuitry and the bit-serial full adder, and \textit{access energy} which includes shifting in racetrack memory and writing the final result back to memory. We observe that the access energy per bit remains constant for all bit-widths and is not affected by the shift distance. As we perform shift multiplication by enabling/disabling the shift circuitry rather than shifting all tracks together (as in \cite{DWMAcc2019}), each activation is only shifted and read $N_b$ times which does not depend on the shift distance. Hence, our design allows access energy per bit to be maintained, even with higher precision values and greater shift-distance.

\begin{figure}
	\centering
		\includegraphics[scale=.24]{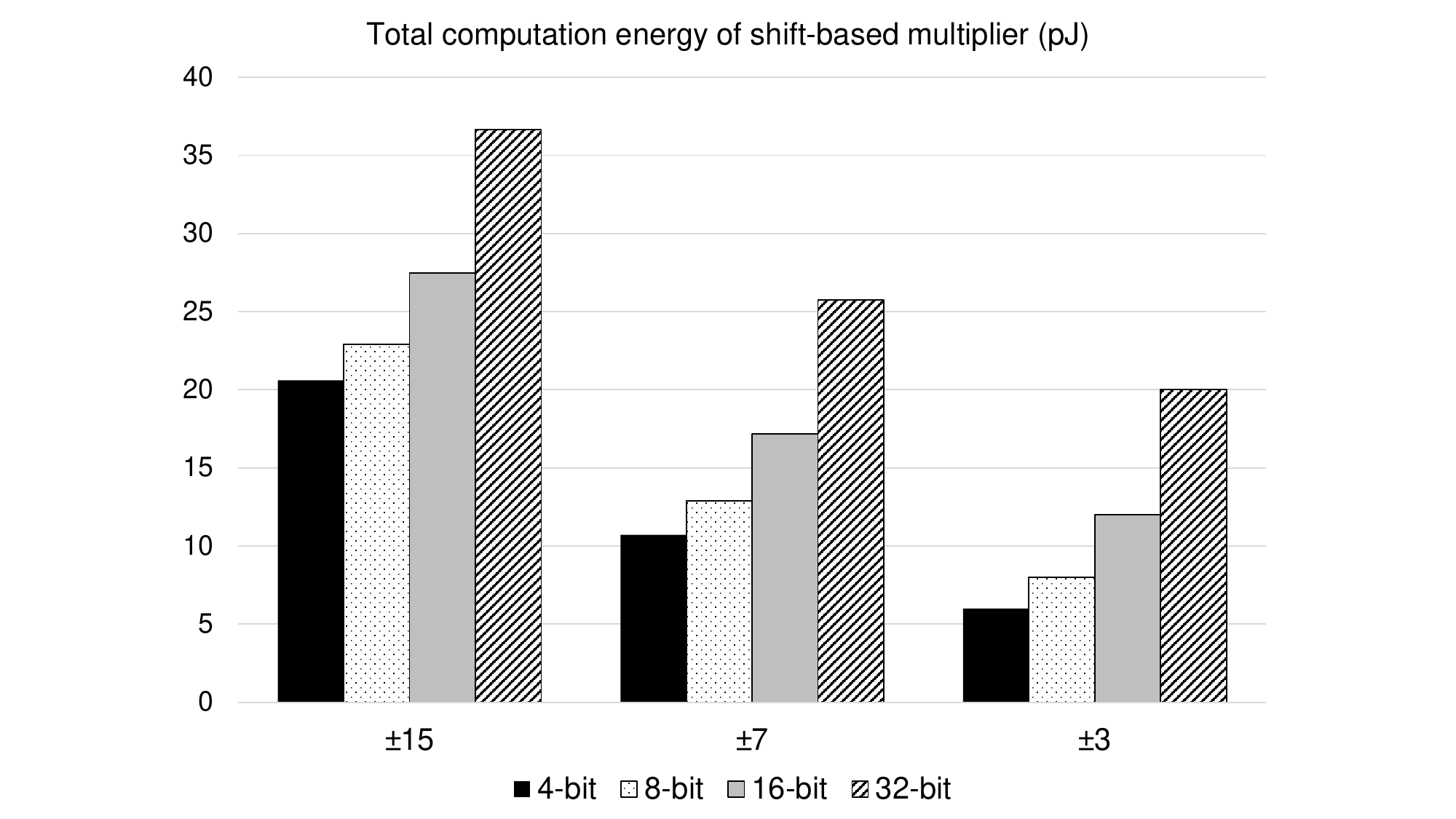}
	\caption{Total energy consumed (pJ) by the shift-based multiplier to perform a shift-and-add on two weights with two activations.}
	\label{fig:ShiftEnergyTotal}
\end{figure}

From Fig. \ref{fig:ShiftEnergy}, we observe that the computation energy per bit increases with increasing shift distance. This result is because the number of cycles during which the computation logic operates is given by $(N_b + 2d_{s})$, where $N_b$ is the activation bit-width and $d_{s}$ is the shift distance. A clearer visualization is shown in Fig. \ref{fig:ShiftEnergyTotal} which shows the total computation energy against the same parameters. We can observe that when the shift distance is kept constant, the increase in energy by moving from one bit-width to another is the same (for example, the difference in total energy between 8-bit and 16-bit inputs for $d_{s} = \pm 15$ is equal to the that between 8-bit and 16-bit inputs for $d_{s} = \pm 7$). Hence, we can consider both input bit-width and shift distance to be independent opportunities for increasing energy efficiency, where both terms should be as small as possible. However, in CNN applications, reducing these two parameters can result in a loss of accuracy, and so the amount to which both parameters can be reduced is limited.    

Regarding area, the shift-based multiplier reuses the shifting circuitry of racetrack memory and a single full adder for bit-serial accumulation. The only additional area costs is the CMOS decrementing counter used to enable and disable the shift circuitry. If a counter with sufficient bit-width is used, the same counter can be used to support different bit-widths and various shift distances. Hence, the additional area cost is minimal and does not vary significantly with the parameters.

Lastly, we evaluate the latencies of different multipliers in performing a MAC operation. The latency of a MAC is considered in cycles. For fair comparison of the Booth multiplier to the operation of the shift-based ones, we consider that two Booth multipliers are available to generate the two products in parallel, which are then accumulated. 

\begin{figure}
	\centering
		\includegraphics[scale=.24]{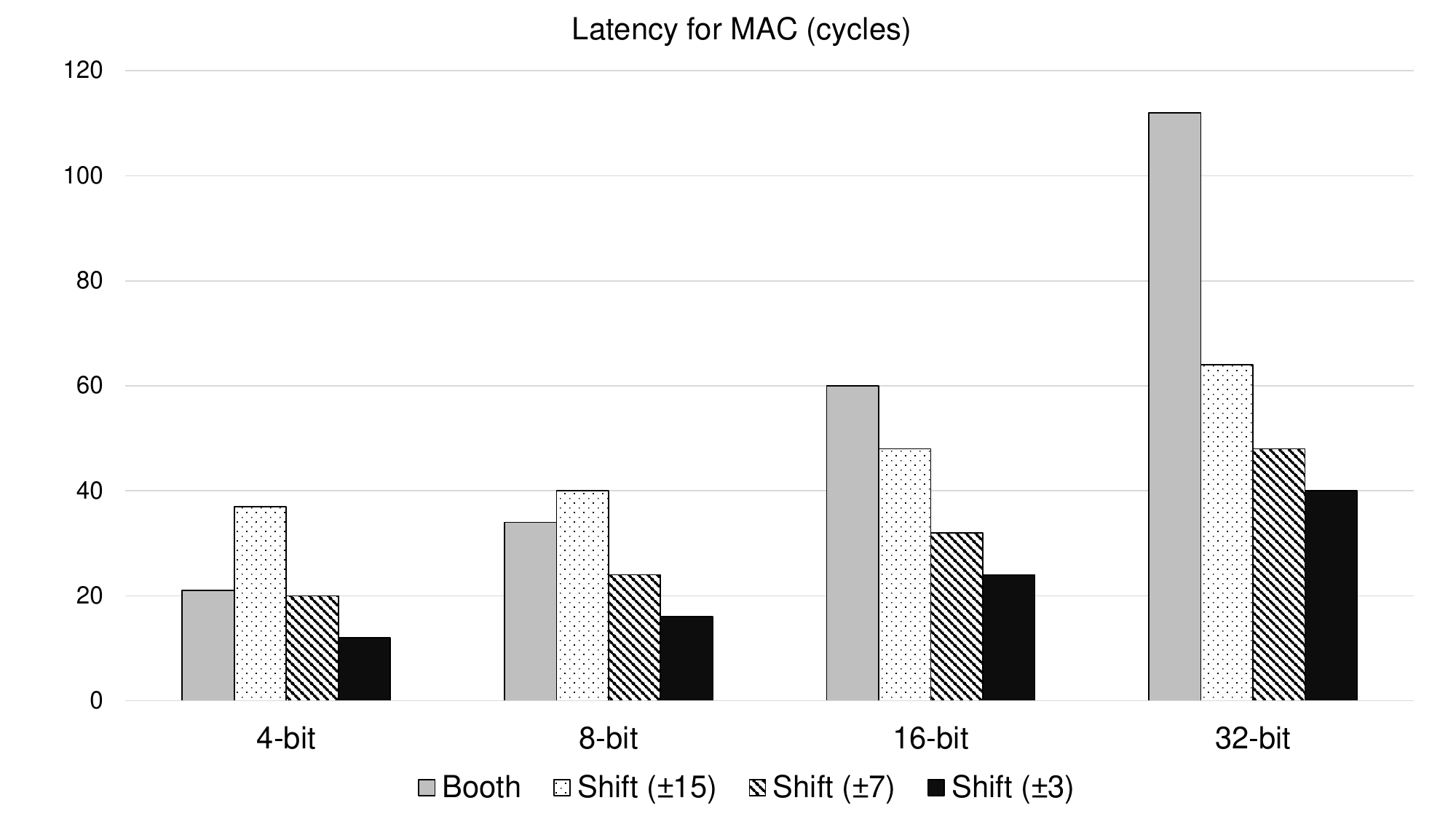}
	\caption{Latencies in cycles of different multipliers performing a multiply-and-accumulate (MAC) operation.}
	\label{fig:MultiplierLatencies}
\end{figure}

Fig. \ref{fig:MultiplierLatencies} presents the latencies of a MAC using different multipliers. From the figure, we observe that at low bit-widths, the Booth multiplier can achieve a lower latency than the shift-based multiplier with $d_{s} = \pm15$, as the latter has a minimum latency due to cycles needed for its shift distance. However, the Booth multiplier is more sensitive to operand bit-width, and increases in latency at a much faster rate with increasing bit-width than the shift-based multipliers. Furthermore, when using a shift-distance of $\pm 7$ or less, the shift-based multiplier always achieves a lower latency than the Booth multiplier across all bit-widths. As previous works in logarithmic quantization \cite{LightNN2018, LogNet2017} have found a shift-distance of $\pm 7$ to be sufficient to achieve reasonable accuracy in CNN inference, our findings suggest that shift-based multiplication can bring considerable improvements in terms of energy, area and latency over regular fixed precision of Booth multiplication in CNN applications.

%%%%%%%%%%%%%%%%%%%%%%%%%%%%%%%%%%%%%%%%%%%%%%%%%%%%%%%
\subsection{Model-System Co-exploration}
In this subsection, we simulate the acceleration of three CNN models: LeNet-5 \cite{LeNet5}, ResNet-20 \cite{ResNet20}, and VGG-16 \cite{VGG16} using our RM-based in-memory accelerator. The three CNN models are used for image classification, and represent models of different sizes and complexities. The model characteristics and datasets used for training are tabulated in Table \ref{table:10}.

\begin{table}
\small
\begin{center}
\caption{CNN models and datasets used for evaluation.}
\label{table:10}
\begin{tabular}{|c|c|c|c|}
\hline
 & LeNet-5 & ResNet-20 & VGG-16 \\
\hline
Dataset & MNIST & CIFAR-10 & ImageNet \\
\hline
No. of parameters & 61.7k & 270k & 138.38M \\
\hline
No. of MACs & 0.416M & 5.948M & 15.47G\\
\hline
FP accuracy (\%) & 98.0 & 91.4 & 94.6 (top-5)\\
\hline
\end{tabular}
\end{center}
\end{table}

Our model-system co-exploration was conducted as follows. First, we evaluated the test accuracy achieved by each CNN model when quantized to the specified bit-widths and across different schemes (linear or logarithmic). This process is described in Section \ref{ModelQuantization}. Next, the overhead costs (latency, area, energy consumption) of our proposed arithmetic circuits and memory banks are obtained using SPICE simulation and NVSim respectively. Using the obtained cost values, we developed a simulator to assess the total energy and latency of an inference pass using our in-memory CNN accelerator. We include all dynamic energy in an inference pass, including both computation energy of arithmetic circuits and memory access energies for reading, writing and shifting MUs as well. Through this process, we evaluate the energy, latency and accuracy achieved during an inference pass to explore key design trade-offs.

\subsubsection{CNN Model Quantization} \label{ModelQuantization}
LeNet-5 and ResNet-20 are built using the parameters and configurations of their respective original papers. LeNet-5 is implemented and trained on the MNIST dataset containing binary images of handwritten digits \cite{MNIST}. For ResNet-20, we follow the methodology of \cite{ResNet20} in building a 20-layer model for CIFAR-10, and train the model for 200 epochs. For VGG-16, we use a pre-trained model on ImageNet from the model zoo of PyTorch \cite{PyTorch}, and validate our model quantization methods on the validation images of ImageNet for ILSVRC2012 \cite{ILSVRC15}. 

The linear quantization scheme proposed in \cite{LogNet2017} is applied to convert activations and weights to fixed-point format. In this conversion, values are mapped to their closest value in the set $X_{lin}$ of Eq. \ref{eq:15}, where $x_{min}$ and $x_{max}$ are selected minimum and maximum values respectively, and $\Delta$ is the level size to yield a total of $2^{N_b}$ levels for $N_b$ bits. For experiments involving logarithmically-quantized weights, we similarly apply the "LogQuant" method in \cite{LogNet2017} to convert weights to their closest power-of-two value. 

\begin{equation} \label{eq:15}
\small
X_{lin} = \{x_{min}, x_{min}+\Delta , ... , x_{max}-\Delta , x_{max}\}
\end{equation}

In our implementation, the models are first trained with linear quantization on activations, while maintaining weights in floating-point format. We use the value of $x_{min}=0$ and $x_{max}=1$ for the magnitude of activations, giving activations $N_b$ bits of precision while preserving their sign. After training is complete, the weights are quantized to either linear or logarithmic quantization, after which the accuracy of the model is evaluated. For linear quantization of weights, we set $x_{min}=0$ and experiment with different values of $x_{max}$ from powers-of-two (1, 2, ... 32) and select the $x_{max}$ that yields the highest validation accuracy. 

\subsubsection{Model-specific Mapping} \label{ModelMapping}
As shown in Table \ref{table:10}, LeNet-5 and ResNet-20 are relatively small models, requiring approximately 61.7 kB and 270 kB of parameter storage for 8-bit parameters. As each bank in our accelerator dedicates 1 MB to parameter storage, a single bank is sufficient to store and execute both models. To minimize costly inter-bank data transfers, we map both models to perform inference entirely within a single bank. LeNet is executed across 8 mat groups (16 parallel multiplications), whereas ResNet-20 is executed across all 16 mat groups in the bank (32 parallel multiplications). We use these models to explore the trade-offs in efficiency, performance and inference accuracy in a single-bank system.

VGG-16 is a large model with approximately 138.4 MB of 8-bit parameters and computing 15.47 billion MAC operations in an inference pass. The model is used for complex classification of 1000 image classes. While such large and complex models are rarely used in embedded systems, we use VGG-16 as a corner case example to consider the scalability of our accelerator for very large models.

For VGG-16, we consider a memory system with 16 accelerator banks. The 16-bank system has a capacity of 32 MB and an area of 14.74 $mm^2$, which is selected to match those of previous works on in-memory computing \cite{DWMAcc2019, Eckert2018}. The 16 MB of dedicated parameter storage in this system is insufficient to hold the 138.4 MB of model parameters. Hence, we investigate the implementation of our system with external DRAM storage, where model data and parameters are stored in and transferred from DRAM when on-chip capacity is insufficient. Along with this setup, we consider the use and efficiency of our accelerator with batching inputs.

In mapping VGG-16, we partition the model into two parts: the \textit{convolution sub-model} and the \textit{fully-connected sub-model}. The convolution sub-model includes all layers of VGG-16 before the fully-connected layers, which consists of convolution layers and max pooling layers. The final three fully-connected layers before classification are included in the fully-connected sub-model. Table \ref{table:11a} presents details on the memory capacities needed for the two parts. From the table, we observe that it is possible to fit all parameters of the convolution sub-model in the dedicated weight storage (16MB) of the memory system. Conversely, the weights of the fully-connected sub-model are not reused and so occupy huge parameter size. Furthermore, we observe that for both sub-models, the maximum storage capacity needed for activations at a time is less than the dedicated activation capacity (16MB). Hence, we perform inferences of VGG-16 in batches to reduce external DRAM transfers of model parameters.

\begin{table}
\small
\begin{center}
\caption{Memory capacity required to store parameters and activations of VGG-16 sub-models on-chip.}
\label{table:11a}
\begin{tabular}{|p{0.35\linewidth}|p{0.2\linewidth}|p{0.27\linewidth}|}
\hline
& Convolution & Fully-connected \\
\hline
Parameter size (MB) & 14.7 & 123.6 \\
\hline
Max activation storage for a layer (MB) & 6.42 & 0.029 \\
\hline
\end{tabular}
\end{center}
\end{table}

We map an inference pass on a batch of images (with batch size $B$) as follows. First, model weights for the convolution sub-model are transferred from DRAM and stored on-chip. Following this, $B$ input images are transferred to the accelerator for computation of the convolution sub-model layers. Depending on the batch size, the activations generated in specific layers may be unable to fit within dedicated activation memory. For example, a batch of 4 images would require 25.7MB of activation storage for the second convolution layer of VGG-16, which cannot fit within the 16MB activation capacity. In these cases, activations are transferred to external DRAM until they are needed for the next layer.

After completely processing the convolution sub-model, a batch of input activations for the fully-connected sub-model is already on-chip. Hence, we stream the fully-connected sub-model parameters from DRAM and process a batch of activations in parallel, allowing weight reuse across batches. We explore the changes in overheads and savings with batch sizes in Section \ref{VGG16}.   

\subsubsection{Model Acceleration: Fixed Point}
\begin{figure}
	\centering
		\includegraphics[scale=.25]{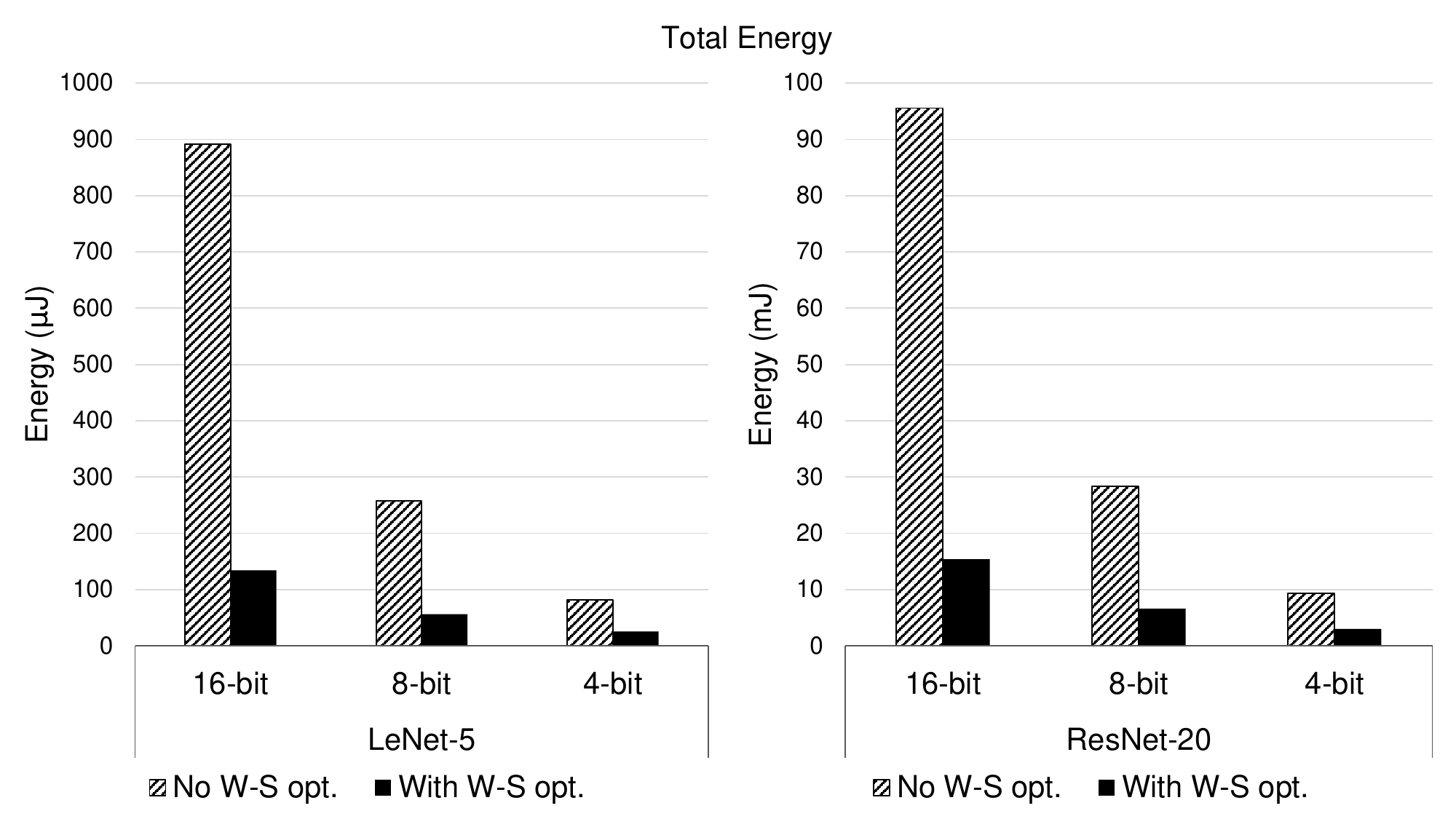}
	\caption{Comparison of total energy for LeNet-5 and ResNet-20 without the write-shift optimization to the energy after write-shift transformation is applied.}
	\label{fig:EvaluateWOpt}
\end{figure}

First, we evaluate the effectiveness of the proposed write-shift optimization at the system level in reducing energy consumption. Fig. \ref{fig:EvaluateWOpt} shows the total energy consumed by the accelerator in performing an inference pass for LeNet-5 and ResNet-20.

As observed in the figure, the write-shift transformation significantly improves energy efficiency at the system-level as well. The proportion of energy saved increases with increasing bit-widths, with 67\% of total energy reduced for 4-bit models to 83\% for 16-bit models in our experiments. As the write-shift optimization is applied only to arithmetic units, these high proportions suggest that computation energy is a significant proportion of our baseline design. From our analysis, applying the write-shift optimization greatly reduces the computation energy, such that access energy becomes the greater proportion of energy costs in the energy-optimized system. For the remainder of our discussion, we use the write-shift optimized system for exploring the wider design space.

\begin{figure}
	\centering
		\includegraphics[scale=.24]{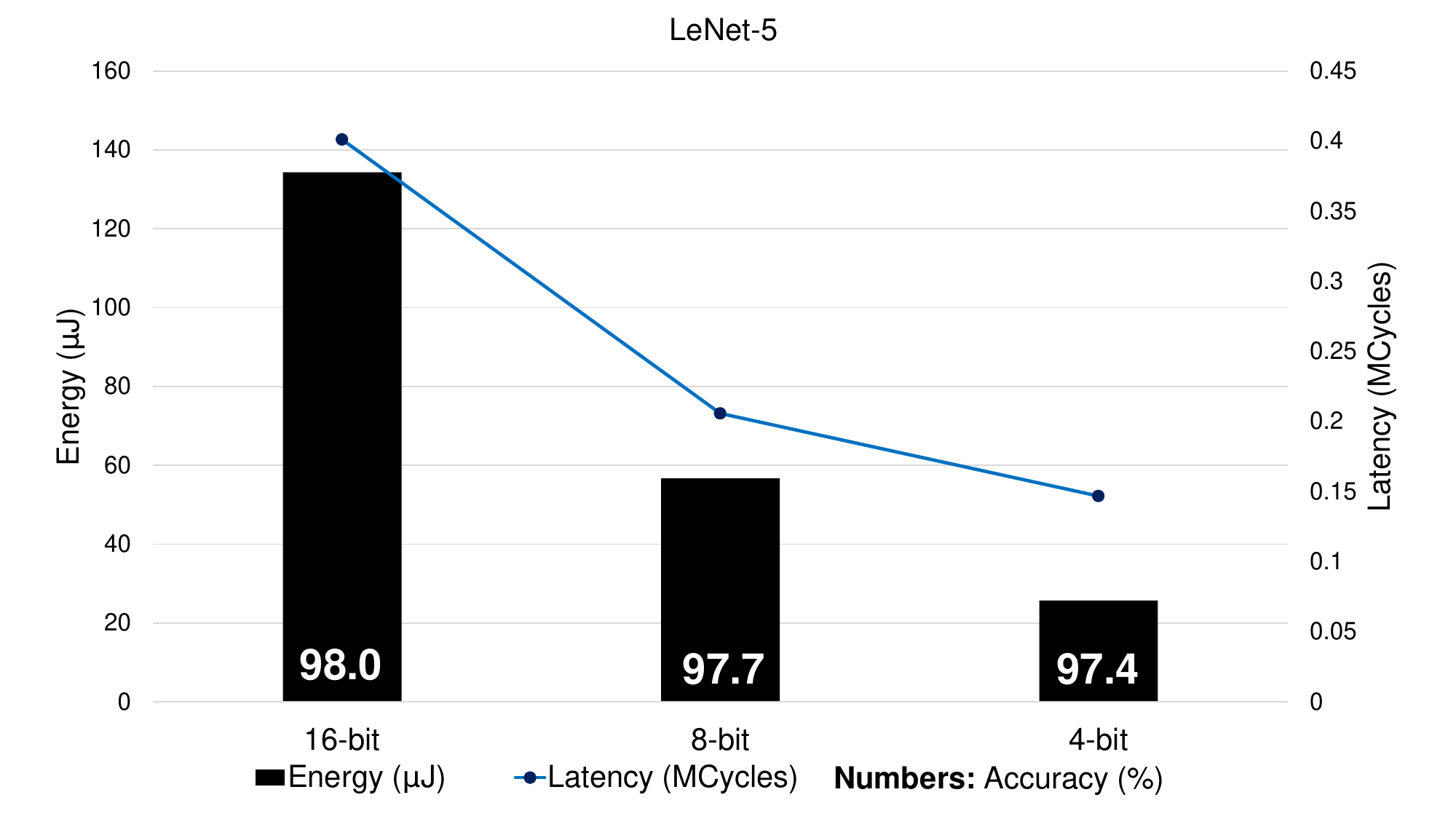}
	\caption{Total energy, latency, and accuracy of a LeNet-5 inference on our RM-based accelerator.}
	\label{fig:LeNetFix}
\end{figure}

\begin{figure}
	\centering
		\includegraphics[scale=.24]{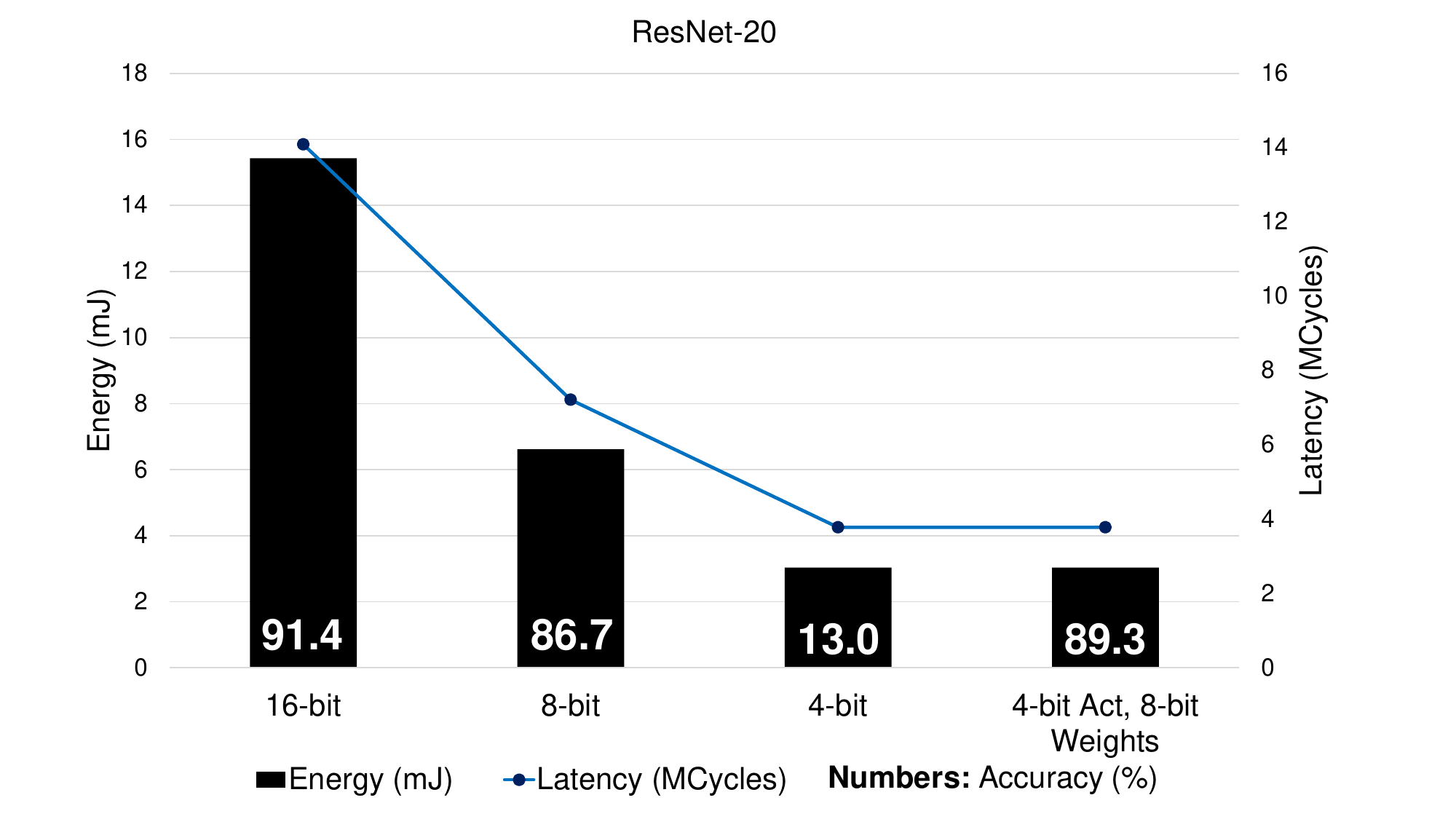}
	\caption{Total energy, latency, and accuracy of a ResNet-20 inference on our RM-based accelerator.}
	\label{fig:ResNetFix}
\end{figure}

Fig. \ref{fig:LeNetFix} and Fig. \ref{fig:ResNetFix} present the energy consumption, latency and accuracies of the LeNet-5 model on MNIST and the ResNet-20 model on CIFAR-10 respectively. We present the model accuracies to explore the trade-off between overheads and accuracy during prediction. We consider only bit-widths between 4 to 16 bits, which has been found in literature to be sufficient precision for CNN applications. All fixed-point models are computed using the Booth multiplier with write-shift transformation applied for better energy efficiency.

From both figures, we observe that both total energy consumed and latency scales with the number of operand bits. The scaling in latency is expected as our accelerator operates in a bit-serial manner, in which activations are accessed and processed one bit per cycle. The total energy consumed more than doubles with twice the bit-width as there is additional energy consumed over two fronts: hardware replication for more partial products, and increased number of write operations with each data access from memory. Hence, we observe that while write operations in arithmetic logic have been reduced from those of \cite{Trinh2013}, data access energy (specifically write energy) is still the dominant source of energy consumption in the entire system after applying the write-shift energy optimization. 

Following this, we consider the model accuracy achieved with varying bit-widths. For smaller models with less complex applications such as LeNet-5, less precision is required to achieve higher accuracy. For LeNet-5, accuracy loss is less than 1\% even when reduced to 4-bit weights and activations. Considering the energy and latency savings attained using lower precision, the 4-bit weight and activation configuration is optimal for implementing LeNet-5 on our accelerator. 
For more complex models such as ResNet-20, performing linear quantization on model weights without retraining can result in significant loss. Applying the post-training linear quantization scheme of \cite{LogNet2017}, we observe that model accuracy is sensitive to the bit-width used: reducing precision results in insignificant accuracy loss for 16-bit (91.4\%), 4.7\% loss for 8-bit (86.7\%), and 78.4\% loss upon quantizing to 4-bit values (13.0\%). 

In addition to fixed precision scheme above, we consider a varying precision configuration by using 4-bit activations while maintaining weights at a higher precision of 8 bits. From Fig. \ref{fig:ResNetFix}, this configuration is able to maintain high accuracy of 89.3\% (2.1\% loss) while incurring comparable energy and latency overheads to that of the 4-bit activation and weight configuration. From our experiments, the varying-precision model achieves even higher accuracy than the 8-bit model, which is counter-intuitive as precision is reduced. The higher accuracy could be because the model parameters adapt to a lower precision of activations during training, resulting in less precision required during post-training quantization of weights. More importantly, this observation suggests that the overheads of our bit-serial accelerator are influenced largely by activation bit-widths. We can afford to reduce activation bit-widths to reap the benefits of precision reduction, while maintaining higher precision of weights to maintain prediction accuracy with minimal additional overheads. 

\subsubsection{Model Acceleration: Logarithmic Quantization}
Next, we explore the hardware-software co-design space with the use of logarithmic quantization for shift-based CNN models. 

We first consider the changes in model accuracy due to logarithmic quantization of model parameters. In LeNet-5, we apply linear quantization with a fixed bit-width to activations during training. After training is complete, we perform logarithmic quantization to all weights and biases in the model (including both convolution and fully-connected layers). We investigate supporting different shift distances including $\pm15$, $\pm7$, and $\pm3$ (a shift distance of $\pm7$ supports the operation of left/right-shifting the activation by 7 positions, corresponding to "log4b" in \cite{LogNet2017}). 

The model accuracies for different activation bit-widths and different shift distances are shown in Table \ref{table:11}. We observe that for LeNet-5, a shift distance of $\pm7$ is sufficient to cover the range of weight values, as a greater shift distance of $\pm15$ brings no change to model accuracy. However, there is significant accuracy loss when shift-distance is reduced to $\pm3$.

\begin{table}
\small
\begin{center}
\caption{Model accuracies (\%) in applying logarithmic quantization to LeNet-5 with varying activation bit-widths and shift distances ($\pm d_{s}$).}
\label{table:11}
\begin{tabular}{|c|c|c|c|c|}
% \hline
% \multicolumn{1}{c}{} & \multicolumn{3}{c}{Shift distance} \\
\hline
\multicolumn{1}{|c|}{} & Fixed point & $\pm15$ & $\pm7$ & $\pm3$ \\
\hline
16-bit & 98.0 & 97.5 & 97.5 & 62.0 \\
\hline
8-bit & 97.7 & 97.7 & 97.7 & 70.3 \\
\hline
4-bit & 97.4 & 97.7 & 97.7 & 78.5 \\
\hline
\end{tabular}
\end{center}
\end{table}

For ResNet-20, we similarly apply linear quantization to activations during training. Next, we apply logarithmic quantization to only the weights and biases of convolution layers and fully-connected layers. The parameters of batch normalization layers are maintained at fixed point, as applying logarithmic quantization to batch normalization layers results in great accuracy loss, with all accuracy levels less than 64\%. As batch normalization layers are maintained at fixed point, these layers are processed using the Booth multiplier unit while convolution layers are computed using shift-based multiplication. In this manner, the flexibility of the accelerator in multiplication types (Booth or shift-based) allows more model layer types to be supported to maintain accuracy, while reaping the savings of shift-based multiplication where appropriate.

The model accuracy for ResNet-20 under logarithmic quantization are tabulated in Table \ref{table:12}. In applying logarithmic quantization, there is a 3.2\% loss in accuracy from the 8-bit fixed point model (86.7\%) to that of the shift-based model (83.5\%), and a 10.0\% loss in accuracy from the 16-bit fixed point model (91.4\%) to the shift-based model (81.4\%). The accuracy levels achieved from logarithmic quantization agree with the findings in \cite{LogNet2017}. Similar to LeNet-5, a shift distance of $\pm3$ is found to be insufficient to support the required model complexity, attaining a completely inaccurate model at 10\% accuracy (equivalent to a random guess for CIFAR-10 with 10 prediction classes).

\begin{table}
\small
\begin{center}
\caption{Model accuracies (\%) in applying logarithmic quantization to ResNet-20 with varying activation bit-widths and shift distances ($\pm d_{s}$).}
\label{table:12}
\begin{tabular}{|c|c|c|c|c|}
% \hline
% \multicolumn{1}{c}{} & \multicolumn{3}{c}{Shift distance} \\
\hline
\multicolumn{1}{|c|}{} & Fixed point & $\pm15$ & $\pm7$ & $\pm3$ \\
\hline
16-bit & 91.4 & 80.9 & 81.4 & 10.0 \\
\hline
8-bit & 86.7 & 83.5 & 83.5 & 10.1 \\
\hline
4-bit & 89.3 & 80.3 & 80.4 & 10.1 \\
\hline
\end{tabular}
\end{center}
\end{table}

In both models, the shift distance has significant impact on the accuracy of inference. Hence, the required shift distance for accurate inference can be determined by model validation beforehand. As we use a decrementing counter to control the shift-based multiplication process, our design can provide flexible support for multiple bit-widths by multiplexing between the desired counter bits as the control signal. The shift-based multiplier can thus be configured accordingly once the required shift distance is obtained.  

\begin{figure}
	\centering
		\includegraphics[scale=.26]{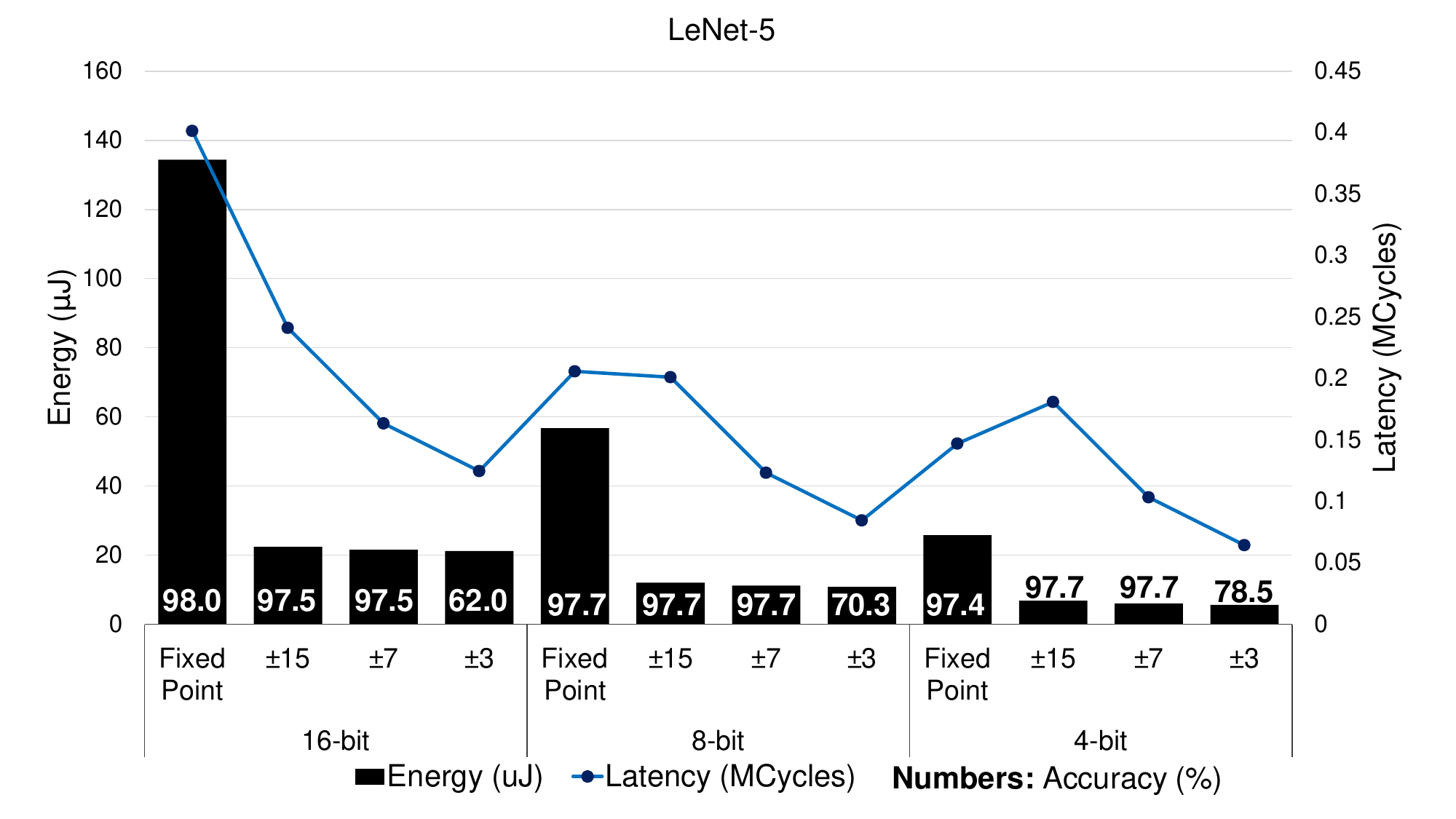}
	\caption{Total energy, latency, and accuracy of a LeNet-5 inference on our RM-based accelerator (fixed point and shift-based).}
	\label{fig:LeNetLog}
\end{figure}

\begin{figure}
	\centering
		\includegraphics[scale=.25]{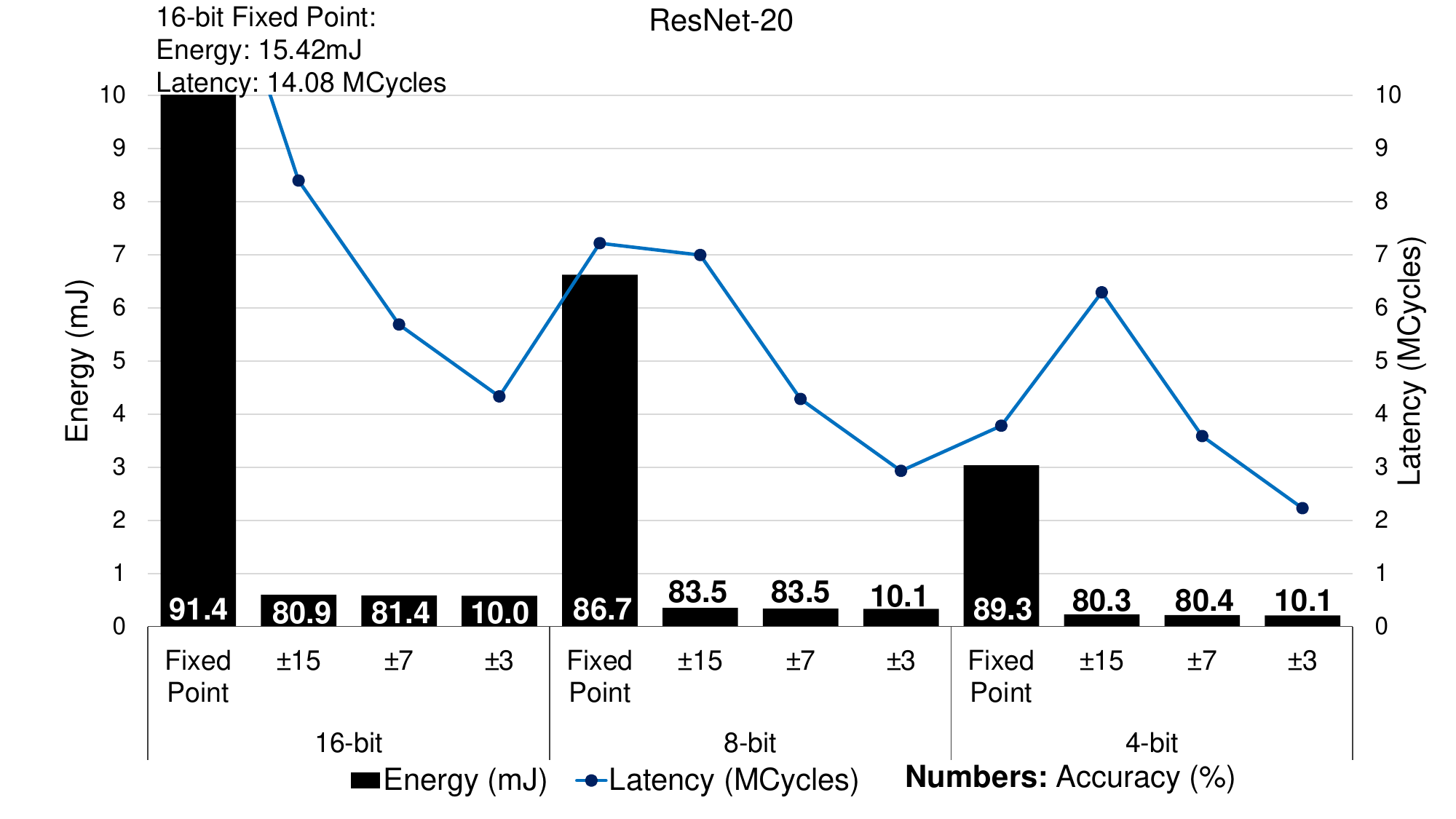}
	\caption{Total energy, latency, and accuracy of a ResNet-20 inference on our RM-based accelerator (fixed point and shift-based).}
	\label{fig:ResNetLog}
\end{figure}

Lastly, we explore the trade-offs between energy consumption, latency, and model accuracy for the shift-based CNN models in our accelerator design. Fig. \ref{fig:LeNetLog} and Fig. \ref{fig:ResNetLog} present the variation of overheads and accuracy levels of LeNet-5 and ResNet-20 respectively. For ResNet-20, estimations of energy and latency are made accounting for the execution of batch normalization layers using the Booth multiplier. Despite the accuracy loss for a shift distance of $\pm3$ in both models, we present their system overheads for more comparison insights. 

We observe several key trends from the figures. Firstly, the use of shift-based multiplication over Booth multiplication results in significant energy savings for both models, with over 70\% less energy consumed for LeNet-5 and over 90\% for ResNet-20. Furthermore, the energy savings are more significant in larger models that compute more MAC operations per inference. This result is due to Booth multiplication performing $\frac{N_b}{2}$ times more write operations for the $\frac{N_b}{2}$ partial products generated (where $N_b$ is activation bit-width), as well as the additional accumulation of these partial products as well. As write operations dominate energy consumption, the reduction of writes needed is expected to save energy dramatically, as is observed in the figures.

In addition, when considering the same activation bit-width, there is little change in total energy consumed with varying shift distance. Although the energy of the shift-based multiplication logic is related to both shift-distance and bit-width (as elaborated in Section \ref{EvaluationMultipliers}), the racetrack memory access energy (which depends on bit-width only) ultimately dominates energy consumption at the system-level once the write-shift optimization is applied. This observation further underscores the importance of developing new optimizations to improve to access overheads of racetrack memory technology.

The latency of an inference is more sensitive to varying shift-distance than bit-width. For both models, we observe that latency increases by only a small proportion across bit-widths for a fixed shift distance. However, the increase in latency across shift distances for a fixed bit-width is approximately twice that of increasing bit-widths instead. Hence, reducing the required shift distance brings performance improvements, but only to the point where the loss in accuracy is reasonable. In general, we observe that in the cases of LeNet-5 and ResNet-20, using a logarithmically quantized model with shift distance of $\pm7$ brings significant energy savings and reasonable improvements in latency, while maintaining an accuracy loss of less than 10\% compared to its corresponding fixed point model.

As a result of our model-system co-exploration, we compare the savings attained by our selected configurations for LeNet-5 and ResNet-20. For LeNet-5, we select the 4-bit shift-based model with shift distance of $\pm7$ which maintains a high accuracy of 97.7\%. Compared to the 8-bit fixed point implementation (commonly used in accelerators in literature), this 4-bit shift-based model yields 89.3\% energy savings and 49.8\% latency reduction on our accelerator with minimal accuracy loss. For ResNet-20, we use the 8-bit shift-based model with a shift distance of $\pm7$, which achieves 94.8\% less energy and 40.6\% lower latency, with an accuracy loss of 3.2\% from 86.7\% to 83.5\%.

\subsubsection{Scalability for Large Models} \label{VGG16}
We simulate VGG-16 to evaluate the performance and efficiency of our accelerator on large models. We use an 8-bit fixed point model which has been found in previous works to maintain sufficient accuracy \cite{ESSA2020}. In addition, we perform logarithmic quantization on the weights using a shift distance of $\pm7$, yielding a top-5 accuracy of 85.3\% (loss of 4.5\%) which is similar to the findings in \cite{LogNet2017}. These two models are used to compare between fixed point and shift-based models when scaled up to large model sizes.

\begin{figure}
	\centering
		\includegraphics[scale=.24]{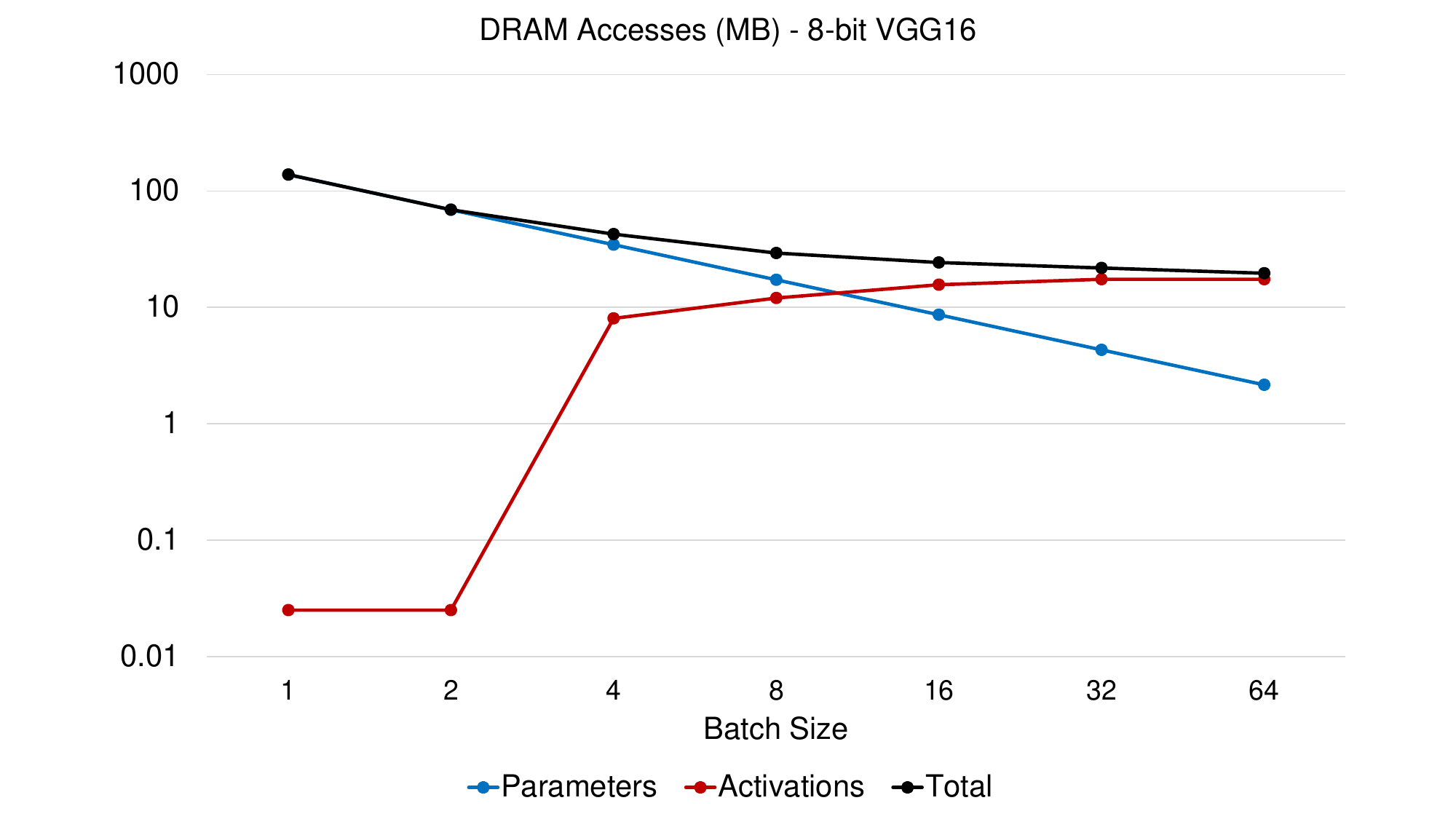}
	\caption{DRAM accesses per frame (image) for 8-bit fixed point VGG-16 by batch size.}
	\label{fig:VGGdram8}
\end{figure}

\begin{figure}
	\centering
		\includegraphics[scale=.24]{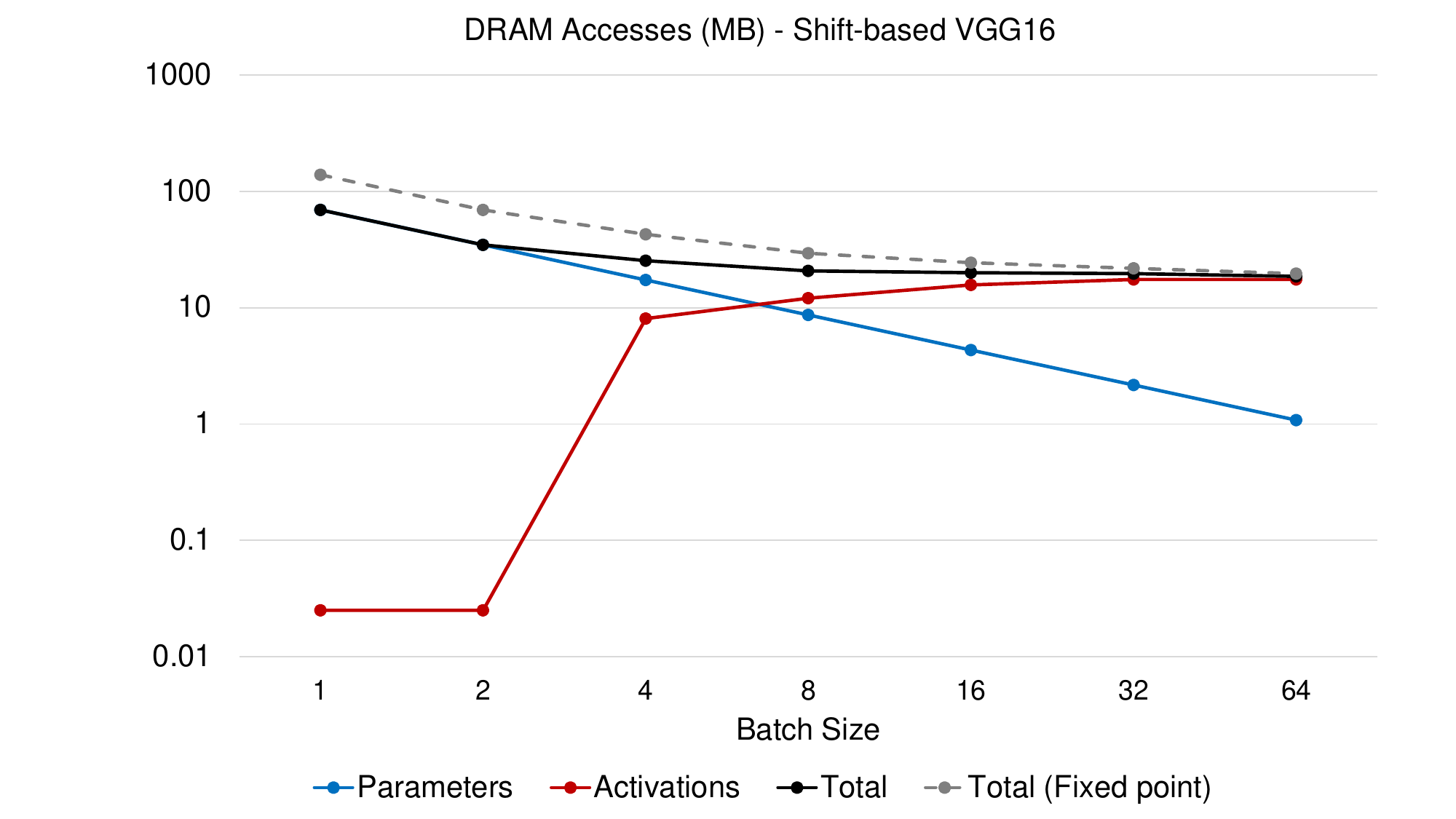}
	\caption{DRAM accesses per frame (image) for 8-bit shift-based VGG-16 by batch size.}
	\label{fig:VGGdramlog}
\end{figure}

As elaborated in Section \ref{ModelMapping}, we map VGG-16 to a 16-bank accelerator with 32 MB capacity. We perform inferences using power-of-two batch sizes from 1 (no batching) to 64. Fig. \ref{fig:VGGdram8} and Fig. \ref{fig:VGGdramlog} show the number of DRAM accesses required for 8-bit fixed point VGG-16 and shift-based VGG-16 respectively. Batching distributes the DRAM access cost of model parameters across multiple input images, allowing parameter accesses per image to be reduced as the parameters are reused on-chip. In our estimates, total accesses are reduced by more than 78\% for fixed point model and over 70\% for the shift-based model when using a batch size of 8 or greater.  

The shift-based model demonstrates similar DRAM access trends to the fixed point model. As parameters of the shift-based model have shorter bit-width than fixed point, the shift-based model has fewer accesses at lower batch sizes when parameter transfers dominate. However, the number of accesses per image for both models eventually converge at larger batch sizes when activation accesses are more significant.

When a batch size of 1-2 is used, all activations can fit within the accelerator's storage capacity for all layers; the reduction in accesses from batch size 1 to 2 is only from parameter reuse. However, beyond a batch size of 2, activations generated in convolution layers that exceed the on-chip storage are transferred to and from DRAM as well, reducing overall DRAM access savings. Moreover, we observe that DRAM transfers of activations surpasses that of parameters beyond a batch size of 16 for fixed point VGG-16, and beyond a batch size of 8 for the shift-based version. While batching still brings overall access savings in our considered case, we observe that activation accesses can easily offset these savings without applying an optimal mapping strategy and batch size.  

\begin{figure}
	\centering
		\includegraphics[scale=.24]{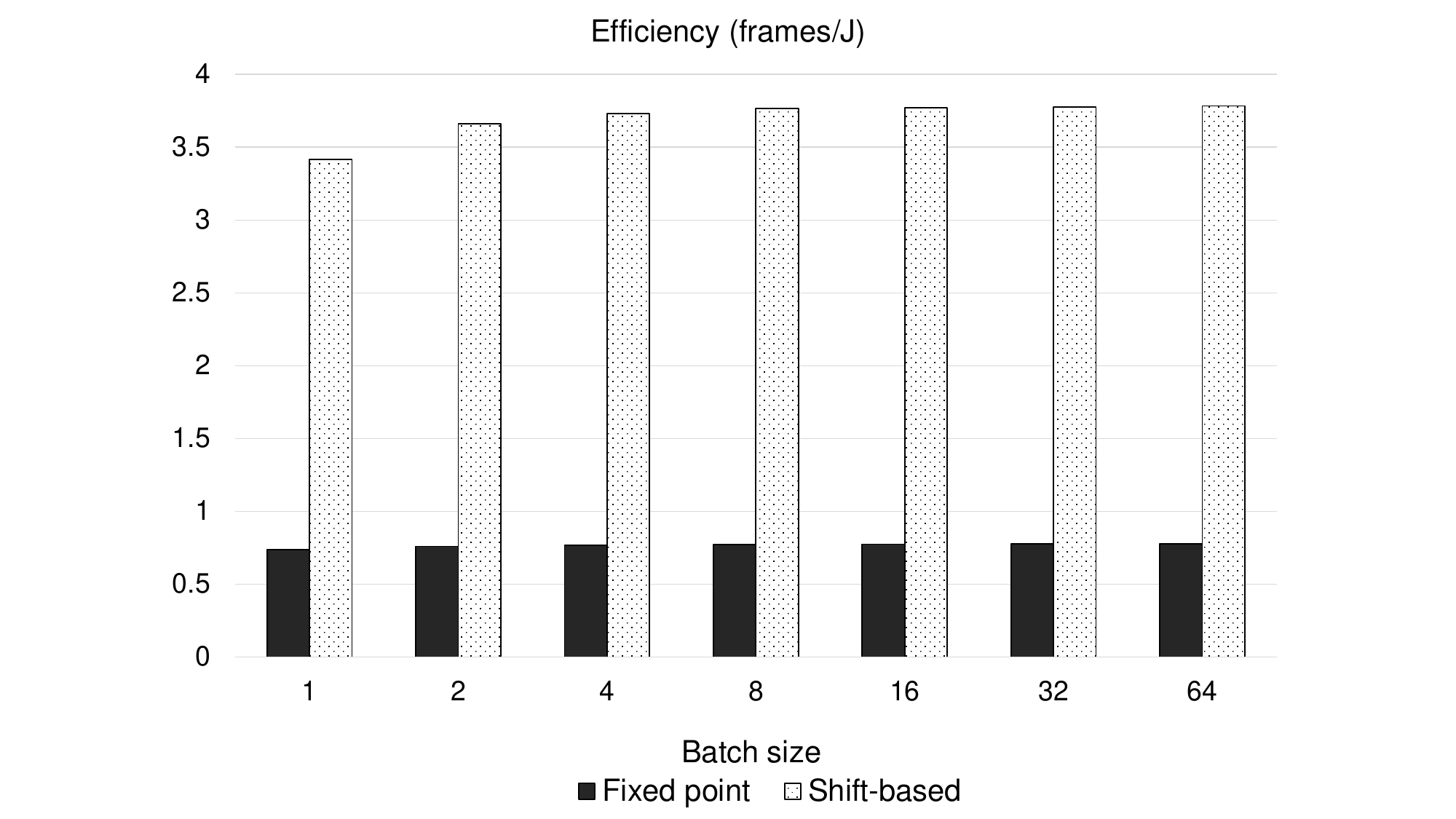}
	\caption{Energy efficiency of accelerating VGG-16 in a 16-bank 32 MB system (higher is better). The estimate includes energy consumed for DRAM accesses of model parameters and activations.}
	\label{fig:VGGEfficiency}
\end{figure}

Fig. \ref{fig:VGGEfficiency} presents the energy efficiency of both the fixed point and shift-based models at varying batch sizes. In this estimate, we include computation energy and dynamic access energy on-chip, as well as energy consumed for DRAM accesses estimated at 70 pJ/bit \cite{ESSA2020, Malladi2012}. From the figure, the energy efficiency of the shift-based model is significantly higher than the 8-bit model when considering large models with DRAM accesses as well. This observation is expected as the shift-based model consumes significantly less computation energy, and reduces energy costs of DRAM accesses as well due to smaller parameter bit-widths as well. Furthermore, we observe that the energy efficiency of the accelerator system does not vary significantly with increasing batch size. This result is due to considerable energy consumed for on-chip data access, which does not depend on batch size. From our analysis, we find that the write energy of racetrack memory continues to be the dominant energy cost for larger models and systems as well. At the same time, DRAM accesses that traditionally dominate energy costs are reduced as in-memory computing provides larger memory capacity directly accessible by computation logic.        

From our results on batching, we select the batch size of 8 for our implementation of VGG-16. The batch size of 8 achieves similar reduction in DRAM accesses as those higher than 8, while reducing latency of transferring activations between layers.

Lastly, we evaluate the efficiency and performance of our accelerator system across our selected configurations of LeNet-5, ResNet-20, and VGG-16. In this evaluation, we use our findings of hardware-software co-exploration to select the optimal activation precision for each model. We then compare the models across three configurations: fixed point without write-shift optimization, fixed point with write-shift optimization, and shift-based model.

\begin{figure}
	\centering
		\includegraphics[scale=.26]{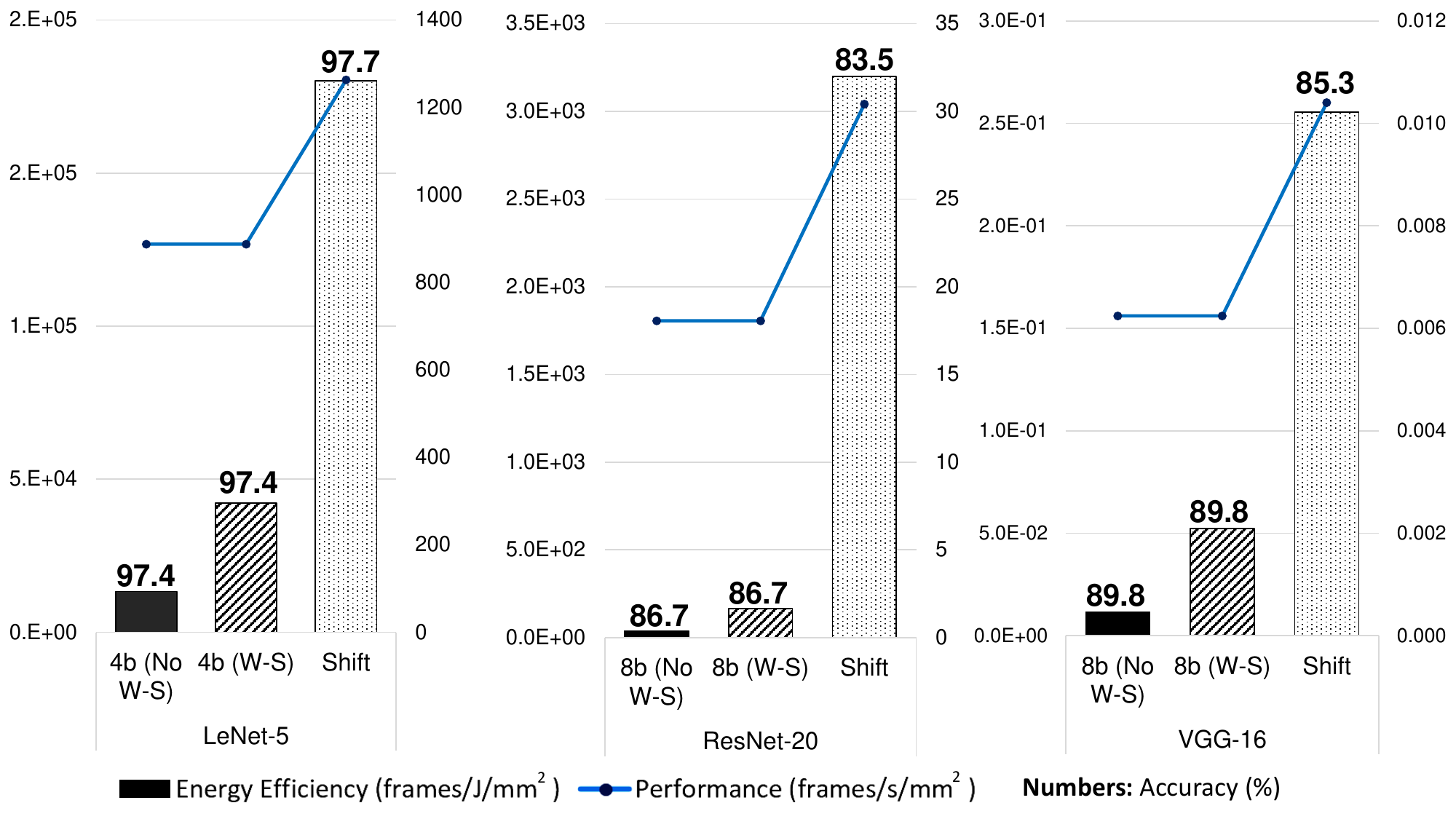}
	\caption{Energy efficiency, performance, and accuracy of the three CNN models across different configurations.}
	\label{fig:EfficiencyPerformance}
\end{figure}

Fig. \ref{fig:EfficiencyPerformance} presents the efficiencies and performance of our accelerator on the three models. The efficiency and performance metrics are scaled according to accelerator area (0.92 $mm^2$ for single-bank, 14.74 $mm^2$ for 16-bank). As previously shown, the write-shift energy optimization and conversion to shift-based multiplication influence energy efficiency significantly, while shift-based multiplication brings performance improvements as well. The difference across configurations is more pronounced in ResNet-20 compared to LeNet-5, as the former requires more MAC operations per inference. Although VGG-16 is the largest model with most computations performed among the three models, the large volume of on-chip accesses and additional need for DRAM transfers reduce the savings that the optimizations bring. We consider large models as VGG-16 to be the corner case, as CNN models developed for embedded systems are typically smaller and less parameterized. Overall, the application of write-shift optimization and using shift-based multiplication achieves greatest improvements on ResNet-20, with 83.5$\times$ better energy efficiency and 1.68$\times$ higher performance on our RM-based accelerator, with 3.2\% accuracy loss.

%% file: tex/6_Related_Works.tex
\section{Related Works}

% SECTION ON MEMORY TECHNOLOGIES
\textit{Non-Volatile Memory Technology.}
Emerging nano-scale non-volatile memories (NVM) are increasingly attracting research interest due to their ultra-low leakage power and high data density. Many works propose to exploit the advantages of these emerging memory technologies in design perspectives, including technologies such as resistive random access memory (RRAM), phase-change memory (PCM), spin-transfer-torque RAM (STT-
RAM), racetrack memory, etc. \cite{kang2017compact, luo2017novel, song2017stt, wang2018ultra}. Among these works, racetrack memory has become a research focus because of its higher data density compared with other NVM technologies. Numerous previous works employ racetrack memory~\cite{luo2018fpga, sun2013cross, wang2014energy, xu2015multilane, zand2017energy}, and propose using the technology for various purposes and applications. The works of \cite{mao2014exploration, mao2017energy, venkatesan2014stag} proposed using racetrack memory to overcome the limit of GPGPU scaling, whereas the works in \cite{wang2016dw, wang2014energy} computed AES encryption in-memory by integrating logic as well. \citet{zhao2013racetrack} and \citet{venkatesan2013dwm} proposed a shift-based energy optimization to the racetrack memory access at the device level. Their work reduces the energy in domain wall nucleation in RM-based cache, which is possibly complementary to our write optimization approach in developing an in-memory computing system.

% SECTION ON CIRCUITS
\textit{In-Memory Computing Logic.}
The adder is the most basic building block of arithmetic logic in digital computing. Although designs of CMOS adders have been comprehensively explored, these designs usually can't be migrated to in-memory technologies and designs directly, as it results in large area and energy consumption for high speed. Thus, a number of magnetic adders are developed and optimized for energy and area efficiency \cite{matsunaga2008fabrication, meng2005spintronics}. 
For example, \citet{Trinh2013} proposed a magnetic full adder (MFA) based on racetrack memory in 2013. Their full adder design utilizes 14 transistors and 16 MTJs, consuming significant energy to transfer input data due to the many MTJs used.
\citet{riente2021parallel} presented a conceptual racetrack array in which tracks can be reused as both memory and reconfigurable logic gates. Parallel bit-wise logic functions are achieved by superposing the magnetic fringe fields of tracks with an external Zeeman field, which nucleates the logical result on an output track. While their design provides a building block for flexible, bit-parallel in-memory logic, their design relies on racetrack writing as well.
Conversely, our circuits are racetrack-based arithmetic units that reduce the number of MTJs and write operations required for greater energy efficiency.

% SKYRMION RACETRACK MEMORY 
\textit{Skyrmion-based Racetrack Memory.}
Another emerging magnetic technology and a potential replacement for domain walls are \textit{skyrmions}. Skyrmions are "particles" of magnetic spin configurations; their presence or absence in racetrack memory is used to represent binary states. Skyrmions can achieve greater storage density and energy efficiency than domain wall-based memory, but face readability challenges \cite{kang2018comparative}. Similar to domains, skyrmions are transferred down a racetrack by a shifting current, and are read or written using MTJ ports as well \cite{luo2021skyrmion}. As both types of racetrack memory have the same shifting and access interfaces, our proposed MTJ-based arithmetic units can be directly applied to a skyrmion-based memory array as well.

Skyrmion-based logic gates have also been proposed in recent works. These gates use the movement of skyrmion "particles" for computation, preventing energy costly MTJ writing. \citet{zhang2015magnetic} designed spin logic gates using nanowires of varying widths, as well as gates for skyrmion duplication and merging. \citet{Liu2017PIMCNN} used the skyrmion logic gates of \cite{zhang2015magnetic} to implement full adder and multiplier units for in-memory CNN computation as well. 
\citet{chauwin2019skyrmion} presented a different approach by utilizing interactions between moving skyrmions to implement a full adder circuit. While these gate designs avoid MTJ write operations, they require duplication or multi-stage interactions of individual skyrmions for complex logic, incurring greater latency. Our full adder design has relatively fewer stages, while our write-shift transformation also avoids MTJ writes by using shifts instead. Consequently, our bit-serial 8-bit full adder outperforms those reported in \cite{Liu2017PIMCNN} (latency, energy, area) and \cite{chauwin2019skyrmion} (latency - energy and area not reported). However, we believe that skyrmion logic has much potential and could achieve highly efficient arithmetic circuits with further research. 

% SECTION ON IN-MEMORY ACCELERATORS
\textit{Deep Neural Network Acceleration.}
Among memory intensive big data applications, deep neural networks (DNN) have received much attention in the challenges faced to overcome the memory wall \cite{aimar2018nullhop, Chen2017, ESSA2020, kim2020exploiting, kwon2021heterogeneous, moons2016energy, zhu2020efficient}. Among these works, Eyeriss \cite{Chen2017} has been widely accepted as a benchmark for CNN accelerator designs. To tackle the memory bottleneck and costly DRAM transfers, the authors proposed a row stationary reuse strategy to maximize the spatial reuse of on-chip data. In addition, the authors employed run-length compression to compactly encode zero values in sparse activations. The combination of efficient data reuse and aggressive compression significantly reduces DRAM accesses. On the observation that CNN weights and activations are generally sparse (has many zero values), the works of \cite{aimar2018nullhop, moons2016energy, zhu2020efficient} studied sparsity encoding schemes to compress model data in a lossless manner. Sparsity encoding is orthogonal to our work; a carefully-designed scheme could further reduce racetrack memory writing in our accelerator, but requires additional encoding/decoding circuits. Among recent works, \citet{mei2021zigzag} presented a DNN architecture generator using a "Memory-Centric" design space, optimizing the memory hierarchy with dataflow mapping within user-defined constraints. These works fall within the conventional computing paradigm separating computing units and the memory hierarchy, which inevitably requires accesses to various levels of memory, including off-chip data transfers.

The in-memory computing paradigm has been a promising approach to reduce or even avoid inter-chip data transfers altogether during inference. Several works have implemented logic within SRAM \cite{Zhang2017IMCSRAM} or cache \cite{Eckert2018} using available CMOS technology, allowing the memory to behave as both regular storage or perform DNN acceleration in different modes. \citet{Eckert2018} repurposed the existing logic within memory banks to perform in-situ bit-serial computation. The translation of these infrastructures to emerging memory technologies can harness the advantages of new technologies for in-memory systems as well.

\textit{Racetrack-based CNN Acceleration.}
The mechanisms of racetrack memory require new optimization techniques for CNN computation. \citet{wang2020automatic} proposed an efficient data layout in racetrack memory to reduce the overhead of the sequential access mechanism during CNN inference. Their work co-optimized the scheduling of CNN operations and data locality for efficient access patterns, but use an out-of-memory CMOS-based Neural Processing Unit for computation. Conversely, our work exploits the intrinsic capabilities of racetrack memory for in-memory computation, avoiding off-chip data transfers. 
Several works \cite{Liu2017PIMCNN, wang2020automatic, Hao2014DWMImage} similarly integrate logic with racetrack memory for CNN inference. \citet{Hao2014DWMImage} demonstrated the mapping of image-processing machine learning algorithms to compute various CNN layers, while introducing the use of magnetic adder units for efficient integration. \citet{Liu2017PIMCNN} used a domain-wall based memory technology for storage, while combining skyrmion-based racetrack memory logic gates to implement full adders within in-memory computing units.
On the other hand, \citet{DWMAcc2019} proposed to utilize the shifting capabilities of racetrack memory to perform bit-serial shift-based multiplication, designing an accelerator targeted for processing shift-based DNNs. Our accelerator follows similar strategies, but provides the options of both regular binary multiplication and shift-based multiplication, demonstrating flexibility in the types of CNN models that can be accelerated.   

%% file: tex/7_Conclusion.tex
\section{Conclusion}
In this paper, we propose an accelerator architecture for the in-memory computation of CNN inferences using on racetrack memory (RM) technology. We design efficient RM-based arithmetic logic that reduces the number of resources and expensive write operations needed for computation. In addition, we propose an RM-based shift multiplier that exploits the intrinsic shifting capabilities of racetrack memory to efficiently accelerate shift-based multiplication. We integrate the designed computation logic with an RM-based memory system, and perform hardware-software co-design on the CNN model and accelerator system. Our strategies in reducing/transforming write operations, as well as integrating shift-based multiplication capabilities, allow the accelerator to achieve significant improvements to energy efficiency of up to 83.5$\times$, and up to 1.68$\times$ performance improvements (over 40\% latency reduction) compared to not-optimized 8-bit fixed point acceleration. 